\documentclass[twocolumn]{aastex631}
\usepackage{amsmath}
\usepackage{enumitem}  
\usepackage{makecell}

\usepackage{rotating}

\begin{document}

\title{An intermediate-mass stripped star with a dark companion in the Small Magellanic Cloud}

\email{lblomber@caltech.edu}

\author[0009-0004-4475-2121]{Lisa Blomberg}
\affiliation{Department of Astronomy, California Institute of Technology, 1200 East California Boulevard, Pasadena, CA 91125, USA}
\author[0000-0002-6871-1752]{Kareem El-Badry}
\affiliation{Department of Astronomy, California Institute of Technology, 1200 East California Boulevard, Pasadena, CA 91125, USA}
\author[0000-0002-9132-6561]{Peter Senchyna}
\affiliation{The Observatories of the Carnegie Institution for Science, 813 Santa Barbara Street, Pasadena, CA 91101, USA}
\author[0000-0002-6960-6911]{Ylva G\"{o}tberg}
\affiliation{Institute of Science and Technology Austria (ISTA), Am Campus 1, 3400 Klosterneuburg, Austria}
\author[0000-0003-0857-2989]{Bethany Ludwig}
\affiliation{Institute of Astronomy, KU Leuven, Celestijnenlaan 200D, 3001, Leuven, Belgium}
\author[0000-0001-7081-0082]{Maria R. Drout}
\affiliation{The David A. Dunlap Department of Astronomy \& Astrophysics, University of Toronto, 50 St. George Street, Toronto, Ontario, M5S 3H4, Canada}
\author[0000-0002-7846-9787]{Ana Bonaca}
\affiliation{The Observatories of the Carnegie Institution for Science, 813 Santa Barbara Street, Pasadena, CA 91101, USA}
\author[0000-0002-4462-2341]{Warren R.\ Brown} \affiliation{Center for
Astrophysics $\vert$ Harvard \& Smithsonian, 60 Garden Street, Cambridge, MA 02138 USA}
\author[0000-0001-7296-3533]{Tim Cunningham}
\affiliation{Center for
Astrophysics $\vert$ Harvard \& Smithsonian, 60 Garden Street, Cambridge, MA 02138 USA}
\author[0000-0001-6350-8168]{Brenna Mockler}
\affiliation{The Observatories of the Carnegie Institution for Science, 813 Santa Barbara Street, Pasadena, CA 91101, USA}
\author[0000-0001-7016-1692]{Przemek Mroz}
\affiliation{Astronomical Observatory, University of Warsaw, Al. Ujazdowskie 4, 00-478 Warszawa,
Poland}
\author[0000-0001-5207-5619]{Andrzej Udalski}
\affiliation{Astronomical Observatory, University of Warsaw, Al. Ujazdowskie 4, 00-478 Warszawa,
Poland}





\begin{abstract}
We present discovery of SMASH J004005$-$733351, an intermediate-mass stripped helium star in the Small Magellanic Cloud with a dark companion. The system was identified in a new optical color-selected survey for stripped stars and confirmed through spectroscopic follow-up. We measure an orbital period of $P_{\rm orb}=19.95\pm 0.002$ hours and a radial-velocity semi-amplitude of $K=91.7 ^{+3.2}_{-3.1}\,{\rm km\,s^{-1}}$. Simultaneous modeling of the optical spectrum and ultraviolet-to-optical spectral energy distribution yields a stripped star evolutionary mass $M_{\rm stripped} = 3.84 \pm 0.54\,M_\odot$, temperature $T_{\rm eff} = 68.3 \pm 6.4$ kK, surface gravity $\log g= 4.76 \pm 0.12$, luminosity $\log(L/L_\odot)= 4.28 \pm 0.13$, radius $R= 0.99 \pm 0.09\,R_\odot$, and an unusually high surface hydrogen abundance of $X_{\rm H, s}=0.63 \pm 0.04$. The companion is undetected spectroscopically, and the orbital solution and  stripped-star evolutionary mass imply a minimum companion mass of $M_2 \gtrsim 1.2\,M_\odot$. The absence of photometric variability from irradiation rules out a main-sequence companion. The companion is most likely a neutron star (NS), although a massive white dwarf or low-mass stripped star cannot be fully excluded. Binary evolution models predict that the stripped star will refill its Roche lobe after core helium exhaustion, likely producing a stripped or ultra-stripped supernova and ultimately a double NS binary if the system survives. SMASH J004005$-$733351 is among the first known stripped stars with a compact object companion. The system provides a new empirical benchmark for the late stages of binary interaction that lead to (ultra-)stripped supernovae and double NS binaries.

\end{abstract}


\section{Introduction} \label{sec:intro}
A star can lose its hydrogen envelope through binary interaction or stellar winds, leaving behind a stripped helium star. Binary evolution models predict envelope stripping for a wide range of stellar masses. Low-mass ($\lesssim 1\,M_\odot$) stripped stars are commonly referred to as hot subdwarf (sdB/O) stars \citep{Heber2009, Heber2016} and are formed via binary interactions \citep{Han2002, Han2003}, such as stable mass transfer or common-envelope evolution (CEE). Massive stripped stars ($\gtrsim 8\,M_\odot$) can appear as Wolf-Rayet (WR) stars \citep{Crowther2007, Shenar2026}. 
WR stars can be formed either by binary interactions, or by wind-driven mass loss, which is likely most important for the most massive stars. In contrast to their low- and high-mass counterparts, intermediate-mass stripped stars, though predicted for decades \citep[e.g.,][]{Kippenhahn_1967}, have only recently been observationally confirmed \citep{Drout_2023, gotberg_2023, ludwig_2026, mullerhorn_2026, Laroche_2026}; and the confirmed population remains small.

The majority of stripped stars formed via binary interactions are predicted to host main-sequence (MS) companions \citep[e.g.,][]{gotberg_2019, Zapartas_2026}. After the initially more massive star undergoes envelope stripping and core helium ignition, the less massive companion, which has a longer evolutionary timescale, remains on the MS. In some cases, after the stripped star has evolved into a compact object-- a white dwarf (WD), neutron star (NS), or black hole (BH), depending on its mass-- the system can undergo a second phase of binary interaction. This stage strips the originally less massive star, producing a stripped star with a compact object companion \citep[e.g., ][]{Tauris2015, gotberg_2020}. Such systems are particularly interesting because they represent the outcome of two distinct stages of stripping and are predicted to be direct progenitors of (ultra-) stripped-envelope supernovae \citep[e.g., ][]{smith_2011, Drout_2013, eldridge_2013, lyman_2016} and double compact-object binaries \citep[e.g.,][]{tauris_2017}.

However, intermediate-mass stripped stars with compact-object companions remain observationally elusive. PSR J1928+1815 was recently proposed to have an intermediate-mass helium stripped star \citep{Yang_2025}, but deep near-infrared imaging ruled out the stripped-star companion \citep{Nagarajan_2026}. A subset of the confirmed intermediate-mass stripped stars in the MCs identified by \cite{Drout_2023} exhibit SB1-type spectra, with no evidence of a luminous companion. \cite{Laroche_2026} confirmed the binary nature of several of these systems, but the nature of their companions remains uncertain. An additional system from the sample of \cite{Drout_2023}, not included in the analysis of \cite{Laroche_2026}, was reported by \citet{Ludwig_thesis} to host a compact-object companion.  A full analysis of that system will be presented in Ludwig et al. (in preparation).

Confirming compact companions to intermediate-mass stripped stars is important for testing binary evolution models and constraining the formation channels of ultra-stripped supernovae (SNe) and gravitational-wave sources. Currently, there are no empirical constraints on the fraction of binaries that survive the second stripping phase \citep[e.g.,][]{Wu_2020, gotberg_2020}, and observations of such systems are needed to constrain these models.

In this paper, we present a newly discovered intermediate-mass stripped star in a binary with a dark companion: SMASH J004005-733351 (hereafter J0040-7333). We identified J0040-7333 in the Small Magellanic Cloud (SMC) as part of a survey for intermediate-mass stripped stars using an optical blue-excess selection. We conduct spectroscopic and photometric analysis of this system and find that: (i) the system is in a relatively compact orbit with a period of $\sim 20$ hours; (ii) its spectra and broadband spectral energy distribution (SED) are dominated by an intermediate-mass stripped star; (iii) the binary mass function implies a minimum companion mass of $M_{\rm comp}\gtrsim1.2\,{\rm M_\odot}$; (iv) light curve modeling rules out a low-mass MS companion, implying instead a compact-object or stripped-star companion; and (v) evolutionary models favor a NS.

The rest of the paper is organized as follows. In Section \ref{sec: discovery}, we discuss the optical search for stripped stars and the discovery of J0040-7333. In Section \ref{sec: data}, we outline the data used for the analysis of the system, including optical and ultraviolet (UV) photometry, optical spectra, and an optical light curve. In Section \ref{sec: analysis}, we describe our analysis methods and present our results, including the orbital solution and the physical properties of the stripped star and its companion. In Section \ref{sec: nature of the system}, we discuss possible formation channels and future evolution, and discuss the nature of the companion. In Section \ref{sec: comparsion}, we compare the newly discovered system to related populations. Finally, we summarize our results in Section \ref{sec: conclusions}.

\section{Discovery}\label{sec: discovery}
During their long-lived core-helium-burning phase, most stripped stars are predicted to be hot and compact, making them UV-bright compared to MS stars of similar luminosity \citep{Gotberg2017, gotberg_2018}. Motivated by this prediction, \cite{Drout_2023} carried out a UV-excess survey in the Magellanic Clouds (MCs) to search for intermediate-mass stripped stars. Their search yielded 16 spectroscopically-confirmed stripped stars (4 in the SMC and 12 in the LMC). However, population synthesis models predict that many hundreds more intermediate-mass stripped stars should exist in the MCs \citep[e.g.,][]{Yungelson_2024, hovis_2025}, indicating that the initial survey likely only discovered the tip of the iceberg. 

A complete list of photometric candidates identified using the general method described in \cite{Drout_2023} was published in \cite{ludwig_2026}. This revealed an additional 820 photometric candidates. This sample of candidates, however, remains limited by the spatial resolution and coverage \citep{Blomberg_2026} of the \textit{Swift} Ultraviolet Survey of the Magellanic Clouds \citep[SUMaC;][]{Hagen_2017}. To identify stripped star candidates inaccessible to the \cite{ludwig_2026} search, we carried out a complementary search using optical photometry.

We found that the spectroscopically-confirmed stripped stars from \cite{Drout_2023} exhibit blue excess in the optical $u$-band data from the Survey of the Magellanic Stellar History \citep[SMASH; ][]{smash}, suggesting that similar stripped stars can potentially be selected based on optical photometry alone. We show this in a color-magnitude diagram (CMD) in Figure \ref{fig: selection function}. Motivated by this, we conducted an optical blue-excess survey using SMASH photometry. The UV-excess search of \cite{Drout_2023} and \cite{ludwig_2026} used UV photometry from the SUMaC, obtained with the \textit{Swift} Ultraviolet/Optical Telescope \citep[UVOT;][]{swift_UVOT}, and optical photometry from the Magellanic Cloud Photometric Survey \citep[MCPS;][]{MCPS}. Although restricting the search to optical wavelengths limits our sensitivity to only the most extreme UV-excess systems, the SMASH survey has two advantages: (i) larger footprint, and (ii) higher spatial resolution, with PSF FWHM of $\sim1''$ compared to the $\sim2.4''$ PSF of \textit{Swift}-UVOT. This allows us to search for intermediate-mass stripped stars in regions previously unexplored and probe crowded fields where the existing UV photometry is affected by blending. In particular, the newly discovered stripped star we present in this paper (J0040-7333) was not identified in previous UV-excess surveys of the Magellanic Clouds \citep{Drout_2023, ludwig_2026} because it lies outside the SUMaC footprint.

\subsection{Candidate selection}\label{sec: selection function}
\begin{figure*}[htb!]
    \centering
    \includegraphics[width=1\linewidth]{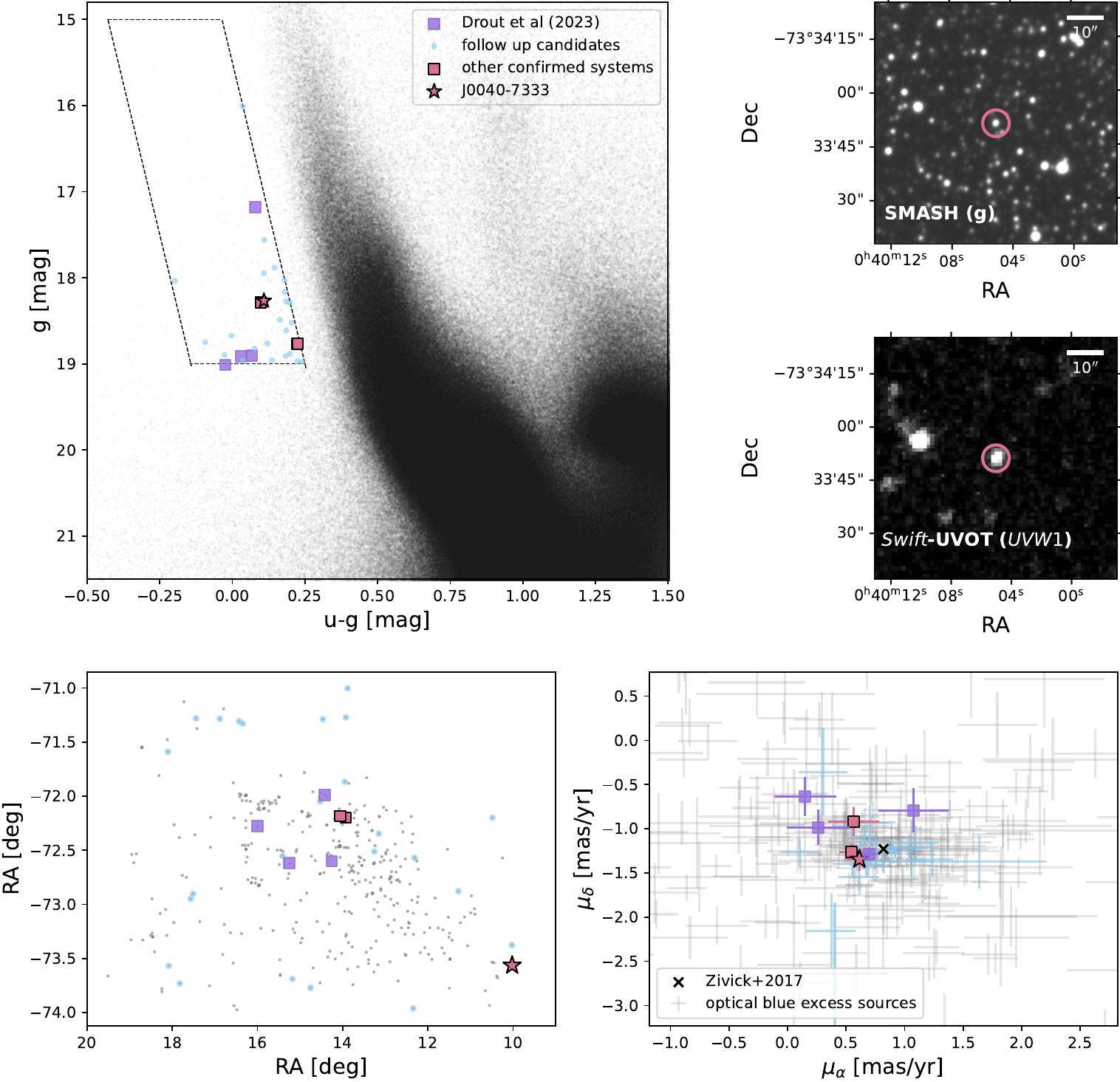}
    \caption{Top left: SMASH CMD of the SMC. Black markers show all SMC point sources in the SMASH catalog with $u-$ and $g-$band photometry. Purple squares show the four previously confirmed intermediate-mass stripped stars \citep{Drout_2023}. Light blue points show the 27 stripped-star candidates selected using the selection criteria described in Section \ref{sec: selection function}. The dashed lines represent the optical color cut. The three new stripped stars confirmed by our search are shown in pink; the star marks J0040-7333. 
    Top right: optical and UV images from SMASH and \textit{Swift}-UVOT. J0040-7333 is marked with a pink circle. Bottom: spatial distribution (left) and \textit{Gaia} proper motions (right).}
    \label{fig: selection function}
\end{figure*}

We carried out a search for intermediate-mass stripped stars in the SMC. We identified candidates in SMASH with $(u-g)$ colors and $g$-band magnitude similar to known intermediate-mass stripped stars. We applied a series of quality cuts to this population and selected 27 of the most promising candidates for spectroscopic follow-up (see Figure \ref{fig: selection function}):
\begin{itemize}
    \item We required the candidate to have $(u-g)$ color and $g$-band magnitudes (AB) consistent with the spectroscopically confirmed intermediate-mass stripped stars: $15 < g< 18.9 $, $14 (u-g) + 15.5 < g < 14 (u-g) + 21 $. We checked all point sources in the SMASH catalog with $9 < {\rm RA \,(deg)} < 24$ and $-74 < {\rm Dec \,(deg)} < -70$ and identified 441 candidates.
    \item Further, we rejected candidates with \textit{Gaia} \citep{Gaia, Gaia_dr3} parallaxes and proper motions inconsistent with SMC membership. We required \texttt{parallax\_over\_error} (i.e., the measured parallax of a source divided by its uncertainty) to be $< 3$. For SMC sources, we expect this quantity to be small, because at SMC distances the true parallax ($\approx 0.02$ mas) is smaller than the Gaia parallax uncertainties for faint sources. We also required proper motions within $2\sigma$ of the bulk motion of the SMC \citep{zivick_2017}. After rejecting possible foreground and background sources, we were left with 238 candidates.
    \item Finally, we selected sources with the highest-quality SMASH photometry. We required each source to pass two SMASH quality flags: \texttt{chi}$<3$ and $-1<$ \texttt{sharp} $<1$. These parameters are outputs of the DAOPHOT PSF-fitting routine reported in the SMASH catalog. The \texttt{chi} parameter measures the scatter in the PSF-fit residuals. The \texttt{sharp} parameter measures the relative width of the best-fit PSF profile compared to the expected stellar PSF. This rejects an additional 151 candidates. We visually inspected the $g$- and $u$-band SMASH images of the remaining candidates and rejected sources with potentially contaminated photometry due to crowding. We selected 27 candidates with the best photometry.
\end{itemize}

\subsection{Spectroscopic follow-up and confirmation}
Most core helium burning stripped stars are predicted to be significantly hotter than typical MS stars of similar luminosity with effective temperatures reaching $T_{\rm eff}\gtrsim 40-100\,{\rm kK}$ \citep{gotberg_2018}. As a result, they are expected to show strong He\,{\sc ii} absorption lines (e.g., He\,{\sc ii} $\lambda\lambda 4200,4542,4686,5412$) which are not expected for MS stars of similar luminosity. These unique spectral features allow systems that contain stripped stars to be identified with a single epoch of spectroscopy, if their photometry is dominated by the stripped star.
We note that intermediate-mass stripped stars usually do not show He\,{\sc ii} emission, unlike their higher-mass counterparts (i.e., WR stars) because they are not sufficiently luminous to drive strong winds that produce detectable wind lines in the optical. This is especially true in low-metallicity environments, such as the SMC \citep[e.g.,][]{Vink2017, gotberg_2023}. However, see \citealt{mullerhorn_2026} for a recently discovered Galactic stripped star that shows emission in He\,{\sc ii}.

We conducted spectroscopic follow-up of the 27 candidates using the Magellan Echellette (MagE; \citealt{MagE}) spectrograph. Of these, three appear to have spectra consistent with spectral models of intermediate-mass stripped stars. One source has optical spectra dominated by the stripped star component (J0040-7333), while the other two exhibit composite spectra containing features of both a stripped star and a luminous companion, similar to the `Class 2' systems presented in \cite{Drout_2023}. In this paper, we focus on J0040-7333. A full description of the survey and analysis of the other stripped stars will be presented in future work.

\section{Data}\label{sec: data}
We use a combination of UV and optical photometry, an optical light curve, and multi-epoch optical spectra to analyze J0040-7333.

\subsection{Photometry}\label{sec: photometry}
J0040-7333 has been observed by several UV and optical photometric surveys over the past few decades. We summarize the photometric data used for our analysis in Table \ref{table: photometry}.
\begin{table*}[htb!]
    \centering
    \begin{tabular}{c c c}
    \hline
    instrument / survey&bandpass & magnitude \\
    \hline
     UVIT & $FUV$& $17.150\pm0.131$ (AB)\\
     \textit{Swift}-UVOT& $UVW1$ & $17.55\pm0.02$ (AB)\\
     MCPS & $U$, $B$, $V$, $I$& $17.344 \pm 0.048$, $18.308\pm0.036$, $18.455 \pm0.042$, $18.781\pm0.048$ (Vega) 
     \\
     SMASH & $u$, $g$, $r$, $i$, $z$& $18.376\pm0.009$, $18.267\pm0.005$, $18.775\pm0.005$, $19.068\pm0.007$, $19.468\pm0.013$ (AB)\\
     OGLE & $V$, $I$ & $18.556 \pm 0.019$, $18.789 \pm 0.048$ (Vega)\\ 
    \hline \\
    \end{tabular}
    \caption{UV (UVIT, \textit{Swift-UVOT}) and optical (MCPS, SMASH, OGLE) photometry for J0040-7333. See Section \ref{sec: photometry} for a description of each survey. The OGLE magnitudes are the mean calibrated magnitudes from archival OGLE images, with uncertainties given by the standard deviation of the individual measurements after outlier rejection.}
    \label{table: photometry}
\end{table*}

\subsubsection{Survey of the Magellanic Stellar History (SMASH)}
The Survey of the Magellanic Stellar History (SMASH; \citealt{smash}) imaged the MCs using the Dark Energy Camera (DECam; \citealt{DeCam}). The catalog provides optical photometry in $ugriz$ bands down to  $\sim 24$ mag.

\subsubsection{Magellanic Cloud Photometry Survey (MCPS)}
The Magellanic Cloud Photometric Survey (MCPS; \citealt{MCPS}) was conducted using the Great Circle Camera \citep{zaritsky_1996} on the $1\,{\rm m}$ Swope telescope at Las Campanas Observatory. The catalog provides photometry in $UBVI$ bands down to $U \sim 21.5$, $B \sim 23.5$, $V \sim 23$ and $I \sim 22$ mag.

\subsubsection{\textit{Swift}-Ultraviolet/Optical Telescope (UVOT)}\label{sec: swift}
The Ultraviolet/Optical Telescope (UVOT) is a UV/optical instrument onboard the \textit{Swift} satellite \citep[UVOT; ][]{Gehrels_2004, swift_UVOT, Poole_2008}. 
We identify 45 archival \textit{Swift}-UVOT $UVW1$-band observations of SMC fields containing J0040-7333. Of these, six images have problems with their WCS solutions. We performed PSF photometry on the remaining images using the photometric pipeline developed by \cite{ludwig_2026}, which uses \textit{The Tractor} \citep{lang_2016}.

\subsubsection{Ultraviolet Imaging Telescope (UVIT)}\label{sec: uvit}
The Ultraviolet Imaging Telescope (UVIT; \citealt{UVIT}) is a UV imager on the Astrosat satellite \citep{astrosat}.  
Several surveys of the SMC have been carried out \citep[e.g., ][]{Hota_smc_shell, Hota_smc}. We downloaded the level-2 image named ``SMC-2'' from the Astrosat-UVIT archive\footnote{\url{https://astrobrowse.issdc.gov.in/astro\_archive/archive/Home.jsp}}, 
and performed PSF-photometry using \textit{The Tractor}. We follow the photometric pipeline developed for the \textit{Swift}-UVOT images by \cite{ludwig_2026} with several modifications. Because no theoretical PSF model is available for UVIT, we build our own empirical PSF model by taking bright and isolated stars from the ``SMC-2'' image and stacking their cutouts using the \texttt{EPSFBuilder} routine in \texttt{photutils}. 
We also modify the background subtraction to be less aggressive by relaxing the sigma-clipping criteria, since a large fraction of pixels in the UVIT image have zero or near-zero counts. We also relax the position prior to account for the asymmetric shape of the empirical UVIT PSF. Finally, we convert the measured detector count rates to magnitudes using the zero points for the appropriate UVIT filter \citep{Tandon_2020}:
\begin{equation}
    m = -2.5\log_{10}(C) + 16.274,
\end{equation}
where $C$ is the count rate given in counts per second measured by \textit{The Tractor}.

\subsubsection{Optical Gravitational Lensing Experiment (OGLE)}
The Optical Gravitational Lensing Experiment (OGLE; \citealt{ogle_1, ogle_ii, ogle_iv}) is a long-running optical time-domain survey conducted with the 1.3 m Warsaw Telescope at Las Campanas Observatory. 
We obtained a long-baseline OGLE light curve for J0040-7333. The $I-$band light curve consists of 3860 epochs taken between 2001 and 2025. The median photometric uncertainty is 0.03 mag per observation.

\subsection{Spectroscopy}\label{sec: spectral data}
We obtained 16 spectra across 10 nights using two different instruments\footnote{All reduced spectra are available here: \url{https://www.dropbox.com/scl/fo/b0z3nrefb316onwau0x88/AMEQz8rWV1KeCLMnY6tFEyY?rlkey=4ekj5cehr1romddaorcb4flra&st=zy0n2r99&dl=0} (to be moved to Zenodo when the paper is accepted).}: 12 epochs with the Magellan Echellette Spectrograph \citep[MagE; ][]{MagE} on the Magellan Baade Telescope at Las Campanas Observatory, and 4 epochs with the Gemini Multi-Object Spectrograph (GMOS-S; \citealt{GMOS}) on the 8.1 m Gemini South Telescope at the Gemini South observatory. All observations are summarized in Table \ref{table: RVs} in Appendix \ref{appendix: rv table}; Figure \ref{fig: spectra all} shows all follow-up spectra. Below, we describe the instrumental setups and data reduction methods used for the follow-up observations.

\subsubsection{Magellan Echellette Spectrograph (MagE)}\label{sec: mage}
We used the 0.7 and 0.85 arcsec slits with $1\times1$ binning for all MagE observations, with exposure times of $600-1800\,\rm{s}$ depending on observing conditions. The resulting spectra have resolution $R\sim 4100$ over a wavelength range of $\lambda \sim 3500 - 7000\,\rm{\AA}$. For each science frame, we subsequently obtained an arc exposure after each observation, which was used by the reduction pipeline for wavelength calibrations. We reduce the data using {\ttfamily PypeIt} \citep{Prochaska_2020a, Prochaska_2020b}. We normalize each spectrum by fitting a polynomial to each echelle order. We reject $2\sigma$ outliers during the fit to mask absorption lines. We repeat this process iteratively until we obtain a stable normalized spectrum in each order. 
After clipping the low-sensitivity edges, we combine the orders by interpolating them onto a common wavelength grid and averaging the overlapping regions, weighting each pixel by the inverse variance.

\subsubsection{Gemini Multi-Object Spectrographs (GMOS-S)}
We used the B480 grating on GMOS-S with a central wavelength of 470 nm. Using the $0.75''$ slit and $1\times1$ binning with 1200 s exposures, we obtain spectra with resolution of $R\sim 900$ for a wavelength range of $\lambda \sim 380 - 600$ nm. Each science observation is supplemented with an arc exposure, which is used for wavelength calibrations. We reduce the data using the standard configuration of the DRAGONS pipeline \citep{dragons, DRAGONS_software} and normalize them using the polynomial-fitting method described in Section \ref{sec: mage}. The GMOS-S spectra contain gaps in wavelength coverage due to space between CCD detectors. We choose a central wavelength such that most of the spectral features of interest fall within the regions covered by the detectors.

\section{Analysis}\label{sec: analysis}

\subsection{Constraining the orbit}
We measure radial velocities (RVs) using the 16 epochs of follow-up spectra. We fit the resulting RVs with a Keplerian orbit.

\subsubsection{RV measurements}\label{sec: RV measurements}
We measure RVs for each of the 16 spectra by cross-correlating the He\,{\sc ii} and Balmer absorption lines with a stripped-star spectral model \citep{gotberg_2023_models}. We perform the $\chi^2$-fit using the model with $T_{\rm eff} = 70$ kK, $\log(g / [\rm{cm s}^{-2}]) = 5.0$, and $X\rm{H}_{\rm s} = 0.6$, which provides a reasonable match to the observed spectra (we carry out a spectral fit in Section \ref{sec: properties of stripped star}). After convolving the model to match the instrument resolution ($R = 4100$ for MagE and $R = 900$ for GMOS-S), we generate a grid of Doppler-shifted spectra with RVs ranging from 40 to 350 km/s. We select 40 $\rm{\AA}$-wide windows centered on several diagnostic lines (He\,{\sc ii} $\lambda\lambda 4542, 4686, 5411$ and He\,{\sc ii}/Balmer blends at He\,{\sc ii} $\lambda 4100$/H$\delta$, He\,{\sc ii} $\lambda 4339$/H$\gamma$, and He\,{\sc ii} $\lambda 4860$/H$\beta$) and compute the $\chi^2$ value for each RV. We visually inspect each line and exclude any problematic lines (e.g., lines affected by problematic background subtraction or located near defective detector pixels that mask a significant portion of the line profile.)

During the initial fitting stages, we tested different combinations and numbers of spectral lines. We restricted our fitting to $4000\,\rm{\AA} \lesssim \lambda \lesssim 5500 \,\rm{\AA}$ because this wavelength range has the highest SNR and contains the most prominent optical features of stripped stars. Although the measured RVs vary somewhat across lines, the RVs measured from different lines are generally consistent within their uncertainties for well-sampled and unproblematic lines; see Appendix \ref{appendix: rv table} for a comparison of RVs measured from pure He\,{\sc ii} lines and He\,{\sc ii}/Balmer blends. We found that the visual inspection step is necessary to exclude lines where the substantial portion of the line profile is affected by artifacts, preventing a reliable fit and RV measurement.

\begin{figure}
    \centering
    \includegraphics[width=1\linewidth]{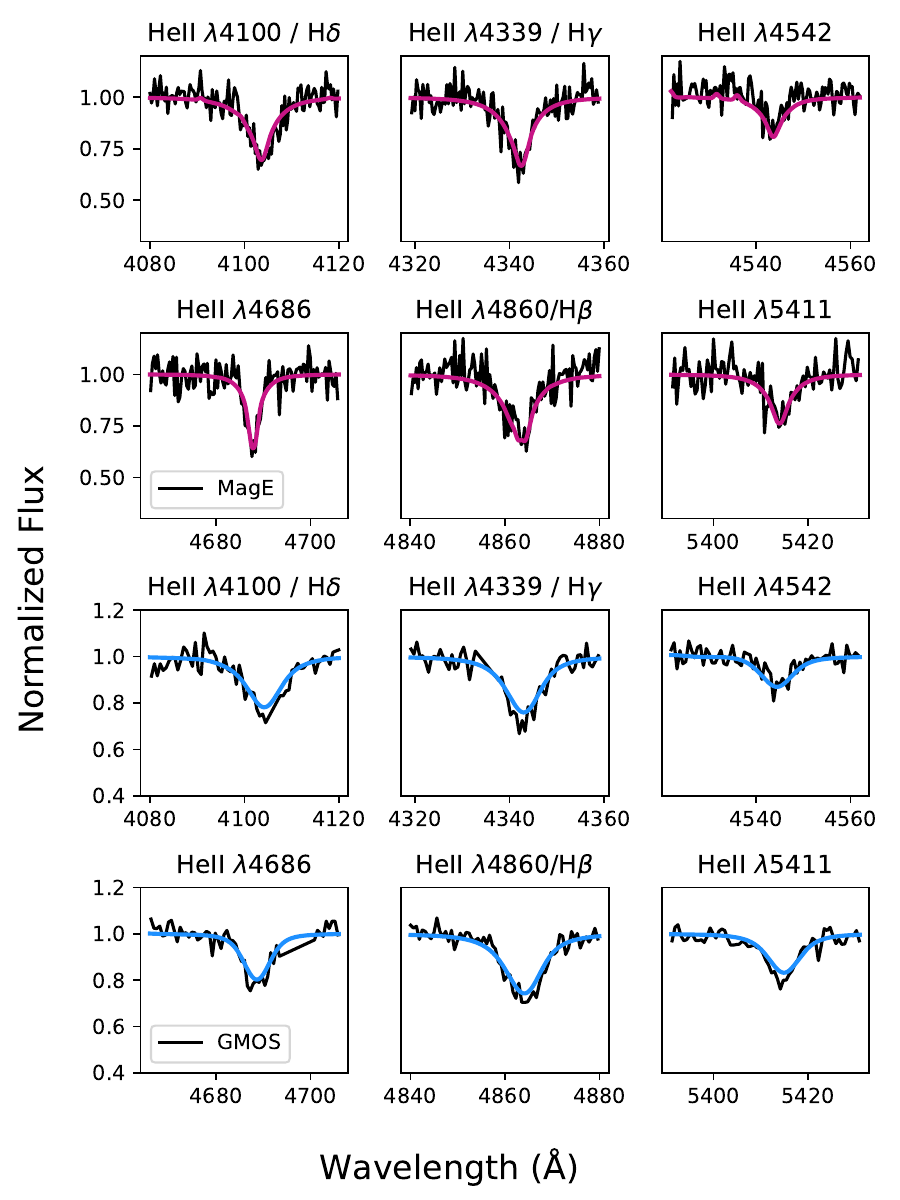}
    \caption{Example RV fits to the observed MagE (top; BJD = 2460941.76) and GMOS (bottom; BJD = 2460994.62) spectra. We cross-correlate synthetic spectra with a set of He\,{\sc ii} and Balmer absorption lines, and determine the best RV using $\chi^2$ minimization (Section \ref{sec: RV measurements}). We show the synthetic spectral models shifted to the best-fit RVs in magenta (MagE; $R = 4100$) and in blue (GMOS; $R=  900$).}
    \label{fig: RV fits}
\end{figure}

We sum the $\chi^2$ contributions from all lines to compute the total $\chi^2$, $\chi_{\rm tot}^2$, for each spectrum, and adopt the velocity corresponding to the minimum $\chi^2_{\rm tot}$ as the best-fit RV. At this stage, we also account for barycentric corrections. We show some example fits in Figure \ref{fig: RV fits}.
We estimate the uncertainties by identifying the range of RVs consistent with the best-fit value at the $1\sigma$ confidence level and round to the nearest integer. We include a $\pm 3\,{\rm km/s}$ systematic uncertainty, added in quadrature with the statistical uncertainty from the $\chi^2$ minimization, to account for potential biases introduced by fixing the spectral model. We report all RV measurements, observation times, and the set of lines used in the fitting procedure, in Table \ref{table: RVs} in Appendix \ref{appendix: rv table}.

\subsubsection{Constraining the orbital solution}\label{sec: orbital solution}

We use \texttt{the joker} \citep[v.1.3.1;][]{joker_2017}, a rejection sampling code designed for binary RV data, coupled with a standard Markov Chain Monte Carlo (MCMC) method to constrain the orbital solution for J0040-7333. We fit for the six free parameters of a Keplerian RV orbit: orbital period ($P_{\rm orb}$), RV semi-amplitude ($K$), systemic velocity ($v_0$), eccentricity ($e$), argument of periastron ($\omega$), and a phase parameter ($\Delta t$). Here, $\Delta t = t_0 - t_{\rm obs, min}$, where $t_{\rm obs, min}$ (BJD = 2460849.808) is the time of the first observation and $t_0$ is the periastron time.

For the orbital period, we use a Jeffreys prior \citep{Jeffreys_1946}, $p(P_{\rm orb}) \propto \frac{1}{P_{\rm orb}}$, with a range of $0.5-5$ days. We select this range after an initial coarse sampling with orbital periods ranging from 1 hour up to 100 days, which showed that the RVs are only consistent with short orbital periods ($\lesssim 5,{\rm days}$). For the RV semi-amplitude, we use a Gaussian prior with $\sigma_{K} = 50$ km s$^{-1}$ centered around $K=0$. For the systemic velocity, we adopt a Gaussian prior with $\mu_{v_0} = 160$ km s$^{-1}$ and $\sigma_{v_0} = 20$ km s$^{-1}$, consistent with the bulk motion of the SMC. For the eccentricity, we assume a uniform prior between 0 and 1. We adopt uniform priors for all phase terms.

\begin{figure*}
    \centering
    \includegraphics[width=\linewidth]{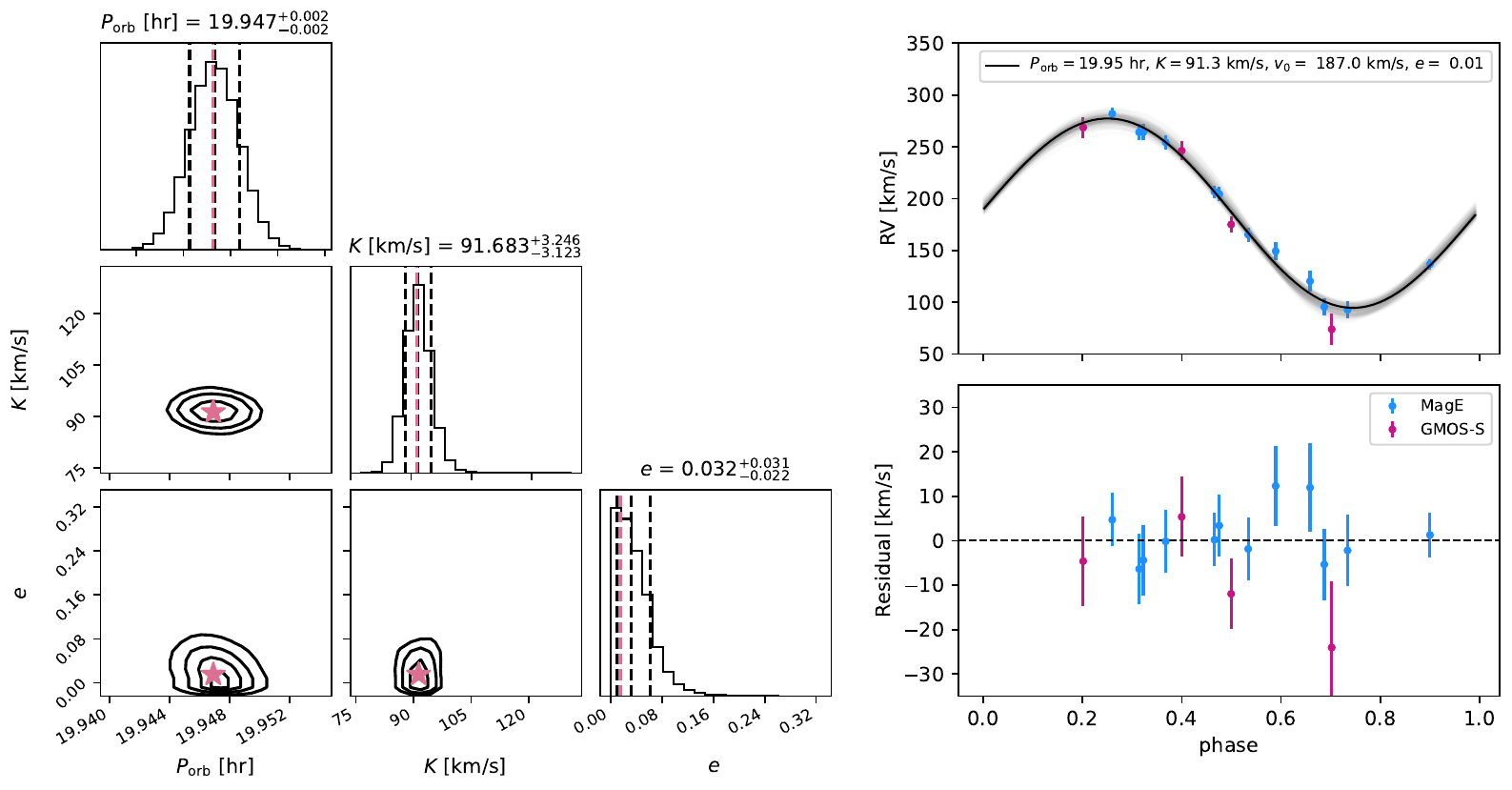}
    \caption{Left: corner plot showing the orbital solution fit posteriors. Pink dashed lines and stars indicate the MAP values. Right: phase-folded RV curve (top). Blue and magenta markers show RVs measured with MagE and GMOS, respectively. The black line represents the best-fit model. Gray lines show 100 random samples from the MCMC posterior distribution. Bottom panel shows the residuals of each RV epoch relative to the MAP best-fit orbital model. Here, we define the orbital phase such that phase = 0 corresponds to the best-fit model crossing the systemic velocity with increasing RV.}
    \label{fig: orbital solution}
\end{figure*}

We first draw $N=50,000,000$ prior samples using \texttt{the joker} and retain them with probability proportional to their likelihood. This results in a single posterior mode with $P_{\rm orb} \approx 20$ hrs, indicating that no other orbital period provides a comparably good orbital solution to the observed RVs. We then use the maximum-posterior sample to initialize a set of MCMC walkers. We define Gaussian priors centered on these values with the following standard deviations: $\sigma_{\rm P_{orb}} = 10$ days, $\sigma_K = 50\,{\rm km \,s^{-1}}$, and $\sigma_{v_0} = 50\,{\rm km \,s^{-1}}$. We sample from the posterior using \texttt{emcee} \citep{emcee_2013} with 50 walkers, each taking 5,000 steps after a burn-in of 100 steps. We report both the median and maximum a posteriori (MAP) constraints in Table \ref{table: orbital solutions}. 
\begin{table}[htb!]
    \centering
    \begin{tabular}{c | c c | c c }
    \hline
     & \multicolumn{2}{c|}{$e=0$}
     & \multicolumn{2}{c}{$e \sim \mathrm{U}(0,1)$} \\     & median & MAP & median& MAP \\
    \hline \hline
     $P_{\rm orb} $ [hr] &$19.947^{+0.002}_{-0.002}$ & 19.948 & $19.947^{+0.002}_{-0.002}$ & 19.947\\
     $K$ [km/s] &$91.2^{+2.4}_{-2.4}$ & 90.9 & $91.7^{+3.2}_{-3.1}$ & 91.3\\
     $v_0$ [km/s] & $186.1^{+2.6}_{-1.7}$& 186.8 & $187.0^{+2.6}_{-2.5}$ & 187.0\\
     $\omega$ [rad] & - & - & $2.512^{+1.003}_{-0.956}$ & 2.407\\
     $e$ & 0 & 0 & $0.032^{+0.031}_{-0.022}$ & 0.015\\
     $\Delta t$ [d] & $-0.005^{+0.006}_{-0.005}$ & $0.006$ & $0.321^{+0.131}_{-0.127}$& 0.307 \\
     $f_m$ [$\rm{M_\odot}$] & $0.066^{+0.005}_{-0.005}$ & 0.064& $0.066^{+0.007}_{-0.007}$& 0.066\\
    \hline
    \end{tabular}
    \caption{Median and MAP constraints of the orbital parameters for a circular and an eccentric orbit (Section \ref{sec: orbital solution}).}
    \label{table: orbital solutions}
\end{table}
Figure \ref{fig: orbital solution} shows the resulting corner plot and the orbital fit.

We find that J0040-7333 has a compact orbit with an orbital period of $P_{\rm orb} = 19.95\,\rm{hours}$ and a RV semi-amplitude of $K \approx 91 \,\rm{km \,s^{-1}}$. We measure a systemic velocity of $v_0 \approx 187 \,\rm{km \,s^{-1}}$, which is consistent with the bulk motion of the SMC \citep[e.g.,][]{Evans_2008}. The inferred eccentricity ($e = 0.032 ^{+ 0.031}_{-0.022}$) is small and consistent with $e=0$ within $2\sigma$. To further test the circular nature of the orbit, we repeat \texttt{the joker} and MCMC sampling with eccentricity fixed to zero (see Table \ref{table: orbital solutions} for results). We find that the  best-fit orbital parameters, including $P_{\rm orb}$, $K$, and $v_0$, obtained from both eccentric and circular orbit fits are consistent.

\subsection{Constraining the properties of the stripped star}\label{sec: properties of stripped star}
We simultaneously fit the normalized stacked spectrum (Section \ref{sec: spectral features}) and broadband SED to constrain the properties of the stripped star. We treat J0040-7333 as an SB1 system and assume that the optical and UV flux is dominated by the stripped star, with negligible contribution from the companion.

\subsubsection{Evolutionary models, synthetic spectra, and synthetic photometry}\label{sec: models}
\textit{Evolutionary models:} we use the evolutionary models for intermediate-mass stripped stars at SMC metallicity (Z = 0.002) presented in \cite{gotberg_2018}. These models were computed using the Modules for Experiments in Stellar Astrophysics \citep[MESA; ][]{paxton_2010, paxton_2013, paxton_2015, Paxton_2018, paxton_2019, jermyn_2023}. The tracks assume Case B mass transfer, which is expected to be the most common channel for producing stripped stars \citep{Sana_2012}, though theoretical models predict that stripped stars can also result from Case C mass transfer \citep[e.g., ][]{Ercolino_2025}. We define the stripped-star lifetime as the core-helium-burning phase, corresponding to $X_{\rm He, c} = 0.001-0.95$. The binary physics assumed in these models produce stripped stars with surface hydrogen abundances of $X_{\rm H,s} \approx0.35-0.4$. We interpolate between tracks with initial masses of $M_i = 9.00 - 16.44 \,{\rm M_\odot}$ to construct continuous evolutionary tracks, including relevant stellar properties (e.g., $T_{\rm eff}$, $L$, $M_{\rm stripped}$) as a function of evolutionary phase.

\textit{Spectral models:} we interpolate a grid of spectral models of intermediate-mass stripped stars \citep{gotberg_2023_models} to produce synthetic spectra for any given combination of $T_{\rm eff}$, $\log(g)$, and $X_{\rm H, s}$. \cite{gotberg_2023_models} provide models spanning $T_{\rm eff} \approx 30-150\,{\rm kK}$, $\log(g) \approx 4-6$, and $X_{\rm H, s} = 0.01, 0.1, 0.3, 0.5, 0.7$. The parameter space is not uniformly sampled, with models absent in some regions, particularly at high $T_{\rm eff}$ and low $\log(g)$.
After resampling $F_{\lambda}$ and $F_{\rm norm}$ onto a finely sampled common wavelength grid with a spacing of approximately $0.2\,{\rm \AA}$, we linearly interpolate the spectral model grid in $T_{\rm eff}$, $\log g$, and $X_{\rm H,s}$ using \texttt{LinearNDInterpolator} from \texttt{SciPy}. We do not extrapolate beyond the available models. 
These spectral models are parameterized solely by the stellar atmospheric properties ($T_{\rm eff}$, $\log(g)$, and $X_{\rm H, s}$) and are not associated with a particular evolutionary stage.

\textit{Synthetic photometry:} we integrate the spectral flux density, $F_{\lambda}(\lambda)$, from our interpolated grid of spectral models with the appropriate UV and optical transmission curves \citep{Rodrigo_2012, Rodrigo_2020, rodrigo_2024} to compute the synthetic flux for each bandpass. In our fitting procedure (Section \ref{sec: mcmc setup no evol} and \ref{sec: mcmc sampling setup}), we scale these synthetic fluxes for a given stellar radius to obtain the observed fluxes. We then convert the flux to an apparent magnitude assuming a distance of 62.44 kpc to the SMC \citep{Graczyk_2020}. To account for reddening, we estimate $A_\lambda$ for each bandpass from $A_V$ using the SMC extinction curve \citep{Gordon_2024}, adopting $R_V = 3.13$. We then apply $A_\lambda$ to the synthetic photometry derived from the spectra to estimate the reddened magnitudes.\footnote{For simplicity, we adopt the SMC extinction curve to model the total extinction toward J0040-7333. This neglects differences between the Milky Way and SMC extinction curves for the foreground dust. By integrating the \cite{edenhofer_2024} 3D dust map out to 2 kpc, we estimate a Milky Way foreground contribution of $A_V \sim 0.12$ mag.}

\subsubsection{Spectral features}\label{sec: spectral features}
We shift the 12 MagE spectra to the stripped star's rest frame, interpolate each spectrum onto a common wavelength grid, and coadd them into a single spectrum. We take a weighted average of the spectra, weighting each pixel by its inverse variance. We show the full coadded spectrum in Figure \ref{fig: spectra}. The stacked spectrum has a signal-to-noise ratio (SNR) of $\sim45$.

As discussed in \cite{gotberg_2023} and demonstrated in the lower left panels of Figure \ref{fig: spectra}, several absorption and emission features are sensitive to effective temperature. For example, the He\,{\sc i} $\lambda4471$ and He\,{\sc i} $\lambda5876$ absorption lines are only present for $T_{\rm eff} \lesssim 50 \,\rm{kK}$ and $T_{\rm eff} \lesssim 60- 70\,\rm{kK}$, respectively, and N\,{\sc v} $\lambda4604$ absorption appears only for stripped stars with $50 \,\rm{kK} \lesssim T_{\rm eff} \lesssim 90 \,\rm{kK}$.

We find that the spectrum of J0040-7333 does not show any He\,{\sc i} $\lambda\lambda4471, 5876$ features but shows absorption in N\,{\sc v} $\lambda 4604$. When fitting for the stellar parameters of the stripped star, we use the presence of N\,{\sc v} $\lambda 4604$ to construct a prior on the atmospheric parameters.

\subsubsection{Joint SED and spectral fits}\label{sec: mcmc setup no evol}
We use the stacked MagE spectrum for spectral fitting, and use photometry from MCPS, \textit{Swift}-UVOT, and UVIT for the SED fitting.\footnote{We do not use photometry from SMASH because we find a systematic offset between the observed and predicted colors of MS stars. Especially in the $u-$band, we find significant offsets which suggest calibration problems. These discrepancies may be related to color-dependent calibration systematics discussed in \citet{smash}.} We simultaneously fit the stacked spectra and SED and fit for four parameters: $T_{\rm eff}$, $\log(g)$, $X_{\rm H, s}$, and $R$.

We do not attempt to simultaneously fit for extinction, but instead perform the fit over a range of $A_V$ values. We find that given the relatively small number of data points in the SED relative to the normalized spectrum, leaving $A_V$ as a free parameter drives the joint fit to a solution with high $A_V$ that provides a poor match to the observed SED. This formally maximizes the likelihood because a higher $T_{\rm eff}$ provides a marginal improvement to the spectral fit but is unlikely to be reliable (see Appendix \ref{apdx: free AV} for further discussion). We therefore separately estimate the reddening for J0040-7333 by comparing the observed UV and optical colors to synthetic colors of stripped stars with properties roughly matched to those of J0040-7333 ($T_{\rm eff} = 60-70\,{\rm kK}$, $\log(g) = 4.7 - 5.0$, $X_{\rm H,s} = 0.65$), reddened over the range of $A_V = 0 - 1.0\,{\rm mag}$. We find that the observed colors best match the synthetic colors for $A_V \approx 0.30$ mag. Because these colors lie on the Rayleigh-Jeans tail of the SED, the synthetic colors, and hence the inferred $A_V$, are only weakly sensitive to the $T_{\rm eff}$ of the stripped star. To assess the systematic uncertainties associated with uncertainty in $A_V$, we repeat our fit over a range of $A_V = 0.25 - 0.35$ mag.

We use the following lines for our fits: He\,{\sc ii} $\lambda 4100$/H$\delta$, He\,{\sc ii} $\lambda 4200$, He\,{\sc ii} $\lambda 4339$/H$\gamma$, He\,{\sc i} $\lambda 4471$, He\,{\sc ii} $\lambda 4542$, N\,{\sc v} $\lambda 4604$, He\,{\sc ii} $\lambda 4859$/H$\beta$, He\,{\sc ii} $\lambda 5412$, and He\,{\sc i} $\lambda 5876$. For each line, we extract a $50\,\rm{\AA}$-wide window centered on the line and renormalize the local continuum. We then use the central $40\,\rm{\AA}$ window around the line for the fitting. We first perform a $\chi^2$ minimization to locate the global best-fit parameters, which we use as the initializations for the MCMC walkers. We place priors on the atmospheric parameters using the presence of the N\,{\sc v} $\lambda4604$ line. For each set of $T_{\rm eff}$, $\log(g)$, and $X_{\rm H,s}$, we convolve the synthetic spectra to $R = 4100$ and measure the equivalent width (EW) of the N\,{\sc v} $\lambda4604$ line. We require the synthetic spectra to show detectable absorption ($EW \geq 0.04\,{\rm \AA}$) to match the observed spectral feature of J0040-7333. We reject models that do not satisfy these criteria by setting their likelihood to negative infinity. Within these bounds, we place uniform priors on all of the fit parameters. We perform the MCMC sampling with 50 walkers, using 1000 burn-in steps followed by a total of 4000 steps of sampling.

From the fit parameters ($T_{\rm eff}$, $\log(g)$, $X_{\rm H, s}$, and $R$), we calculate the bolometric luminosity ($L$) using the Stefan-Boltzmann law: $L = 4\pi \sigma_{\rm SB} R^2T_{\rm eff}^4$. Following \cite{gotberg_2023}, we estimate the stripped-star mass ($M_{\rm stripped}$) in two different ways. First, we estimate an evolutionary mass using a mass-luminosity relation (right panel of Figure \ref{fig: mcmc no evol sed + hr}) derived from the evolutionary models from \cite{gotberg_2018}. The relation is derived by taking the masses and luminosities of stripped stars half-way through helium burning. Second, we compute a spectroscopic mass from the inferred surface gravity and radius.

\subsubsection{Joint SED and spectral fits with evolutionary tracks}\label{sec: mcmc sampling setup}

We also repeat the  SED and spectral fits by coupling the fit to evolutionary models of intermediate-mass stripped stars. This method allows us to constrain the stripped-star mass directly from the evolutionary tracks without relying on a mass-luminosity relation that is derived from a fixed evolutionary stage. However, it also implicitly assumes that the evolutionary tracks are correct, which introduces systematic uncertainties. Specifically, the binary-evolution assumptions adopted by \cite{gotberg_2018}, produce stripped stars with surface hydrogen mass fractions within a rather confined range ($X_{\rm H, s} \approx 0.35-0.4$), which is lower than we infer for J0040-7333.

For these fits, we sample three free parameters: initial mass ($M_i$), core helium mass fraction ($X_{\rm He, c}$), and surface hydrogen mass fraction ($X_{\rm H, s}$). For each combination of these parameters, we obtain effective temperature ($T_{\rm eff}$), surface gravity ($\log(g)$), bolometric luminosity ($L$), radius ($R$), and evolutionary mass ($M_{\rm stripped}$) of the stripped star using the evolutionary models. We then use these stellar properties to generate a model SED and synthetic spectra, which we compare to the observed spectra and photometric SED. We again fix $A_V$ in each fit and repeat the fit for a range of $A_V$.

For each run, we perform an initial fit for $M_i$, $X_{\rm He, c}$, and $X_{\rm H, s}$ using $\chi^2$ minimization to determine the maximum-likelihood solution, which we use to initialize the MCMC sampling. For $M_i$ and $X_{\rm H, s}$, we adopt uniform priors, while for $X_{\rm He,c}$, we adopt an evolutionary-timescale prior derived from the stellar models. We sample the evolutionary tracks uniformly in $X_{\rm He,c}$, and assign each value of $X_{\rm He,c}$ a prior probability that is proportional to the amount of evolutionary time the model spends near that core helium fraction. For each sampled set of model parameters, we use the corresponding $T_{\rm eff}$ and $\log(g)$ to produce the synthetic spectra. Similar to the fit without evolutionary model constraints (Section \ref{sec: mcmc setup no evol}), we require that the atmospheric parameters produce  synthetic spectra with detectable N\,{\sc v} $\lambda4604$ absorption. We run the MCMC sampling with 50 walkers, using 1000 burn-in steps followed by a total of 4000 steps of sampling.

\subsubsection{Results}\label{sec: mcmc sample results}
\begin{figure*}
    \centering
    \includegraphics[width=\linewidth]{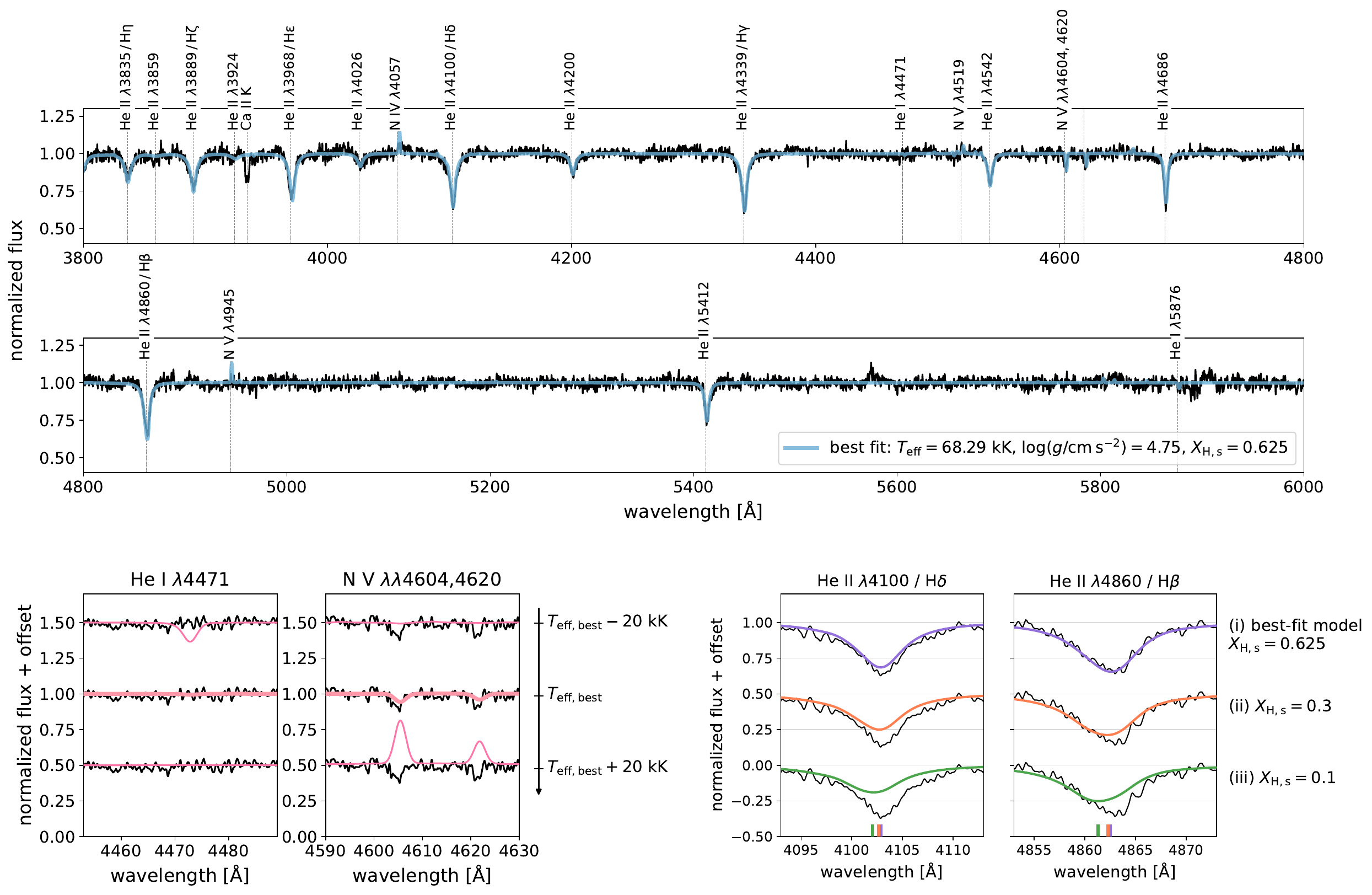}
    \caption{Top: stacked observed spectrum (black) and best-fit spectral model (blue). Bottom left: temperature-sensitive spectral features. Pink lines show spectral models with a range of  effective temperatures, holding the surface gravity and hydrogen fraction fixed.  He\,{\sc i} $\lambda4471$ absorption is  present only for $T_{\rm eff} \lesssim 50$ kK, whereas absorption in the N\,{\sc v} $\lambda4604,4620$ doublet is present for $50 \,{\rm kK}\lesssim T_{\rm eff} \lesssim 90\,{\rm kK}$.  
    Bottom right: effect of surface hydrogen mass fraction on the shape and apparent central wavelength of He\,{\sc ii}/Balmer blended absorption lines. Black curves show the observed line profiles, and colored curves show three models with a range of $X_{\rm H,s}$. $T_{\rm eff}$ and $\log(g)$ are fixed. Tick marks along the x-axis mark the effective line center. As $X_{\rm H,s}$ increases, the central wavelength shifts because the Balmer line contribution becomes stronger relative to the He\,{\sc ii} lines. We show all He\,{\sc ii}/Balmer lines in Figure \ref{fig: He-Balmer all} of Appendix \ref{apdx: hydrogen abundance}.}
    \label{fig: spectra}
\end{figure*}

\begin{figure}[h!]
    \centering
    \includegraphics[width=1\linewidth]{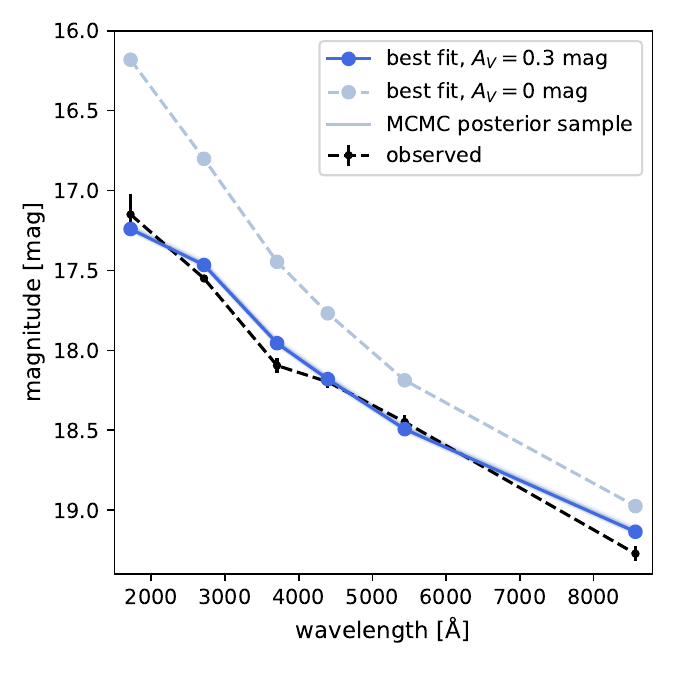}
    \caption{Black points show the observed SED (all bands converted to AB magnitudes). Blue curve shows the best-fit reddened SED model ($A_V = 0.3$ mag). Light blue curves show a random sample of models drawn from the MCMC posterior distribution of the fiducial fit, which is representative of the statistical uncertainty. Dashed blue curve shows the intrinsic best-fit stripped-star SED model ($A_V = 0$ mag).}
    \label{fig: sed fit}
\end{figure}

We report the best-fit parameters and their uncertainties for both methods described in Section \ref{sec: mcmc setup no evol} and \ref{sec: mcmc sampling setup} in Table \ref{tab:mcmc results}. We show the best-fit spectra in Figure \ref{fig: spectra} and best-fit SED in Figure \ref{fig: sed fit}. We adopt the inferred parameters from the SED and spectral fits without evolutionary-track constraints (Section \ref{sec: mcmc setup no evol}) with $A_V = 0.3$ mag as our fiducial best-fit parameters for J0040-7333. We take the statistical uncertainties ($\sigma_{\rm stat}$) from the MCMC posterior of the fiducial fit. We estimate the systematic uncertainty ($\sigma_{\rm sys}$) by taking the difference in the range of the best-fit parameters across all fits (varying both $A_V$ and the fitting method) and dividing by two. We compute the total uncertainty by summing $\sigma_{\rm stat}$ and $\sigma_{\rm sys}$ in quadrature. See Appendix \ref{apdx: mcmc AV} for the full set of best-fit parameters from Section \ref{sec: mcmc setup no evol} and \ref{sec: mcmc sampling setup}, which we use to estimate the systematic uncertainties.
\begin{table*}
    \renewcommand{\arraystretch}{1.3}
    \begin{tabular}{c|c|c|c}
    \hline
         & No evolutionary model & With evolutionary model & Adopted \\\hline\hline
        $T_{\rm eff}$ [K] & 68206 - 73656 & 60900 - 64434 &$ 68288^{+ 731}_{-794}\,(\mathrm{stat})\pm6378\,(\mathrm{sys})$\\
        $\log(g/\rm{[cm\,s}^{-2}])$ & 4.755 - 4.815 & 4.865 - 4.984 & $4.755 ^{+ 0.002}_{-0.002}\,(\mathrm{stat})\pm0.115\,(\mathrm{sys})$\\
        $X_{\rm H,s}$  & 0.604 - 0.626 & 0.652 - 0.674 & $ 0.625^{+ 0.012}_{- 0.011}\,(\mathrm{stat})\pm0.035\,(\mathrm{sys})$\\
        $R$ [$R_\odot$] & 0.949 - 0.998 & 1.002 - 1.132 & $ 0.988^{+0.011}_{- 0.010}\,(\mathrm{stat})\pm0.092\,(\mathrm{sys})$\\
        $\log(L/L_\odot)$ & 4.244 - 4.423 & 4.163 - 4.208 & $4.281^{+ 0.013}_{-0.014 }\,(\mathrm{stat})\pm0.130\,(\mathrm{sys})$\\
        $M_{\rm evol}$ [$M_\odot$] & 3.811 - 4.328 & 3.263 - 3.437 & $ 3.840^{+ 0.046}_{- 0.048}\,(\mathrm{stat})\pm0.533\,(\mathrm{sys})$\\
        $M_{\rm spec}$ [$M_\odot$] & 1.868 - 2.374 & 3.257 - 3.423 & $ 2.025^{+ 0.050}_{- 0.046}\,(\mathrm{stat})\pm0.778\,(\mathrm{sys})$
        \\\hline
    \end{tabular}
    \caption{
    Stripped-star properties constrained from the spectral and SED fitting (see Section \ref{sec: mcmc setup no evol} and \ref{sec: mcmc sampling setup}). We report the range of best-fit parameters for each fitting method when $A_V$ is varied between $0.25-0.35$ mag; see Table \ref{table: full mcmc no evol} and \ref{table: full mcmc} in Appendix \ref{apdx: mcmc AV} for the full list of best-fit parameters for each $A_V$.
    The ``Adopted'' column gives the fiducial best-fit value ($A_V = 0.30$ mag; without evolutionary-model constraints). We report the uncertainty from the MCMC posterior of the fiducial fit as the statistical uncertainty, and we estimate the systematic uncertainty as half the difference between the maximum and minimum best-fit values across all fits.}   
    \label{tab:mcmc results}
\end{table*}

The inferred stripped-star evolutionary mass, $M_{\rm stripped} = 3.840 \pm 0.535\,\rm{M_\odot}$, falls within the expected intermediate-mass range. Using the best-fit luminosity and evolutionary mass, we infer an Eddington factor of $\Gamma_e = \frac{\kappa_e L}{4\pi c G M} \sim 0.12$. Adopting the spectroscopic mass instead yields $\Gamma_e \sim 0.23$. The inferred stellar properties are broadly consistent with theoretical predictions \citep[e.g.,][]{gotberg_2018} and previously observed intermediate-mass stripped stars \citep{Drout_2023, gotberg_2023}. We further compare J0040-7333 to related populations from literature in Section \ref{sec: comparsion}.

\subsubsection{Surface hydrogen mass fraction}\label{sec: hydrogen fraction}
As we demonstrate in the lower right panel of Figure \ref{fig: spectra}, (see also Figure \ref{fig: He-Balmer all} in Appendix \ref{apdx: hydrogen abundance}) the surface hydrogen mass fraction is constrained by the relative strengths and the central wavelengths of blended He\,{\sc ii}/Balmer lines, which determine the relative contributions of the hydrogen and helium to the absorption features. Visual inspection alone suggests that we can rule out a low surface hydrogen mass fraction due to the central wavelength and depth of the He\,{\sc ii}/Balmer blended lines.  

A natural question is whether the inferred high $X_{\rm{H, s}}$ could be due to Balmer absorption by an unresolved hydrogen-rich companion. If the stripped star had a surface hydrogen fraction more typical of hot stripped stars ($X_{\rm H,s} \approx0.3$), reproducing the observed strength of the He\,{\sc ii}/Balmer blends would require a MS companion of $\sim 2\,{\rm M_\odot}$. Such a companion would be too cool to exhibit He\,{\sc i} absorption, which would otherwise make it easily identifiable in the spectrum. However, at this luminosity, it would contribute roughly half of the EWs of the Balmer lines. Such a companion would change the shape of the blended lines between epochs, including producing split line cores at maximum RV difference between the two components. However, we find that the He\,{\sc ii}/Balmer line profiles remain stable with phase. Furthermore, because the companion would have a different RV than the stripped star, it would bias the RVs measured from blended lines. Compared to RVs derived from the pure He\,{\sc ii} lines, we would expect systematically higher RVs near the stripped star's RV minimum and lower RVs near its RV maximum, in both cases drawing the measured RV toward the center-of-mass RV.
However, we find no evidence of such trends; see Figure \ref{fig: rv_compare lines} of Appendix \ref{appendix: rv table}. Significant contamination by a luminous companion is therefore unlikely (and also inconsistent with the light curve; see Section~\ref{sec:light_curve}), so we conclude that the measured hydrogen abundance reflects the true surface composition of the stripped star.

This inferred hydrogen abundance is unusual in the context of known hot stripped stars. Although the temperature and surface gravity inferred for J0040-7333 are consistent with the standard picture of hydrogen-poor core-helium burning stars \citep[e.g., ][]{gotberg_2018}, its surface appears to be more hydrogen rich than the previously observed population of hot and compact intermediate-mass stripped stars \citep{gotberg_2023}. 
The inferred surface hydrogen mass fraction instead is comparable to that of partially-stripped stars \citep[e.g.,][]{Ramachandran_2023}, which are predicted to be cooler and more bloated \citep[e.g.,][]{Klencki_2022, Dutta_2024} compared to what we infer for J0040-7333.

Current binary evolution models predict that retaining a substantial hydrogen-rich envelope after stripping prevents a stripped star from contracting. 
As a result, such stars are generally not expected to reach the compact radii ($R \sim 1\,\rm{R_\odot}$) and high effective temperatures ($T_{\rm eff} \gtrsim 50 \,{\rm{kK}}$) that are typical of their more fully stripped counterparts. 
The combination of a hot, compact structure and relatively high surface hydrogen abundance we observe for J0040-7333 is therefore difficult to reproduce with existing evolutionary models. 

\subsection{Stellar rotation}
\begin{figure}
    \centering
    \includegraphics[width=\linewidth]{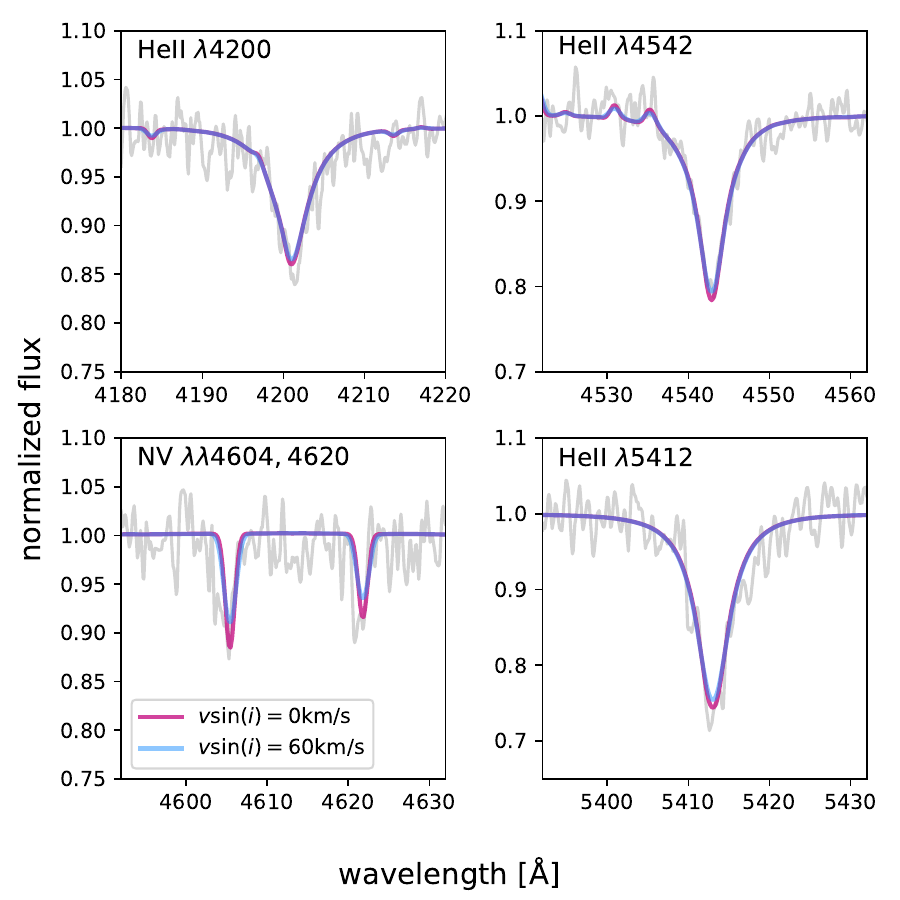}
    \caption{Best-fit synthetic spectral models with (blue) and without (pink) rotational broadening for a set of He\,{\sc ii} and N\,{\sc v} lines. The synthetic model shown here adopts a nitrogen mass fraction of $X_{\rm N}=0.004$. We produced these broadened synthetic spectra using the publicly available code from \cite{Carvalho_2023}. We expect $v\sin(i) \lesssim 60\,\rm{km \,s^{-1}}$ if the stripped star is tidally synchronized. Grey shows the observed spectrum. With the resolution of MagE, we cannot robustly constrain $v\sin(i)$.}
    \label{fig:vin(i)}
\end{figure}

At orbital periods of $\lesssim 1\,{\rm day}$, tides are predicted to efficiently synchronize the rotation of the helium stripped star with the binary's orbital period \citep[e.g., ][]{Ma_2024}. Using the inferred orbital period and the radius of the stripped star, we can infer an equatorial rotational velocity of $2\pi R/ P_{\rm orb} \approx 60\,{\rm km\,s^{-1}}$ for the stripped star for an edge-on, tidally synchronized system. In Figure \ref{fig:vin(i)}, we show the best-fit synthetic spectra with no rotation ($v\sin(i) = 0\,\rm{km\,s^{-1}}$) and with tidally-synchronized rotation ($v\sin(i) = 60\,\rm{km\,s^{-1}}$).
We find that, at the resolution and SNR of the MagE spectra, we are unable to robustly detect rotation with $v\sin(i) = 60 \,{\rm km\,s^{-1}}$. With the resolution of spectra we have, we can rule out high rotation rates and infer an upper limit of $v\sin(i)\lesssim 90\,{\rm km\,s^{-1}}$. However, for a more robust measurement of the stellar rotation, higher-resolution data, e.g., $R\gtrsim10,000$, is necessary. Assuming the stripped star is tidally synchronized, measurement of $v\sin(i)$ would constrain the inclination angle of the binary and provide stronger constraints on the companion mass.

\subsection{Constraining the companion mass}\label{sec: companion mass}
\begin{figure*}[htb!]
    \centering
    \includegraphics[width=\linewidth]{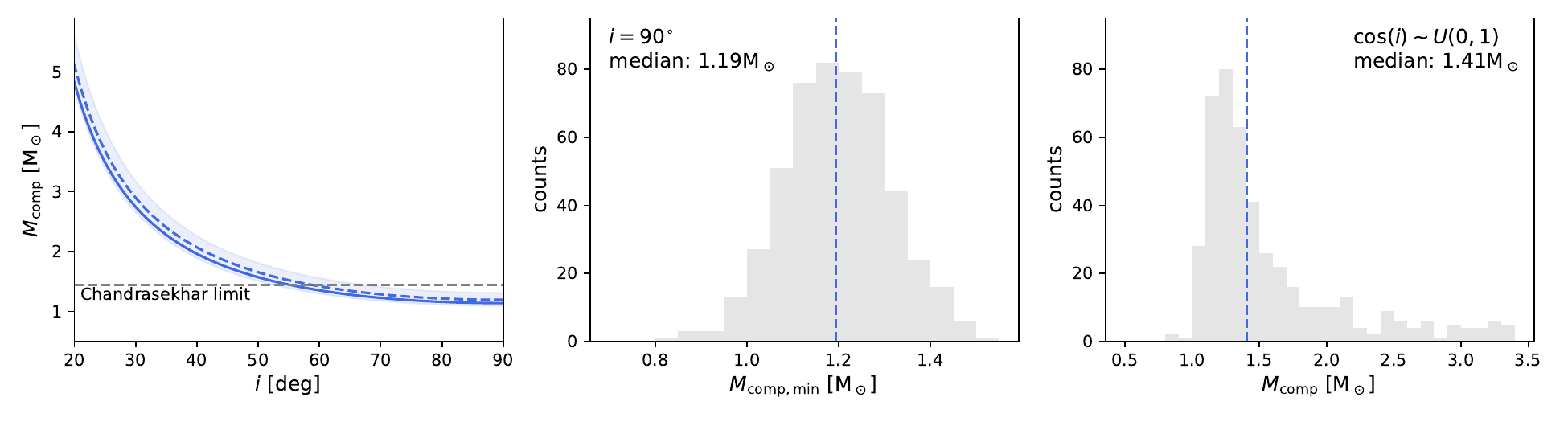}
    \caption{
    Left: inferred companion mass as a function of inclination angle. Solid blue line shows the companion mass at each inclination, computed using the best-fit $P_{\rm orb}$, $K$, $e$, and $M_{\rm stripped}$ values. 
    Dashed blue line shows the median companion masses at each inclination, computed from samples of $P_{\rm orb}$, $K$, and $e$ drawn from their respective MCMC posterior distributions and $M_{\rm stripped}$ drawn from a Gaussian distribution defined by the best-fit evolutionary mass and its total uncertainty. Shaded blue region indicates the $1\sigma$ uncertainty on the companion mass. Gray dashed line marks the Chandrasekhar mass limit for WDs. Center: distribution of minimum companion masses, assuming an inclination of $i = 90^\circ$. Dashed line marks the median. Right: distribution of companion masses obtained by adopting an isotropic inclination distribution, i.e., $\cos(i) \sim U(0, 1)$.}
    \label{fig: comp mass}
\end{figure*}

Using the orbital parameters ($P_{\rm orb}$, $K$, $e$) and the mass of the stripped star, we estimate the mass of the companion. We calculate the binary mass function, $f_m$, and numerically solve it for the companion mass, $M_2$:
\begin{equation}\label{eq: mass func}
    f_m = \frac{M_2^3 \sin^3i}{(M_{\rm stripped} + M_2)^2} = \frac{P_{\rm orb} K^3}{2\pi G} (1-e^2)^{3/2},
\end{equation}
We sample $P_{\rm orb}$, $K$, and $e$ from their respective MCMC posterior distributions and obtain $f_m = 0.066 \pm 0.007\,\rm{M_\odot}$. We sample $M_{\rm stripped}$ from a Gaussian distribution with $\mu = 3.84\,{\rm M_\odot}$ and $\sigma = 0.54\,{\rm M_\odot}$ reflecting the fiducial best-fit evolutionary mass and the total uncertainty. For a given inclination angle $i$, we numerically solve Equation \ref{eq: mass func} to obtain the corresponding companion mass. For each inclination angle, we estimate the companion mass and uncertainties using the median and the 16th/84th percentiles from the resulting companion mass distributions, respectively. We show the inferred companion mass as a function of inclination angle in Figure \ref{fig: comp mass}.

For our inferred stripped-star mass of $M_{\rm stripped} = 3.84\,{\rm M_\odot}$, the inferred minimum companion mass is $M_2 = 1.19 \,\rm{M_\odot}$, assuming an inclination angle of $i=90^\circ$. For an isotropic inclination distribution, the companion mass distribution peaks at $M_2 = 1.2-1.3\,\rm{M_\odot}$, with a median of $M_2 = 1.41\,\rm{M_\odot}$. The inferred companion masses are consistent with a low-mass MS star, a low-mass stripped star, a high-mass WD, or a NS companion. At lower inclinations ($i\lesssim55^{\circ}$), WD companions become increasingly disfavored, since the inferred companion mass exceeds the Chandrasekhar limit.

\subsection{Light curve analysis}
\label{sec:light_curve}
We analyze the OGLE $I-$band light curve of J0040-7333. The light curve contains 3860 epochs of photometry taken between 2001 and 2025. We search for periodic variability and compare the observed light curve to theoretical light curves of a stripped star binary with properties matched to J0040-7333.

\subsubsection{Periodic variability}
\begin{figure*}[htb!]
    \centering
    \includegraphics[width=0.9\linewidth]{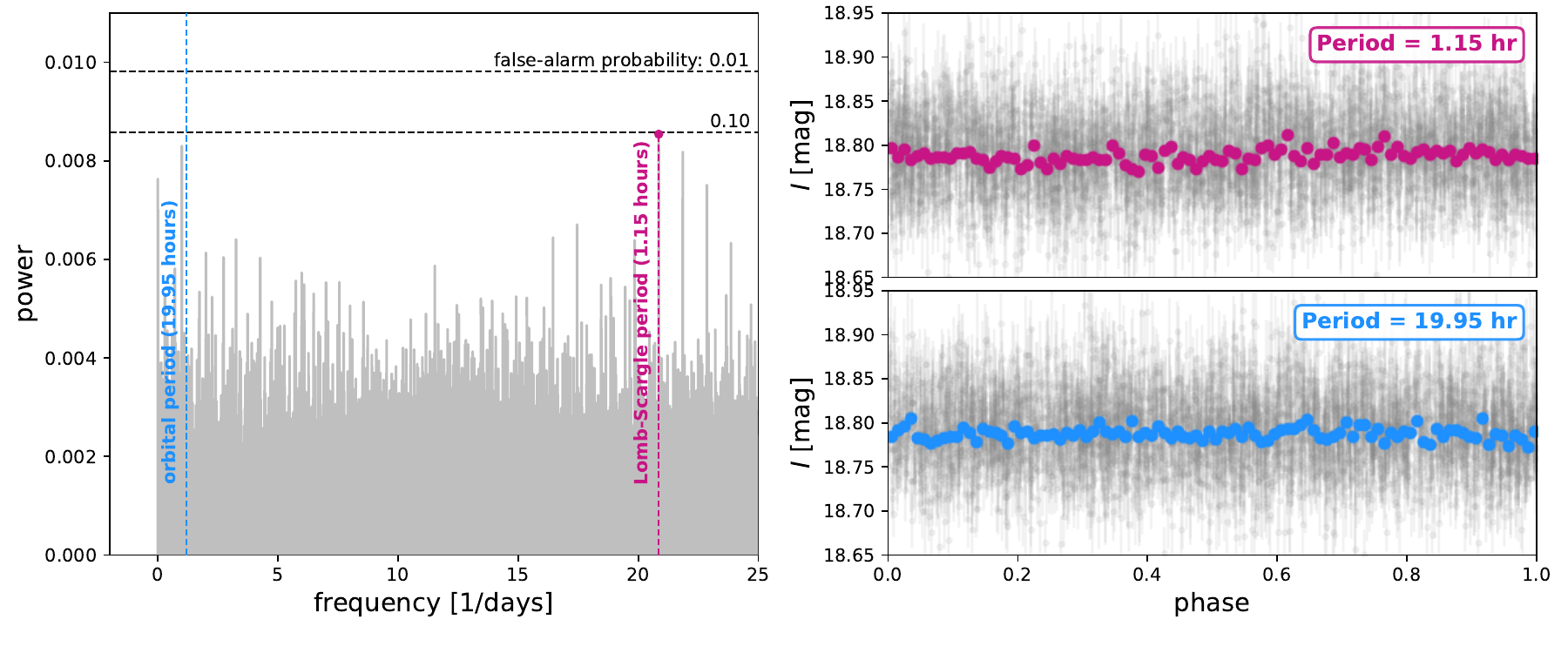}
    \caption{Left: Lomb-Scargle periodogram of the OGLE light curve. The strongest periodicity is detected at 1.15 hours, marked by the magenta dashed line. We detect no significant periodicity at the orbital period (19.95 hours), marked by the blue dashed line. Black dashed lines mark the 1\,\% and 10\,\% false-alarm probability thresholds. Right: OGLE light curves phase-folded to the Lomb-Scargle period (top) and the orbital period (bottom). Gray points show the raw data. Magenta and blue points show phase-folded light curves, binned to $N =100$. We do not detect stable periodicity above $\gtrsim\,1\%$ at either period.}
    \label{fig: lomb-scargle}
\end{figure*}

Using a Lomb-Scargle periodogram \citep{Lomb_1976, Scargle_1982}, we identify the highest-power period at 1.15 hours. Figure \ref{fig: lomb-scargle} shows the periodogram and light curve phased to this period, which does not show convincing periodic variability suggestive of binarity.
Since this period is much shorter than the RV period derived in Section \ref{sec: orbital solution}, it is unlikely that this signal is related to binary-induced variability such as irradiation or ellipsoidal modulation. This suggests that this 1.15 hour variability, if real, is intrinsic to the stripped star. We can rule out rotational modulation, which would imply $v\sin(i) \sim 1000 \,{\rm km s^{-1}}$. Not only is such rapid rotation physically implausible, but a $v\sin(i)$ of this magnitude would produce extreme line broadening inconsistent with the observed spectra. Although pulsational variability remains a plausible explanation, current theoretical models predict only pulsations during the post-mass transfer contraction phase of a stripped star \citep[e.g., ][]{Fadeyev_2025}. Given the small amplitude of the variability and its relatively high false-alarm probability of $10.5\,\%$, the signal could very well be due to instrumental systematics or background variability from nearby sources. 

We also check for periodicity on the orbital period by phase-folding the light curve to $P_{\rm orb} = 19.95\,\rm{hours}$. We again find no stable periodicity. To more robustly assess the sensitivity of the OGLE light curve, we perform an injection-recovery test. We inject a sinusoidal signal with a range of periods near the orbital period of $\sim 20$ hours (e.g., 19, 20, 21 hours) and apply the Lomb-Scargle periodogram to test whether the injected signal is recovered. We find that the injected signals with peak-to-peak amplitude above 1\,\% are recovered with the injected period corresponding to the strongest peak in the periodogram. This suggests that the true binary-related peak-to-peak photometric variability is $\lesssim 1\%$.

\subsubsection{Lack of reflection variability rules out a MS companion}\label{sec: rule out MS}
Binaries containing a hot star and a cool star often exhibit periodic variability on the orbital period due to the reflection effect \citep[e.g., ][]{Wilson_1990}. The hot star irradiates one side of the cooler companion, causing that side of the star to become hotter and brighter. This produces periodic variability matched to the orbital period, caused by the viewing of the day and night side of the star. This effect is observed in e.g., short period extreme mass-ratio binaries \citep[e.g.,][]{naze_2023, naze_2025} and sdB + low-mass MS (e.g., M dwarf) binaries \citep[e.g.,][]{schaffenroth_2023, he_2025}. Given the 20-hour orbital period and the extreme luminosity ratio of J0040-7333, we expect similar irradiation effects to be detectable in the OGLE $I$-band light curve, if the companion is a MS star.

We create synthetic light curves for J0040-7333 for a range of inclination angles, using the PHysics Of Eclipsing BinariEs code \citep[PHOEBE v.2.4.20; ][]{phoebe_2005, phoebe_2016, phoebe_2018, phoebe_2020, phoebe_2020b}. PHOEBE is a light-curve modeling code that accounts for both the geometry (e.g., orbital period, component radii, inclination) and stellar physics (e.g., irradiation, limb darkening, star spots) of the binary. To model the OGLE photometry, we create a custom bandpass for PHOEBE using the OGLE-$I$ band transmission profile.

PHOEBE requires the radius and temperature of each star, as well as the limb darkening and albedo coefficients, to model irradiation effects. It also requires information about the binary's orbital period and separation. We create a light curve model for J0040-7333 using the best-fit orbital period and assuming a circular orbit. We model the primary star as the stripped star using the inferred evolutionary mass, effective temperature, and radius derived from our spectral and SED modeling (Section \ref{sec: properties of stripped star}). We model its atmosphere as a blackbody. We model the companion as a MS star. For a given inclination angle, we compute the companion mass from the best-fit parameters using the binary mass function (Section \ref{sec: companion mass}). We estimate the effective temperature and radius using the stellar model grids from \cite{Pecaut_2013}. We model its atmosphere using the \texttt{ck2004} atmosphere models. \cite{Laroche_2026} showed that some low-mass stellar companions to stripped stars could still be in their late pre-MS stage. Although we do not separately model these systems, such stars are expected to be cooler and more inflated than their MS counterparts. Their larger radii and lower intrinsic optical fluxes would generally make the reflection signal more prominent; thus, modeling the companion as a MS star is sufficient for providing conservative estimates of the expected reflection effect for MS-like companions.

We set the bolometric gravity-darkening and albedo coefficients to $\beta = 1.0$ and $\alpha = 1.0$ \citep{von_Zeipel_1924} for stars with radiative envelopes, and $\beta = 0.32$ and $\alpha = 0.5$ \citep{Lucy_1967} for those with convective envelopes. For both stars, we manually set the bolometric and pass band limb-darkening coefficient to 0.4. We model the stripped star and any MS companion with $T_{\rm eff} \geq7,200\,\rm{K}$ using the radiative-envelopes prescription, and model MS companions with $T_{\rm eff} \leq 7,200\,\rm{K}$ using the convective-envelope prescription \citep[e.g.,][]{Claret_1998}.

\begin{figure}
    \centering
    \includegraphics[width=\linewidth]{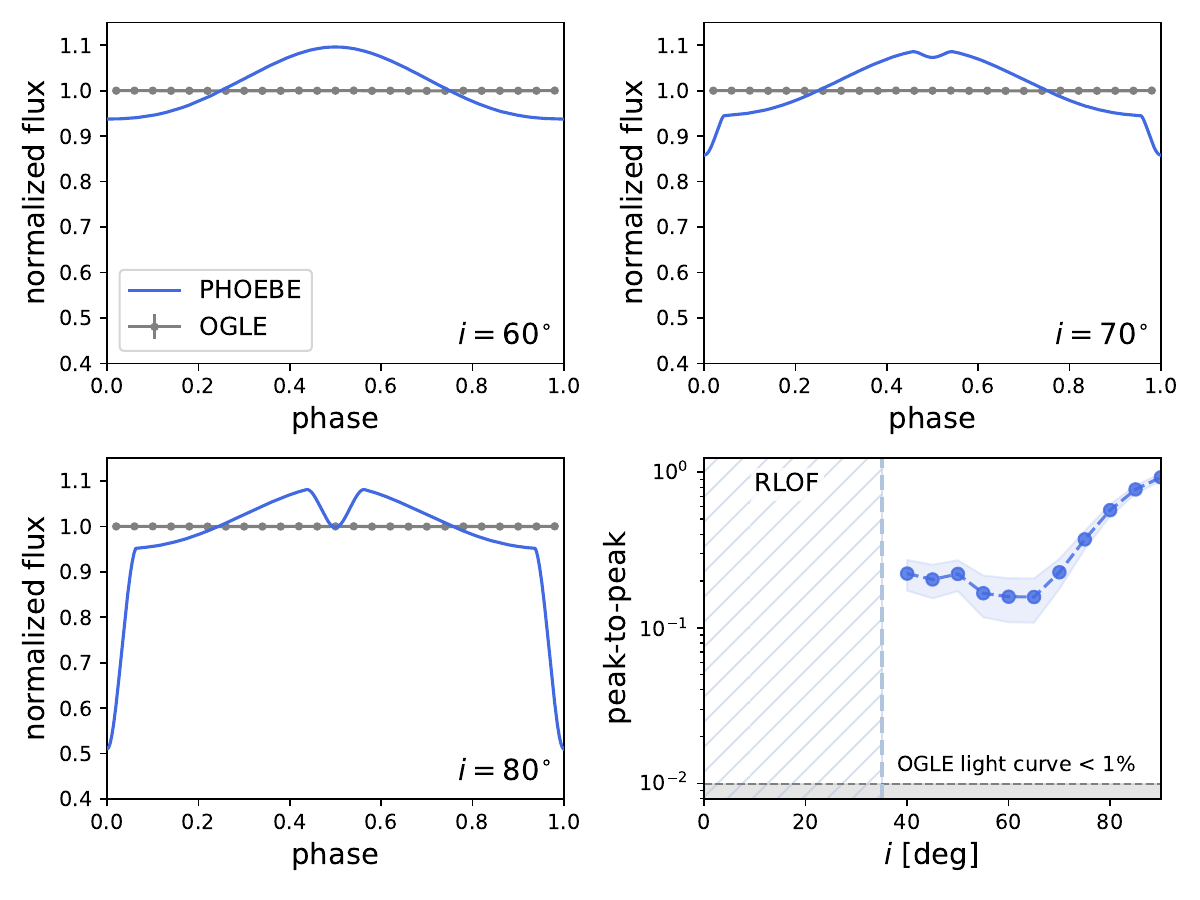}
    \caption{Top \& bottom left: blue curve shows the predicted PHOEBE light curve models for J0040-7333 for inclination angles of $i = 60, 70$, and $80$ degrees, assuming a MS companion. Grey curve shows the phase-folded and binned OGLE $I$-band light curve of J0040-7333. Bottom right: blue curve shows the maximum peak-to-peak photometric variability of the PHOEBE light curve model as a function of inclination angle. The shaded region around the curve represents the systematic uncertainty associated with light curve modeling assumptions. For these models, the non-detection of photometric variability above $\sim1$\% in the OGLE light curve rules out a MS companion. For the mass-radius relation we adopt, the inferred MS companion fills its Roche lobe for $i \lesssim 35^\circ$; we therefore restrict our light curve modeling to inclination angles above $i=40^\circ$.}    
    \label{fig: example light curves}
\end{figure}

Figure \ref{fig: example light curves} shows a set of representative PHOEBE light-curve models.
We find that at low inclination angles, $i \lesssim 70^{\circ}$, the reflection effect dominates the photometric variability, whereas at high inclination angles the eclipse dominates. For a binary with a fixed stripped star and MS companion mass, we expect the reflection variability to decrease with inclination. However, here the companion mass inferred from the binary mass function increases for lower inclination angles. Because more massive MS stars have larger radii, their irradiated surface areas increase with mass, thereby increasing the amplitude of the photometric variability due to reflection. This effect approximately cancels out the $\sin i$ dependence of the irradiation variability, which makes the photometric variability somewhat constant below inclination angles of $i \lesssim 70^{\circ}$. We find that PHOEBE models predict photometric variability of $\gtrsim 15 \,\%$ in the OGLE $I$ band for all inclination angles.

The predicted photometric variability due to reflection/irradiation is only weakly model-dependent. We test the sensitivity of the predicted reflection amplitude to several model assumptions: (i) the \texttt{irrad\_method}, which determines how irradiation is treated by PHOEBE; (ii) limb-darkening and gravity-darkening prescriptions; and (iii) the albedo coefficients. We find that variations in (i) and (ii) affect the predicted photometric variability by $\lesssim1-2\,\%$. The assumed albedo of the companion can have a more significant effect on the predicted variability amplitude. When the albedo is varied over representative ranges of $\alpha =0.4-0.6$ for stars with convective envelopes and $\alpha =0.8-1.0$ for stars with radiative envelopes \citep[e.g.,][]{Claret_2001}, the predicted photometric variability can change by $\sim3$--$5\,\%$. To account for the above effects, we adopt a systematic uncertainty of $5\,\%$ associated with these modeling assumptions. This uncertainty is larger than the variation produced by changing the primary and secondary stellar properties within their inferred uncertainties, but small compared to the difference between model predictions and observations. 
For all inclination angles and modeling assumptions, the PHOEBE models with MS companions predict photometric variability of $\geq 10\%$, at least ten times larger than the observed upper limit of  $1\,\%$. The absence of such variability for J0040-7333 strongly disfavors a MS companion.  

\textit{Compact-object companion: }if J0040-7333 hosts a compact-object companion, the expected dominant photometric signal is ellipsoidal variability. We estimate this effect using PHOEBE by turning off irradiation and treating the companion as a dark, compact source. PHOEBE models the tidal distortion of the stripped star induced by the compact-object companion and computes the resulting ellipsoidal variability. For J0040-7333, the predicted peak-to-peak photometric variability due to ellipsoidal modulation is $\sim 0.3-0.5\,\%$, which is below the sensitivity of the OGLE light curve. Thus, it is not surprising that we do not detect such variability. 

\textit{Stripped-star companion: }for reflection effects to be sufficiently large and thus detectable, the binary must have a sufficient flux ratio. We create a PHOEBE model for the stripped-star companion scenario by modeling both the primary and secondary as a stripped star, taking the $M-R$ and $M-T_{\rm eff}$ relations from \cite{gotberg_2018}. We find that reflection variability is negligible ($\lesssim 1\,\%$) for a stripped-star companion of all possible masses, because the companion itself is hot, resulting in a rather small flux ratio between the primary and secondary stars. This variability is below the detection threshold of the OGLE $I$-band light curve. A stripped-star companion, however, is expected to produce eclipses at high inclinations ($i \gtrsim 80^{\circ}$). The absence of eclipses in the OGLE light curve therefore rules out high orbital inclinations in the stripped-star companion scenario.

We further explore the nature of the companion in Section \ref{sec: nature of companion}.

\subsection{X-ray and radio observations}\label{sec: radio & xray}
We search both radio and X-ray surveys for counterparts associated with J0040-7333.

\textit{Radio: }if the companion is a NS that has been recently spun up via mass accretion, the system may be observable as a radio pulsar. We checked the 1.3 GHz SMC point-source catalog \citep{Cotton2024} compiled from MeerKAT \citep{Meerkat} observations. MeerKAT is a radio interferometer located in the Northern Cape province of South Africa. It is composed of 64 dishes, each with a diameter of 13.5 m. We cross-matched this radio catalog with the SMASH optical catalog using a cross-match radius of 1 arcsec and found no radio counterpart associated with J0040-7333. This indicates that the source is not detected above the MeerKAT sensitivity threshold. The local root-mean-squared noise level of the catalog is $\sim 11 \mu$ Jy beam$^{\rm{-1}}$, corresponding to $3\sigma$ and 5$\sigma$ upper limits of $\sim 33\,\mu$Jy beam$^{-1}$ and $\sim 55\,\mu$Jy beam$^{-1}$, respectively, at 1.3 GHz. 

Galactic pulsars exhibit radio luminosities at 1.4 GHz ($L_{1.4\,{\rm GHz}} = 4\pi d^2 S_{1.4\,{\rm GHz}}$) that approximately span $L_{1.4\,{\rm GHz}} = (10^{18.5} - 10^{21.5}) \,{\rm \,erg\,s^{-1}\,Hz^{-1}}$ \citep[e.g.,][]{Manchester2005}, 
with recycled pulsars typically occupying the lower end of the luminosity distribution \citep[e.g.,][]{Kramer_1998, Gitika_2023}. At SMC/LMC distances, pulsars with such luminosities produce flux densities near or below the detection thresholds of current surveys. Consequently, no millisecond pulsars have been detected in the MCs, and the currently known SMC pulsar population is biased towards the most radio-luminous young pulsars \citep[e.g.,][]{McConnell_1991, Crawford_2001, Manchester_2006, Titus_2019}. A non-detection of J0040-7333 is therefore not unexpected, even if the system hosts a recycled pulsar or a normal young pulsar.

\textit{X-ray: }if the companion were currently undergoing active accretion, the system may be detectable in X-rays. We used the XMM-Newton \citep{Jansen_2001} upper limit server\footnote{\url{https://xmmuls.esac.esa.int/upperlimitserver/}} to estimate upper limits on the X-ray emission for J0040-7333. For the count-rate to flux conversion, we assume $N_{\rm H} = 6\times 10^{20}\,{\rm cm^{-2}}$ based on the Galactic foreground \citep{Dickey_1990}, and adopt the default power law spectrum with a photon index  $\Gamma = 2$. We obtain 3$\sigma$ upper limits for the fluxes: $\lesssim 9.4 \times10^{-15}\,\rm{erg\,s^{-1}\,cm^{-2}}$, $\lesssim4.3 \times 10^{-14} \,\rm{erg\,s^{-1}\,cm^{-2}}$, and $\lesssim 1.9 \times 10^{-14}\,\rm{erg\,s^{-1}\,cm^{-2}}$ for the $0.2-2$ keV, $2-12$ keV, and $0.2-12$ keV bandpasses, respectively.
At a SMC distance of 62.44 kpc \citep{Graczyk_2020}, this corresponds to X-ray luminosities of $L_x\lesssim 4 \times10^{33}\,\rm{erg\,s^{-1}}$, $L_x\lesssim2 \times 10^{34} \,\rm{erg\,s^{-1}}$, and $L_x\lesssim 9 \times 10^{33}\,\rm{erg\,s^{-1}}$, respectively.

These X-ray luminosity upper limits correspond to mass-accretion rates of $\dot{M}_{\rm acc}\lesssim 10^{-12} - 10^{-13}\, \rm{{M_\odot\, yr^{-1}}}$, assuming a radiative efficiency of $\eta \approx 0.1$. 
To assess whether the companion could be accreting from the stripped star's wind, we use an order-of-magnitude Bondi-Hoyle-Lyttleton (BHL) wind-accretion calculation. We adopt an accretion efficiency parameter $\alpha_{\rm BHL} = 1$, assume a circular orbit, and estimate the wind velocity as $v_{\rm wind} \approx \left(1 - \frac{R_{\rm stripped}}{a} \right)^{\beta} v_{\rm \infty}$, taking $\beta = 1$ and $v_{\rm \infty} / v_{\rm esc}\sim 2.5$ \citep{Lamers_1995}. We take the inferred best-fit $R_{\rm stripped}$ from the spectral and SED modeling, and estimate the orbital semi-major axis using the best-fit $P_{\rm orb}$, $M_{\rm stripped}$, and a NS mass of $M_{\rm NS} = 1.4\,{\rm M_\odot}$. For accretion rates of $\dot{M}_{\rm acc}\lesssim 10^{-12} - 10^{-13}\, \rm{{M_\odot\,yr^{-1}}}$, this yields a wind-mass loss rate of $\dot M_{\rm wind} \lesssim 5\times10^{-9}- 5\times10^{-8}\,{\rm M_\odot\,yr^{-1}}$. These upper limits are broadly consistent with upper limits on the wind mass loss rates for similar intermediate-mass stripped stars in the MCs \citep{gotberg_2023}. We conclude that the lack of detected X-rays is not in tension with the presence of a NS companion.

While millisecond pulsars also produce faint X-ray emission, the fact that none are known in the MCs speaks against their detectability in X-rays at $\sim60 \,{\rm kpc}$ distances. Therefore, these X-ray upper limits also do not constrain this scenario.

\section{Nature of the system}\label{sec: nature of the system}
Here we discuss the past and future evolution of the J0040-7333 using binary evolutionary models from the literature. We will also discuss the nature of the companion in context of its formation channel.

\subsection{Formation channel and formation history}\label{sec: past evolution}
As discussed in Section \ref{sec: rule out MS}, J0040-7333 could be hosting either a compact-object or stripped-star companion. Here, we discuss the possible evolutionary channels for each of the scenarios.

First, we consider the case where J0040-7333 contains a stripped star and a compact-object companion. The most straightforward evolutionary channel for producing such binaries includes two stages of stripping \citep[e.g.,][]{Tauris2015, tauris_2017, gotberg_2020}. In most models, the first interaction is stable and produces a stripped star + MS binary with a relatively wide orbit. The stripped star subsequently explodes, leaving behind a compact object. The second interaction occurs when the MS star evolves, leading to an episode of CE evolution, producing a close stripped star + compact object binary.

The nature of the first mass-transfer stage for J0040-7333 is difficult to constrain directly from observations, since a second episode of interaction and a SN likely occurred between then and now.

\begin{figure}[h!]
    \centering
    \includegraphics[width=\linewidth]{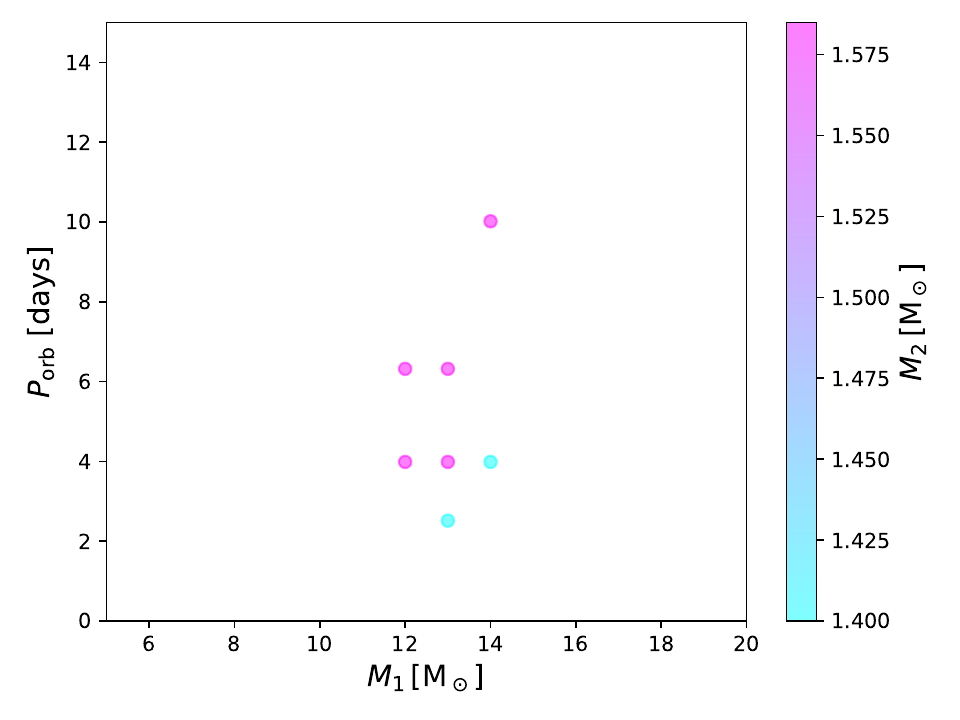}
    \caption{Initial masses and periods of models in the BPASS library that produce a binary similar to the current state of J0040-7333 at some point in their evolution. These initial conditions represent the binary properties after the formation of the compact object but before the CEE that formed the stripped star. Here, $M_1$ represents the mass of the secondary stripped star prior to the CE phase, and $M_2$ represents the birth mass of the compact-object companion.}
    \label{fig: BPASS models}
\end{figure}

The evolutionary history since the formation of the compact object can be better constrained. The short period and unequal mass ratio of J0040-7333 suggests that the system recently underwent a CE phase, which likely ejected the envelope of the stripped star \citep{Ivanova_2013}. We search for evolutionary tracks representing this interaction using the grid of binary models from Binary Population and Spectral Synthesis \citep[BPASS v2.2.1; ][]{Eldridge2017}. We use the grid of models with a metallicity matched to the SMC ($Z = 0.004$) and search for models that approximately match the inferred properties of J0040-7333 at any point in their evolution. We select only from the ``type 2'' models, which track the evolution of the secondary star after the primary star has finished its evolution. Specifically, we require: $T_{\rm eff, 1} = 60-80$ kK, $\log(g_1) = 4.5 - 6$, $\log(L_1/\rm{L_\odot}) = 4 - 5$, $M_1 = 3-4\,\rm{M_\odot}$, $M_2 = 1.2-1.6\,\rm{M_\odot}$, and $P_{\rm orb} = 0.5- 1.5$ days. Here, star 1 is the donor in the second mass transfer stage and star 2 is the compact object. We find seven models that satisfy the above criteria. Figure \ref{fig: BPASS models} shows all of the corresponding models. 
We do not impose any constraints on the surface hydrogen abundance when selecting the models, but note that the seven selected models have $X_{\rm H, s} \approx 0.18-0.35$.

These models predict that prior to the second mass transfer stage (i.e., the CE phase), the system had an orbital period of $P_{\rm orb} = 2.5 - 10$ days, the progenitor of the stripped star had a mass of $12.0-14.0\,\rm{M_\odot}$, and the compact-object companion had a mass of $M_2 = 1.4-1.6\,\rm{M_\odot}$. The initial mass of the stripped star predicted by  BPASS is slightly larger than our inferred  value of $M_{\rm init} \approx10\,{\rm M_\odot}$ from the SED and spectral fits with evolutionary tracks (Table \ref{table: full mcmc}). This discrepancy likely arises from differences in stellar and binary physics adopted by the two models, including, e.g., their treatment of convective overshooting. Nonetheless, the inferred initial masses are broadly consistent with expectations for progenitors of intermediate-mass stripped stars \citep[e.g.,][]{gotberg_2018}.

We now consider the case where the companion is a stripped star. If the companion is a stripped star, J0040-7333 likely formed via a different channel, which is not fully captured by the BPASS models. The BPASS models do not self-consistently evolve both stellar components in detail simultaneously, but instead treat the mass-transfer evolution in two distinct stages. In the first mass-transfer stage (`type 1' models), BPASS models the evolution of the primary, while the secondary is treated using rapid stellar-evolution prescriptions. The secondary is only modeled in detail during second mass-transfer stage (`type 2' models), which begins after the death of the primary star. As a result, the BPASS models do not explicitly model systems where both components are simultaneously stripped stars. However, several binary-evolution models predict the formation of such systems. Two possible formation channels considered in literature are: (i) a two-stage stripping scenario, in which the secondary is stripped before the first stripped star reaches the end of its life \citep[e.g.,][]{Toonen_2018}; and (ii) a double-core CE scenario, in which the envelopes of both stars are simultaneously ejected via CEE \citep[e.g.,][]{Vigna_2018}.

\subsection{Nature of the companion}\label{sec: nature of companion}

\subsubsection{Compact-object companion: WD or NS}
Following Section 8 of \cite{Laroche_2026}, we estimate the fraction of the companion mass distribution (obtained by assuming isotropic inclination; right panel of Figure \ref{fig: comp mass}) falls within typical masses of different compact objects. We find that 50\,\% of the distribution lies within typical WD masses ($M_{\rm WD} = 0.2 -1.44\,{\rm M_\odot}$) and 87\,\% fall within typical NS masses ($M_{\rm NS} = 1 - 3\,{\rm M_\odot}$). Only 11\,\% of the distribution exceeds $\sim 3\,{\rm M_\odot}$, suggesting a BH companion is unlikely. Below, we consider the WD and NS scenarios in context of their evolutionary channels.

Because more massive stars evolve faster, the compact-object companion we observe today is likely the remnant of  the initially more massive star in the binary, and the stripped star we observed today likely formed from the secondary.
Since the stripped star has an evolutionary mass of $M_{\rm stripped} \sim 3.8\,\rm{M_\odot}$, this indicates that its progenitor (i.e., the secondary) must have been sufficiently massive \citep[$\gtrsim 10-12\,{\rm M_\odot}$; ][]{gotberg_2018} to produce such a system. If the progenitor of the compact object was initially more massive than this, it very likely produced a NS. The situation, however, is somewhat more complex than these considerations suggest, because the progenitor of the stripped star may have gained significant mass from the compact object progenitor via stable mass transfer. For example, for a configuration such as $M_{\rm 1, ZAMS} \sim 8-9 \,{\rm M_\odot}$ and $M_{\rm 2s, ZAMS} \sim 6-7 \,{\rm M_\odot}$, the primary is sufficiently low-mass to avoid core-collapse after envelope stripping, yet the secondary could plausibly gain enough mass to reach a pre-stripping mass of $10-12\,{\rm M_\odot}$ and produce a $3.8\,{\rm M_\odot}$ stripped star. Indeed, there are observed WD-NS binaries in which the WD is thought to have formed first \citep[e.g.,][]{vanKerkwijk_1999}; see \citet{Tauris_2000} and \citet{Toonen_2018} for theoretical studies of their formation channels. 

For J0040-7333 to be produced by such a scenario and host a WD, the initial conditions are rather constrained. The primary must have had an initial mass near $8-10\,M_\odot$, the mass ratio must have been close to unity, and mass transfer must have been quite conservative. A NS companion scenario requires less fine-tuning, and the properties of J0040-7333 are consistent with canonical NS-NS binary formation channels and with the observed binary pulsar population \citep[e.g., ][]{Tauris2015, tauris_2017}.

Quantifying the probability that the companion is a NS or a WD would require population synthesis simulations, which we have not attempted. Although a NS scenario appears favorable, we emphasize that a WD companion cannot be fully ruled out without further observational constraints.

\subsubsection{Stripped-star companion}
Although binary evolutionary models predict the existence of double stripped-star binaries, their formation requires a relatively fine-tuned set of initial conditions. Both formation channels discussed in Section \ref{sec: past evolution} require the two stars to have similar evolutionary timescales, thus favoring binaries with similar initial masses. In the two-phase stripping scenario, the secondary must initiate mass transfer before the primary completes its stripped-star lifetime; population synthesis models predict that initial mass ratios of $q_i \sim 0.85-1$ are required \citep{Toonen_2018}. The double-core CE scenario requires even more closely matched evolutionary timescales, favoring initial mass ratios of $q_i \approx 1$ \citep[e.g.,][]{Dewi_2006, Vigna_2018}. Additionally, this evolutionary pathway produces stripped stars with comparable masses \citep[e.g.,][]{Dewi_2006, Vigna_2018}, which does not match the inferred properties of J0040-7333. Given that J0040-7333 contains a $3.8\,{\rm M_\odot}$ stripped star and a likely substantially less massive companion ($\sim 1.5\,{\rm M_\odot}$), the inferred mass disparity is difficult to explain with the above formation scenarios.

Additionally, observational evidence suggests these systems are rare. No intermediate-mass stripped star with a stripped-star companion has been confirmed, and among the low-mass counterparts, only one sdB+sdB binary, PG 1544+488, has been observationally confirmed \citep{Ahmad_2004, scener_2014}. Thus, although we cannot fully rule out a stripped-star companion, the conditions required for the above theorized formation channels make it unlikely.

In the following section, we consider the future evolution of the system under the assumption that the companion is a NS. Regardless of the nature of the companion, J0040-7333 is expected to undergo another phase of mass transfer as the stripped star expands following core-helium depletion. The $\sim 3.8 \,{\rm M_\odot}$ stripped star will ultimately undergo core collapse and produce a NS. If the binary survives the natal kick, it can form a double compact-object binary. The resulting system will be a WD-NS binary if the current companion is a WD, or a NS-NS binary if the companion is a NS. If instead the companion is a stripped star, it may ultimately form a WD or NS-- depending on its core mass-- and produce either a NS-WD or NS-NS binary.

\subsection{Future evolution}\label{sec: future evolution}
Based on the stellar properties and the companion mass inferred from the spectra and SED, the stripped star currently fits well within its Roche lobe, with a filling factor of $R/R_{\rm Roche\,lobe} \sim 0.3$. However, after the stripped star exhausts its helium, it is expected to expand rapidly, up to tens of solar radii \citep{gotberg_2018, Laplace_2020, Nagarajan_2026}, which will lead to a subsequent phase of mass transfer. 

\begin{figure}[h!]
    \centering
    \includegraphics[width=\linewidth]{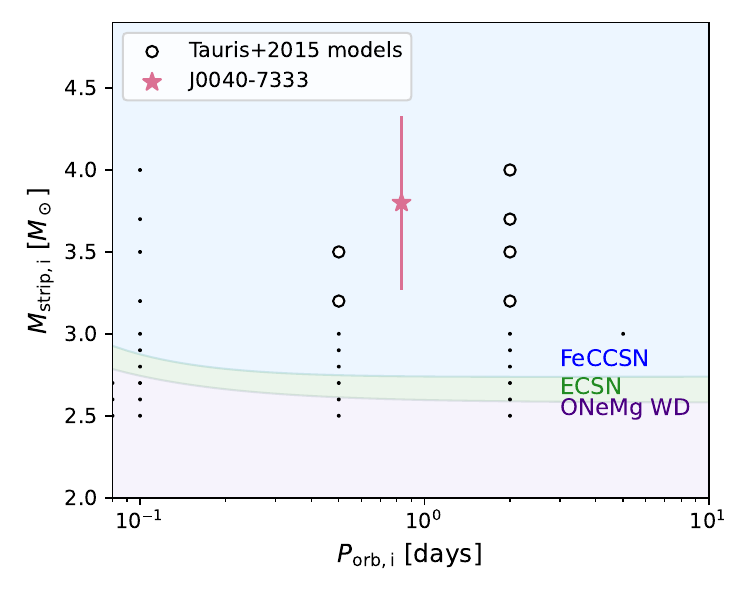}
    \caption{Predicted end-of-life outcome for stripped star + $1.35\,\rm{M_\odot}$ NS binaries as a function of stripped star mass and orbital period at the birth of the secondary stripped star. Shaded blue region marks the initial conditions of binaries which are predicted to undergo iron core-collapse supernovae (FeCCSNe). Shaded green region marks systems which are predicted to undergo electron-capture supernovae (ECSNe). Purple region marks the systems which do not ignite late stage burning and thus form ONeMg WDs. The pink star marks the inferred properties for J0040-7333. These models predict that the stripped star will explode and leave behind a NS+NS binary. Black points show all models available in \cite{Tauris2015}. White circular markers show those that are approximately matched to J0040-7333, which we use to predict the fate of J0040-7333 in Section \ref{sec: future evolution}.}
    \label{fig: future evolution}
\end{figure}

We use previously constructed evolutionary tracks of stripped star + NS binaries from \cite{Tauris2015} to explore possible future evolution of J0040-7333. 
These models follow the evolution of fully stripped helium stars with canonical NS companions ($M_{\rm NS} = 1.35\,M_\odot$) through silicon burning. We select models from their grid that approximately match the inferred orbital period ($P_{\rm orb} =  0.5-2 \,{\rm d}$) and stripped-star mass ($M_{\rm stripped, i} = 3.2 - 4.0\,\rm{M_\odot}$) of J0040-7333. We consider both models with and without stellar winds. Figure \ref{fig: future evolution} shows the distribution of stripped star masses vs. orbital periods and their final outcomes; we highlight J0040-7333 and show the \cite{Tauris2015} models with similar properties that we use for comparison.

These models assume naked helium stars without hydrogen-rich envelopes,  whereas J0040-7333 appears to have retained a substantial amount of surface hydrogen. This additional hydrogen layer can affect the radial expansion after core helium exhaustion \citep[e.g.,][]{Laplace_2020}, which in turn can influence how the binary will further evolve. Here we use the grid of models from \cite{Tauris2015} to gain a qualitative understanding of the future evolution of J0040-7333, while keeping this caveat in mind.

We also note there are other works that model the end-of-life of stripped star + NS systems \citep[e.g., ][]{2002_Dewi, Wu_2018, Wu_2020, gotberg_2020, Jiang_2021, Wu_2022}. Each of these models makes slightly different binary physics assumptions (e.g.,  mass transfer rates, duration of mass transfer, stellar physics); however, we find that qualitatively they all predict similar outcomes. For our discussion, we consider the results from \cite{Tauris2015} because it provides the densest grid of models which best matches the observed parameter space of J0040-7333.

\textit{Mass transfer and further stripping:}
\cite{Tauris2015} predict that binaries similar to J0040-7333 will undergo another stage of mass transfer (case BB or case BC), with maximum mass-transfer rates of $\dot M_{\rm strip} \sim 1.0 \times 10^{-4} -3.9 \times 10^{-4}\,\rm{M_\odot\,yr^{-1}}$. This mass transfer will further strip the envelope of the stripped star, which affects the type of SN that it will produce. Since the mass-transfer rate will exceed the Eddington luminosity, most of the mass would likely be lost from the system rather than accreted onto the NS. The system could be observable as a luminous X-ray binary during this stage.

\textit{Recycling of the NS companion:} 
although the NS is expected to accrete only a small amount of mass during the above mass-transfer phase, it can spin up the NS and produce a recycled pulsar \citep[e.g., ][]{Bhattacharya1991}. In the most conservative case where accretion is limited by the Eddington luminosity, the NS can be spun up to spin periods of $10-100$ milliseconds \citep{Tauris2012}. If the accretion is more efficient by a factor of a few, the NS can be spun up to even shorter periods. The mildly recycled NS would likely be observable as a radio pulsar in the future \citep[e.g., ][]{Kramer_1998, Gitika_2023}, but would not be easily detectable at SMC distances.

\textit{Final fate of the stripped star and SN:}
evolutionary models from \cite{Tauris2015} predict that stripped stars with masses $M_{\rm stripped} \gtrsim 2.5 - 3\,{\rm M_\odot}$ will undergo Fe core collapse and produce NSs (Figure \ref{fig: future evolution}). This collapse is accompanied by a SN, whose observable properties depend on the mass and the composition of the remaining pre-SN envelope. In particular, it is predicted that a helium envelope mass of $\gtrsim 0.15 \,\rm{M_\odot}$ is required to produce a Type Ib SN with visible helium features in its spectrum \citep[e.g.,][]{Hachinger2012}. The binary evolutionary models from \cite{Tauris2015} predict that, after additional stripping during Case BB mass transfer, the stripped star will retain an envelope of $\sim0.15-0.5\,\rm M_\odot$. This suggests that J0040-7333 will likely produce a Type Ib SN. However, the lower end of the predicted envelope masses is also consistent with an ``ultra-stripped'' SN \citep[Type Ic; e.g.,][]{Tauris2013}.

\textit{Double NS binary:}
the FeCCSN will impart a natal kick on the newly formed NS \citep[e.g.,][]{Janka2017}. The post-SN fate is sensitive to both the amount of mass lost during the explosion and the magnitude and the direction of the kick. The binary can either: (i) survive the kick and produce a NS binary or (ii) become unbound due to the kick. Quantifying the survival probability would require modeling of the SN dynamics and kick distribution, which is beyond the scope of this paper. Simple calculations predict that of order half of all systems will survive \citep[e.g., ][]{Valli_2025}. If the binary survives, the source will continue to lose angular momentum via gravitational-wave emission which shrinks its orbit, and will eventually lead to a merger detectable via gravitational waves \citep[e.g., ][]{Abbott_2017}.

\section{Comparison to related populations}\label{sec: comparsion}
We compare J0040-7333 to the four other published intermediate-mass stripped stars in the SMC \citep{Drout_2023, gotberg_2023, Laroche_2026}. Our candidate selection was motivated by the observed colors of the known stripped-star population. Consequently, J0040-7333 exhibits similar UV and optical colors to the previously identified stripped stars (top-left panel of Figure \ref{fig: selection function}).

\begin{figure*}
    \centering
    \includegraphics[width=0.8\linewidth]{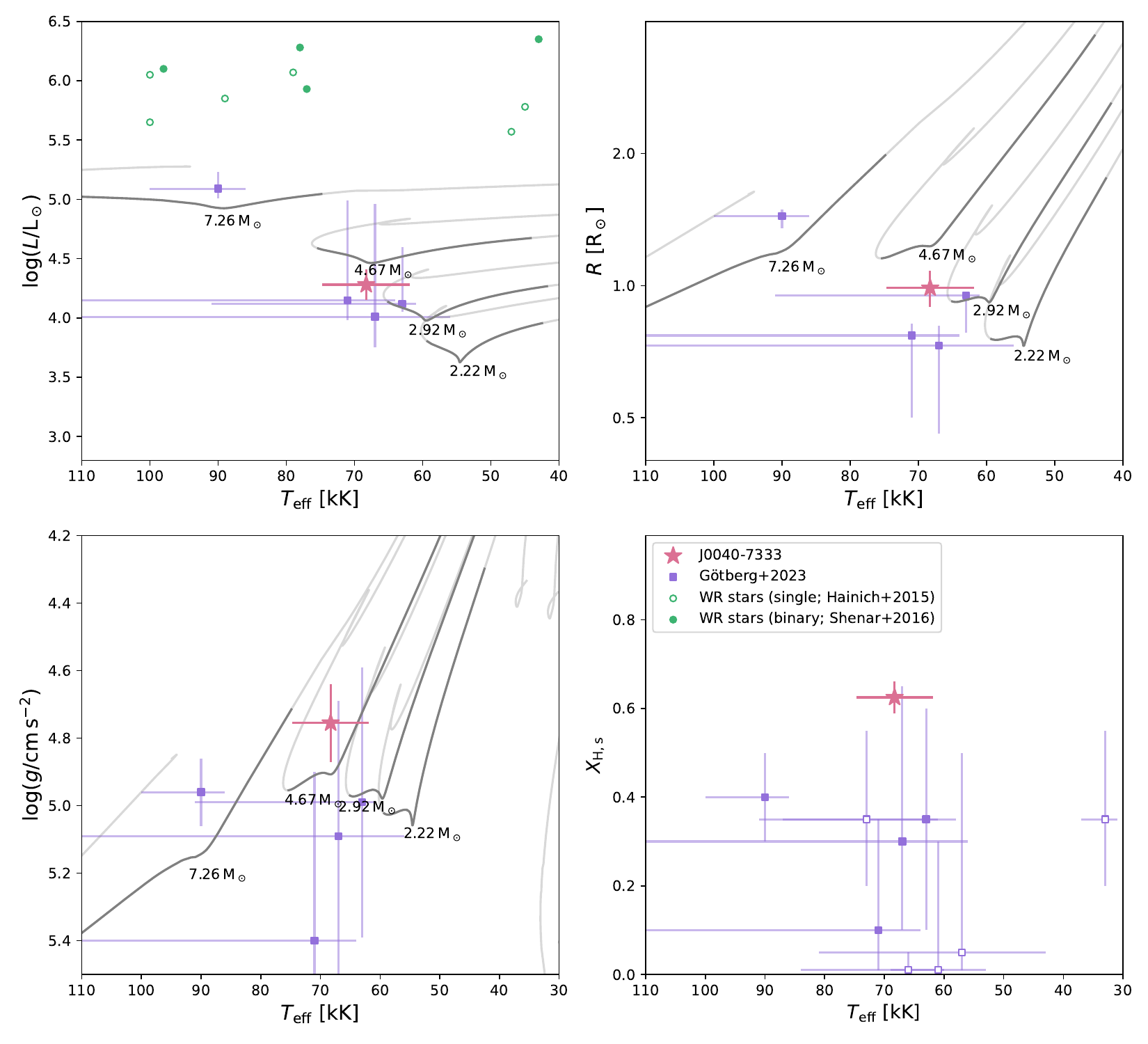}
    \caption{Comparison of J0040-7333 to other intermediate-mass stripped stars (purple) and WR stars (green) in the MCs. The solid pink star shows the best-fit stellar parameters for J0040-7333 with uncertainties (Table \ref{tab:mcmc results}). Gray curves show evolutionary models from \cite{gotberg_2018}, labeled by stripped star mass halfway through core helium burning. The dark gray region marks the central core helium burning stage ($0.05 \leq X_{\rm He,c}\leq 0.95$). We find that three of the four spectroscopically confirmed intermediate-mass stripped stars in the SMC \citep{Drout_2023, gotberg_2023} have stellar properties similar to J0040-7333. Top left: HR diagram. Top right: $T_{\rm eff}$-$R$ diagram. Bottom left: Kiel diagram. Bottom right: $T_{\rm eff}$ vs. $X_{\rm H, s}$. The hollow purple squares mark the intermediate-mass stripped stars in the LMC.}
    \label{fig: compare_d23}
\end{figure*}

Figure \ref{fig: compare_d23} compares J0040-7333 to other stripped stars on a Hertzsprung-Russell (HR) diagram, a $T_{\rm eff}$–$R$ plot, a Kiel diagram, and a $T_{\rm eff}-X_{\rm{H, s}}$ plot.
We find that the atmospheric parameters ($T_{\rm eff}$ and $\log(g)$) and stellar properties (e.g., $M_{\rm stripped}$, $R$, $\log(L)$) of J0040-7333 are broadly consistent with previously observed intermediate-mass stripped stars.\footnote{\cite{gotberg_2023} estimated uncertainties as the range in parameter values that produces a change in reduced $\chi^2$ of order unity. This results in much larger uncertainties than would be obtained using the conventional definition of a change in \textit{total} $\chi^2$ of order unity. In contrast, we estimate statistical uncertainties using MCMC sampling with a likelihood that is tied to the total $\chi^2$, which yields substantially smaller uncertainties. We additionally estimate systematic uncertainties by comparing the inferred parameters obtained from different fitting approaches. The true uncertainties likely lie somewhere between those inferred from the two approaches. Although the uncertainties reported by \cite{gotberg_2023} neglect systematic uncertainties, their adopted method likely overestimates the statistical uncertainties.}

The one notable discrepancy between J0040-7333 and the previously identified hot intermediate-mass stripped stars is the surface hydrogen mass fraction. J0040-7333 has $X_{\rm H,s} = 0.625 \pm 0.037$, which is higher than value ($X_{\rm H,s} \sim 0.1- 0.4$) measured for the previously observed stripped stars. This suggests that J0040-7333 has retained more of its hydrogen envelope than is typical for hot, compact stripped stars (see also Section \ref{sec: hydrogen fraction} for additional discussion).

It is interesting to note that the stripped stars in the SMC (J0040-7333 and \citealt{gotberg_2023}) typically have higher inferred hydrogen mass fractions ($X_{\rm H,s} \sim 0.1- 0.65$) compared to those in the LMC ($X_{\rm H,s} \sim 0.01- 0.35$). 
\cite{Laroche_2026} found that the stripped stars with lowest surface hydrogen abundances ($X_{\rm H, s} \approx0.01$) in the LMC may represent an entirely different population of stripped stars that are not currently residing in a binary. Regardless, the currently known SMC stripped stars appear to have systematically larger $X_{\rm H,s}$ than their LMC counterparts. One possible explanation is that stars typically have weaker stellar winds in lower-metallicity environments \citep{Vink2017}. Lower metallicity donors are predicted to contract and detach earlier during mass transfer, leaving a thicker hydrogen envelope. However, the observed differences could also arise from other factors -- e.g., different stripped-star formation channels -- or may simply reflect small-number statistics. We also note that J0040-7333 is located toward the outskirts of the SMC, whereas the previously identified stripped stars are clustered toward the central regions of the SMC (bottom-left panel of Figure \ref{fig: selection function}). The lower local metallicity in this region compared to the central star-forming regions \citep[e.g.,][]{Choudhury_2018, Lah_2024} may also play a role in  explaining the unusually high $X_{\rm H, s}$ of J0040-7333  compared to other SMC stripped stars.

A subset of the confirmed intermediate-mass stripped stars in the MCs have orbital solutions and constraints on their companion masses \citep{Laroche_2026}. These systems span orbits with periods from $P_{\rm orb} \leq \rm day$ to $P_{\rm orb} \approx300\,{\rm days}$ and have RV semi-amplitudes of $K = 18 - 76\,{\rm km\,s^{-1}}$. The short-period systems have minimum companion masses of $0.5-1\,{\rm M_\odot}$, while the long-period system has a substantially higher minimum companion mass of $\sim 2.9\,{\rm M_\odot}$. The nature of these companions has not yet been determined: their inferred masses are consistent with MS, stripped star, or WD/NS companions for the short-period systems, and a MS, stripped star, or BH companion for the long-period system.

\begin{figure}[h!]
    \centering
    \includegraphics[width=\linewidth]{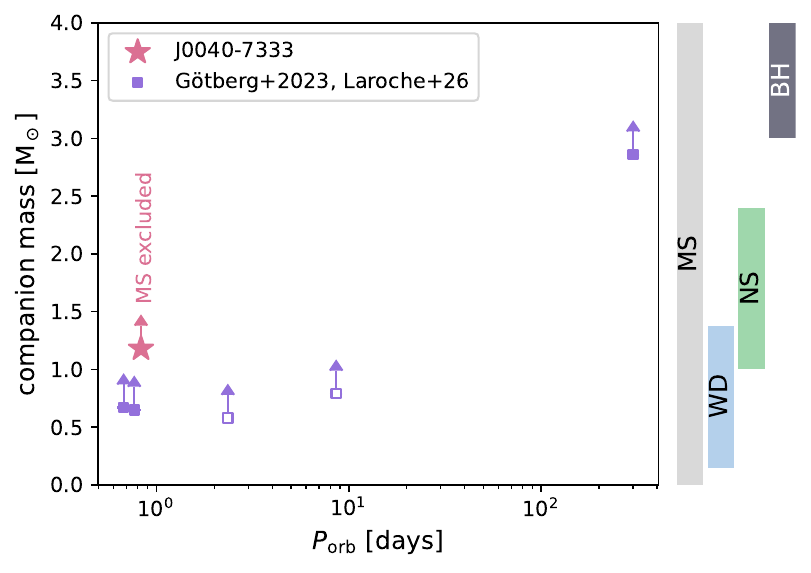}
    \caption{Orbital period vs. minimum companion mass. Pink star marks J0040-7333. Purple squares show previously identified intermediate-mass stripped stars from literature \citep{Drout_2023, gotberg_2023, Laroche_2026}; the filled and open markers represent the SMC and LMC populations, respectively. The colored bars on the right indicate the typical masses corresponding to different types of companions. A MS companion is ruled out for J0040-7333 using light-curve modeling (Section \ref{sec: rule out MS}). The nature of the companions for the remaining sources has not yet been constrained, and their inferred masses are consistent with MS, stripped-star, and compact-object companions.}
    \label{fig: orb vs comp}
\end{figure}

Figure \ref{fig: orb vs comp} compares the orbital periods and minimum companion masses of this population with those inferred for J0040-7333.
We find that the orbital period of J0040-7333 is similar to those of Star 2 and Star 4. However, J0040-7333 has a comparatively high minimum companion mass, consistent with its large RV semi-amplitude ($K\sim 90\,{\rm km\,s^{-1}}$) relative to those of the other two systems ($K\sim 65\,{\rm km\,s^{-1}}$). It remains to be determined whether this reflects a more massive companion or simply a more edge-on inclination.

\section{Summary and conclusions}\label{sec: conclusions}
We present the discovery of an intermediate-mass stripped star in a binary with a compact-object companion. We discovered SMASH J004005-733351 during our search for stripped stars using an optical blue-excess selection with photometry from SMASH. We summarize our main results below:

\begin{enumerate}[label=(\roman*), wide, labelwidth=!, labelindent=0pt]
\item \textit{RV follow-up and orbital solution:} we carried out spectroscopic RV monitoring, obtaining a total of 16 epochs of RVs using Magellan/MagE and Gemini/GMOS-S spanning a total baseline of $\sim6$ months (Figure \ref{fig: spectra all}; Table \ref{table: RVs}).  Using \texttt{the joker} and MCMC sampling, we constrained the system's orbital solution (Figure \ref{fig: orbital solution}). We find the system is in a relatively compact orbit with an orbital period of $P_{\rm orb} = 19.947\pm 0.002\,{\rm hours}$ and an RV semi-amplitude of $K = 91.7^{+3.2}_{-3.1}\,{\rm km \,s^{-1}}$. The eccentricity of the system is low, consistent with a circular orbit (Table \ref{table: orbital solutions}).

\item \textit{Spectral \& SED modeling: }to constrain the stellar properties of the stripped star (Table \ref{tab:mcmc results}), we simultaneously fit the observed spectra (Figure \ref{fig: spectra}) and photometric SED (Figure \ref{fig: sed fit}). We infer an evolutionary mass of $M_{\rm stripped} = 3.84 \pm 0.54\,M_\odot$. The inferred effective temperature, $T_{\rm eff} = 68.3 \pm 6.4$ kK, surface gravity, $\log g= 4.76 \pm 0.12$, and luminosity, $\log(L/L_\odot)= 4.28 \pm 0.13$, are consistent with the previously observed population in the SMC (Figure \ref{fig: compare_d23}). 
However, we find that the newly discovered stripped star has a larger surface hydrogen mass fraction ($X_{\rm H, s} = 0.63 \pm 0.04$) than the previously identified hot and compact stripped stars in the MCs. 
The unusually high $X_{\rm H,s}$ is not reproduced by standard  binary evolutionary models, suggesting that these models may not fully capture the relevant binary physics (Section \ref{sec: hydrogen fraction}).

\item \textit{Nature of the companion: }we infer a minimum companion mass of $M_{\rm comp} \approx 1.19\,{\rm M_\odot}$ using the binary mass function (Figure \ref{fig: comp mass}). We search the OGLE $I$-band light curve for periodic variability and rule out photometric variability above $\sim 1\,\%$ at the orbital period (Figure \ref{fig: lomb-scargle}). From the lack of a reflection effect, we rule out a MS companion (Figure \ref{fig: example light curves}). A NS companion is most likely given evolutionary-timescale arguments, though it is difficult to definitively rule out a massive-WD companion. A low-mass stripped star companion (e.g., sdO) cannot be ruled out either, but is unlikely (Section \ref{sec: nature of companion}). We find that J0040-7333 is not detected in radio or X-ray; however we do not expect the system to be detectable even for a NS companion (Section \ref{sec: radio & xray}.) 

\item \textit{Formation history and future evolution}: if J0040-7333 hosts a NS companion, standard binary evolution scenarios suggest that the system likely formed via two stages of mass transfer. Most models predict that the first interaction is stable, whereas the second is unstable and likely involves CEE. Although the nature of the first mass-transfer stage is difficult to constrain, the compact and nearly circular orbit we observe today is consistent with a system that has recently gone through CEE. Using the BPASS grid of models, we infer a pre-CE orbital period of $2.5-10$ days and a progenitor mass of $12.0-14.0\,\rm{M_\odot}$ for the stripped star (Section \ref{sec: past evolution}). Evolutionary models of stripped star + NS binaries predict that, following core helium depletion, the stripped star will expand and undergo a subsequent phase of mass transfer. This interaction could sufficiently spin up the NS companion, producing a mildly recycled pulsar. At the end of its evolution, the stripped star will likely undergo core-collapse as a Type Ib SN and form a second NS. If J0040-7333 remains bound following the natal kick, gravitational-wave radiation will shrink the orbit, eventually leading to a double-NS merger (Section \ref{sec: future evolution}).

\end{enumerate}
Higher resolution optical spectral follow-up could provide robust constraints on the rotation of the stripped star, $v\sin(i)$.
If tidal synchronization is assumed, stellar rotation velocity, $v$, can be estimated from the orbital period and stripped-star radius. Combining this with a measurement of $v\sin(i)$ can provide constraints on inclination angle and companion mass. Further UV spectroscopic follow-up could provide constraints on the wind-mass loss rate. Deeper X-ray and radio observations may be able to detect a NS companion. Nonetheless, the current data already provide a rare observational anchor for a progenitor system to a double NS binary.

J0040-7333 is one of several systems identified through our search for optical blue-excess sources using photometry from SMASH. Multi-epoch spectroscopic observations of the remaining candidates are ongoing. We plan to publish these in a future paper. Continued efforts to identify and characterize stripped stars both in the MCs and in the Milky Way are crucial for expanding the observed population and placing stronger constraints on binary evolution models and understanding binary interactions.



\section{acknowledgments}
We thank Alex Laroche and Dandan Wei for helpful discussions.

This research was supported in part by NSF grant AST-2540180 and in part by Scialog grant \#SA-LSST-2024-114c from Research Corporation for Science Advancement. T.C. was supported by NASA through the NASA Hubble Fellowship grant HST-HF2-51527.001-A awarded by the Space Telescope Science Institute, which is operated by the Association of Universities for Research in Astronomy, Inc., for NASA, under contract NAS5-26555. M.R.D. acknowledges support from NSERC through grant RGPIN-2025-06224, the Canada Research Chairs Program (CRC-2023-00127), the Ontario ERA program (ER22-17-164) and the Dunlap Institute at the University of Toronto.
 
This paper includes data gathered with the 6.5 meter Magellan Telescopes located at Las Campanas Observatory, Chile. Support for the design and construction of the Magellan Echellette Spectrograph was provided by the Observatories of the Carnegie Institution of Washington, the School of Science of the Massachusetts Institute of Technology, and the National Science Foundation through a collaborative Major Research Instrumentation grant to Carnegie and MIT (AST-0215989). These observations were obtained under NOIRLab program ID 2025A 226700. This research made use of \texttt{PypeIt},\footnote{\url{https://pypeit.readthedocs.io/en/latest/}} a Python package for the semi-automated reduction of astronomical slit-based spectroscopy \citep{Prochaska_2020a, Prochaska_2020b}.

Based on observations obtained at the international Gemini Observatory, a program of NSF NOIRLab, which is managed by the Association of Universities for Research in Astronomy (AURA) under a cooperative agreement with the U.S. National Science Foundation on behalf of the Gemini Observatory partnership: the U.S. National Science Foundation (United States), National Research Council (Canada), Agencia Nacional de Investigaci\'{o}n y Desarrollo (Chile), Ministerio de Ciencia, Tecnolog\'{i}a e Innovaci\'{o}n (Argentina), Minist\'{e}rio da Ci\^{e}ncia, Tecnologia, Inova\c{c}\~{o}es e Comunica\c{c}\~{o}es (Brazil), and Korea Astronomy and Space Science Institute (Republic of Korea). Data was obtained as part of the GS-2025B-FT-107 program, and processed using DRAGONS (Data Reduction for Astronomy from Gemini Observatory North and South).

We acknowledge the use of public data from the Swift data archive.

This research is based in part on results obtained from the AstroSat mission of the Indian Space Research Organisation (ISRO) and archived at the Indian Space Science Data Centre (ISSDC).

The OGLE project has received funding from the Polish National Science Centre grant OPUS 2024/55/B/ST9/00447 awarded to A.U.

This research has made use of the SVO Filter Profile Service ``Carlos Rodrigo,'' funded by MCIN/AEI/10.13039/501100011033/ through grant PID2023-146210NB-I00.


%

\vspace{5mm}


\software{Astropy, \textit{The Tractor}, \texttt{pypeit}, DRAGONS,  \texttt{The joker}, \texttt{emcee}, PHOEBE}



\appendix

\section{Normalized spectra}\label{appdx: normalized spectra}
\begin{figure*}[htb!]
    \centering
    \includegraphics[width=\linewidth]{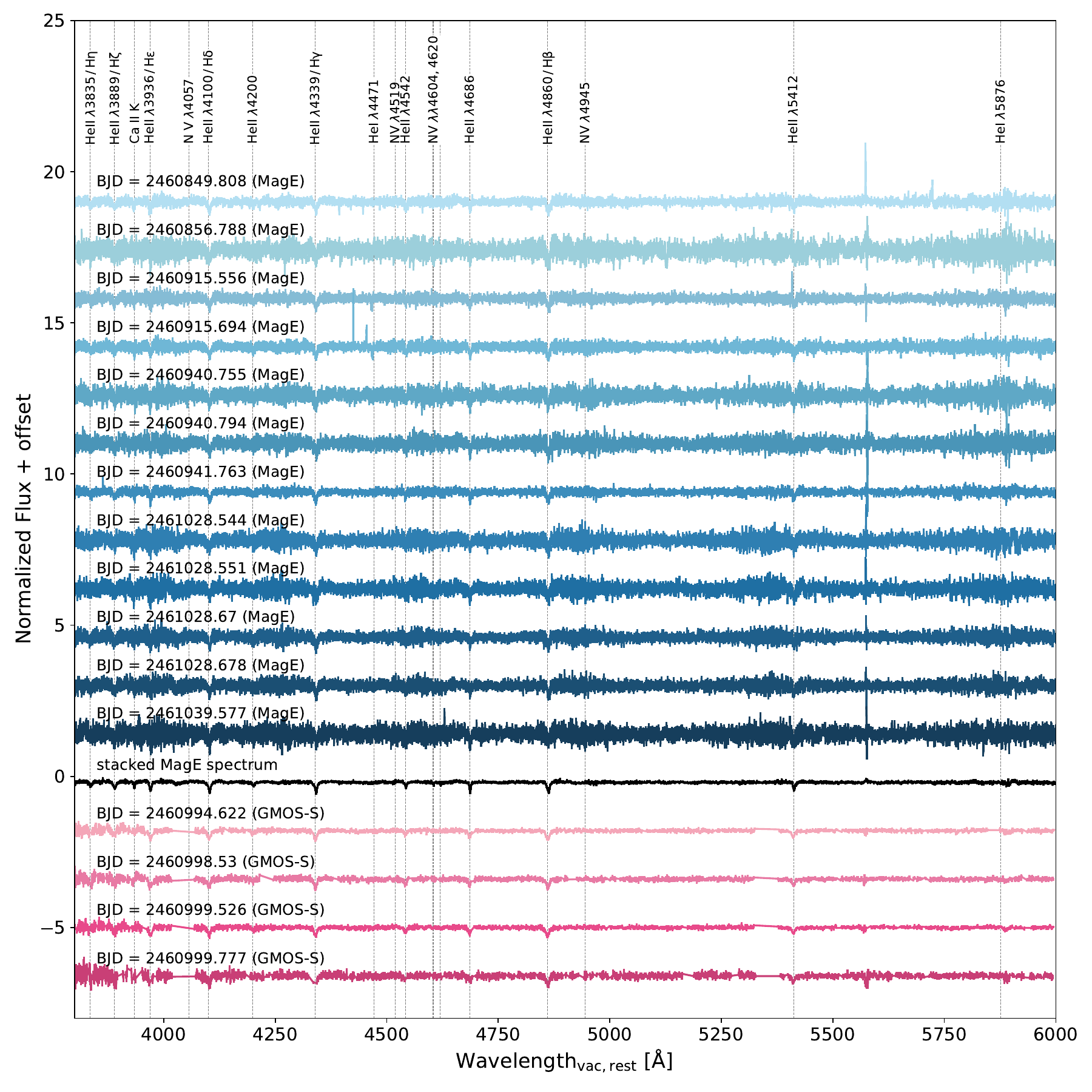}
    \caption{
    All follow-up spectra taken on MagE (blue) and GMOS-S (pink), shown with their uncertainties. Labels indicate observation time and instrument. Black curve shows the high-SNR spectrum produced by stacking all MagE observations. Two features near $4450\,{\rm \AA}$ only appear in the two spectra obtained on BJD = 2460915; we identify these as systematic artifacts because they are not consistent with any expected stellar features and are not present in the spectra from other nights. We mask these regions for these two epochs when constructing the stacked spectrum. We also identify residual sky-subtraction features near $5570\,\rm{\AA}$ in several epochs, which arise from imperfect sky subtraction in the reduction pipeline.}    
    \label{fig: spectra all}
\end{figure*}

In total we obtained 16 epochs of spectra: 12 on MagE and 4 on GMOS-S (Section \ref{sec: spectral data}). We show all normalized spectra and the high SNR spectrum produced by stacking all of the MagE observations in Figure \ref{fig: spectra all}.

\section{Radial velocity measurements}\label{appendix: rv table}
\begin{figure}[h!]
    \centering
    \includegraphics[width=0.35\linewidth]{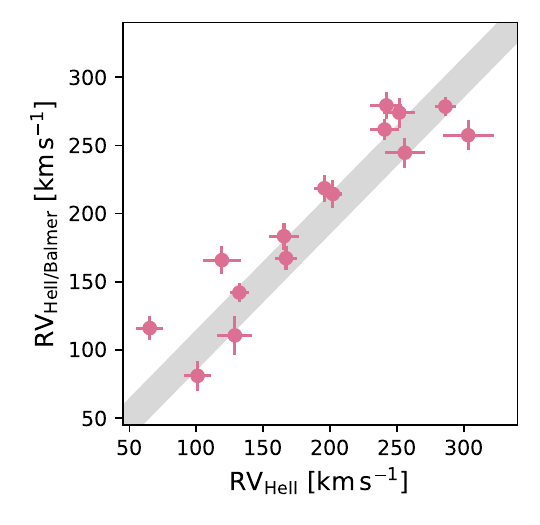}
    \caption{Comparison of RV measurements from two independent sets of lines: (i) pure He\,{\sc ii} lines and (ii) He\,{\sc ii}/Balmer blends. For each epoch, we estimate RVs using the methods described in Section \ref{sec: RV measurements}. The gray shaded region shows the 1:1 relation with $\pm15\,{\rm km\,s^{-1}}$ deviations. If the He\,{\sc ii}/Balmer blends were contaminated by a luminous companion, its contributions would bias the measured RVs towards the companion's velocity. Compared to the RVs measured from the pure He\,{\sc ii} lines, this would draw the RVs measured from blended lines toward the center-of-mass RV. We find no evidence for such a trend.}
    \label{fig: rv_compare lines}
\end{figure}

Table \ref{table: RVs} lists all spectral observations, including the instrument, time of observation, instrument setup, and the measured RVs.
\begin{sidewaystable}[h!]
\begin{tabular}{c c c c c c c}
\hline\hline
observation time [BJD] & RV [$\rm{km\,s^{-1}}$] & instrument & slit-width & exposure time [s] & SNR & lines used for RV fitting \\
\hline
\hline
     2460849.808& $281.96\pm 6$ & MagE &$0.85^{\prime\prime}$&1800&$\sim 15$& \makecell[l]{He\,{\sc ii} $\lambda$4100/H$\delta$, He\,{\sc ii} $\lambda$4339/H$\gamma$, He\,{\sc ii} $\lambda$4542, \\He\,{\sc ii} $\lambda$4686,He\,{\sc ii} $\lambda$4860/H$\beta$, He\,{\sc ii} $\lambda$5411\\}\\
     2460856.788& $120.44 \pm 10$ & MagE & $0.85^{\prime\prime}$&1800&$\sim 8$&\makecell[l]{He\,{\sc ii} $\lambda$4100/H$\delta$, He\,{\sc ii} $\lambda$4339/H$\gamma$, He\,{\sc ii} $\lambda$4686, \\He\,{\sc ii} $\lambda$4860/H$\beta$}\\
     2460915.556& $254.35\pm7$ & MagE &$0.85^{\prime\prime}$&900&$\sim 13$&\makecell[l]{He\,{\sc ii} $\lambda$4100/H$\delta$, He\,{\sc ii} $\lambda$4339/H$\gamma$, He\,{\sc ii} $\lambda$4686, \\He\,{\sc ii} $\lambda$4860/H$\beta$}\\
     2460915.694& $165.27\pm7$ & MagE &$0.85^{\prime\prime}$&900&$\sim 13$&\makecell[l]{He\,{\sc ii} $\lambda$4100/H$\delta$, He\,{\sc ii} $\lambda$4339/H$\gamma$, He\,{\sc ii} $\lambda$4686, \\He\,{\sc ii} $\lambda$4860/H$\beta$, He\,{\sc ii} $\lambda$5411}\\
     2460940.755 & $95.57\pm 8$ & MagE &$0.85^{\prime\prime}$&900&$\sim 10$&\makecell[l]{He\,{\sc ii} $\lambda$4100/H$\delta$, He\,{\sc ii} $\lambda$4339/H$\gamma$, He\,{\sc ii} $\lambda$4686, \\He\,{\sc ii} $\lambda$4860/H$\beta$, He\,{\sc ii} $\lambda$5411}\\
     2460940.794 & $92.71\pm8$ & MagE &$0.85^{\prime\prime}$&900&$\sim 10$&\makecell[l]{He\,{\sc ii} $\lambda$4100/H$\delta$, He\,{\sc ii} $\lambda$4339/H$\gamma$, He\,{\sc ii} $\lambda$4686,\\ He\,{\sc ii} $\lambda$4860/H$\beta$}\\
     2460941.763& $136.81 \pm 5$ & MagE& $0.85^{\prime\prime}$&1000&$\sim 17$&\makecell[l]{He\,{\sc ii} $\lambda$4100/H$\delta$, He\,{\sc ii} $\lambda$4339/H$\gamma$, He\,{\sc ii} $\lambda$4542, \\He\,{\sc ii} $\lambda$4686, He\,{\sc ii} $\lambda$4860/H$\beta$, He\,{\sc ii} $\lambda$5411}\\
     2460994.622 & $175.12\pm 8$ & GMOS-S &$0.75^{\prime\prime}$&1200&$\sim 24$& \makecell[l]{He\,{\sc ii} $\lambda$4100/H$\delta$, He\,{\sc ii} $\lambda$4339/H$\gamma$, He\,{\sc ii} $\lambda$4542, \\He\,{\sc ii} $\lambda$4686, He\,{\sc ii} $\lambda$4860/H$\beta$, He\,{\sc ii} $\lambda$5411}\\
     2460998.530 & $268.60\pm10$ & GMOS-S &$0.75^{\prime\prime}$&1200&$\sim 20$& \makecell[l]{He\,{\sc ii} $\lambda$4100/H$\delta$, He\,{\sc ii} $\lambda$4339/H$\gamma$, He\,{\sc ii} $\lambda$4860/H$\beta$, \\He\,{\sc ii} $\lambda$5411}\\
     2460999.526 & $246.29 \pm 9$ & GMOS-S &$0.75^{\prime\prime}$&1200&$\sim 23$&\makecell[l]{He\,{\sc ii} $\lambda$4100/H$\delta$, He\,{\sc ii} $\lambda$4339/H$\gamma$, He\,{\sc ii} $\lambda$4542, \\He\,{\sc ii} $\lambda$4860/H$\beta$, He\,{\sc ii} $\lambda$5411}\\
      2460999.777 & $74.06\pm 15$ & GMOS-S & $0.75^{\prime\prime}$&1200&$\sim 15$&\makecell[l]{He\,{\sc ii} $\lambda$4100/H$\delta$,  He\,{\sc ii} $\lambda$4860/H$\beta$, He\,{\sc ii} $\lambda$5411}\\
      2461028.544 & $264.13 \pm 8$ & MagE& $0.70^{\prime\prime}$&600&$\sim 10$&\makecell[l]{He\,{\sc ii} $\lambda$4100/H$\delta$, He\,{\sc ii} $\lambda$4339/H$\gamma$, He\,{\sc ii} $\lambda$4686, \\He\,{\sc ii} $\lambda$4860/H$\beta$, He\,{\sc ii} $\lambda$5411}\\
     2461028.551 & $264.18 \pm 8$ & MagE& $0.70^{\prime\prime}$&600&$\sim 10$&\makecell[l]{He\,{\sc ii} $\lambda$4100/H$\delta$, He\,{\sc ii} $\lambda$4339/H$\gamma$, He\,{\sc ii} $\lambda$4542,\\ He\,{\sc ii} $\lambda$4686, He\,{\sc ii} $\lambda$4860/H$\beta$, He\,{\sc ii} $\lambda$5411}\\
     2461028.670 & $206.77\pm 6$ & MagE& $0.70^{\prime\prime}$&600&$\sim 12$&\makecell[l]{He\,{\sc ii} $\lambda$4100/H$\delta$, He\,{\sc ii} $\lambda$4339/H$\gamma$, He\,{\sc ii} $\lambda$4542,\\ He\,{\sc ii} $\lambda$4686, He\,{\sc ii} $\lambda$4860/H$\beta$, He\,{\sc ii} $\lambda$5411}\\
      2461028.678 & $204.46\pm 7$ & MagE&$0.70^{\prime\prime}$&600&$\sim 12$& \makecell[l]{He\,{\sc ii} $\lambda$4100/H$\delta$, He\,{\sc ii} $\lambda$4339/H$\gamma$, He\,{\sc ii} $\lambda$4542,\\ He\,{\sc ii} $\lambda$4686, He\,{\sc ii} $\lambda$4860/H$\beta$, He\,{\sc ii} $\lambda$5411}\\
       2461039.577 & $149.37\pm9$ & MagE& $0.70^{\prime\prime}$&1200&$\sim9$&\makecell[l]{He\,{\sc ii} $\lambda$4100/H$\delta$, He\,{\sc ii} $\lambda$4339/H$\gamma$, He\,{\sc ii} $\lambda$4686,\\ He\,{\sc ii} $\lambda$4860/H$\beta$}\\
\hline
\end{tabular}
\caption{All spectroscopic follow-up observations and the measured RVs for each epoch.}
\label{table: RVs}
\end{sidewaystable}
To assess whether the RV measurements are sensitive to the choice of spectral line, we independently repeat the RV measurements described in Section \ref{sec: RV measurements} using two subsets of lines. In particular, we compare RVs measured from pure He\,{\sc ii} lines and those measured from He\,{\sc ii}/Balmer blends, since contamination from a luminous companion would introduce a systematic offset in the Balmer lines. We compare the resulting RVs in Figure \ref{fig: rv_compare lines}. We find that the two sets of RVs broadly follow a 1:1 relation.

\section{Best-fit parameters for all $A_V$}\label{apdx: mcmc AV}
We repeat the fitting procedures described in Section \ref{sec: mcmc setup no evol} and \ref{sec: mcmc sampling setup} for a range of $A_V$ values to assess systematic uncertainties associated with fixing the reddening value.

\subsection{SED and spectral fit without evolutionary tracks}\label{apdx: no evol model fit}
\begin{figure}[h!]
    \centering
    \includegraphics[width=\linewidth]{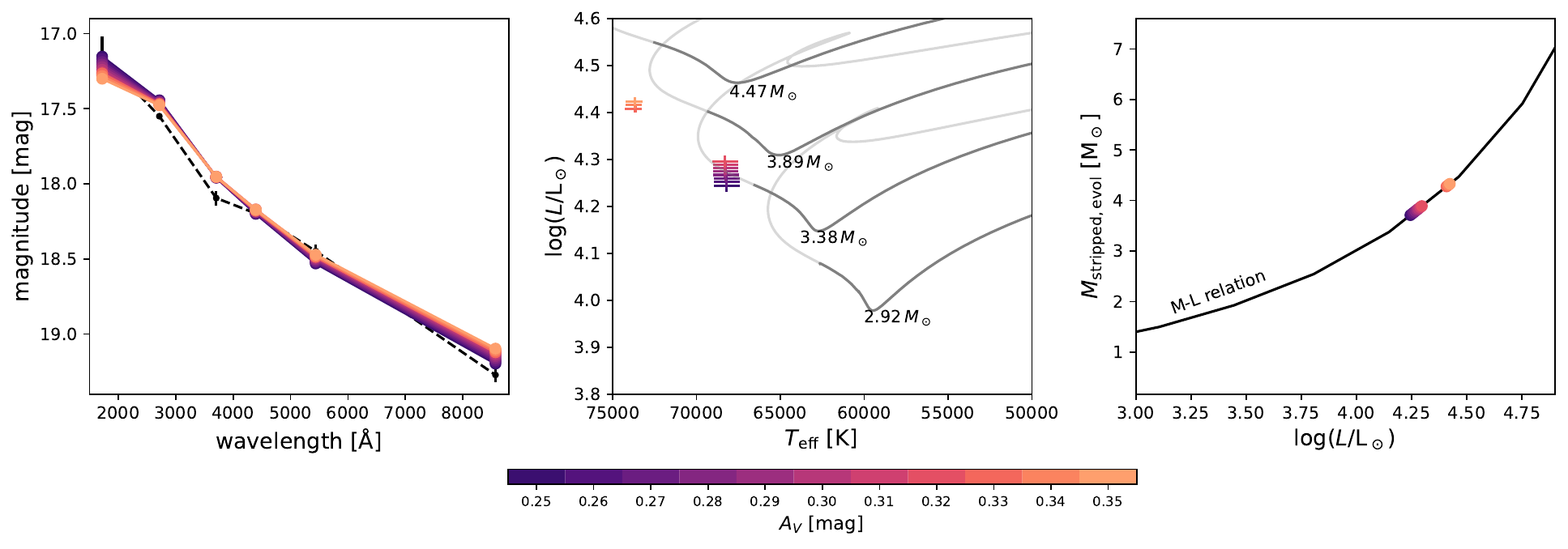}
    \caption{Results for the SED and spectral fits without evolutionary model constraints for a range of $A_V$ values (Section \ref{sec: mcmc setup no evol}; Table \ref{table: full mcmc no evol}). Left: black points show the observed SED constructed from optical and UV photometry from the MCPS catalog, \textit{Swift}-UVOT, and UVIT. 
    Colored curve show the best-fit SED model for a range of $A_V$ values.
    Middle: HR diagram. The colored markers show the best-fit $T_{\rm eff}$ and $\log(L)$ for each $A_V$ with their statistical uncertainties from their MCMC posteriors. Gray lines show evolutionary tracks for intermediate-mass stripped stars taken from \cite{gotberg_2018}. The thick segments indicate the stripped-star lifetime ($0.05 \leq X_{\rm He, c} \leq 0.95$). Right: black curve shows the $M-\log(L)$ relation for stripped stars, using the intermediate-mass stripped-star masses and luminosities at halfway through He burning \citep{gotberg_2023}. Colored markers show the best-fit $\log(L)$ for each $A_V$, which we use to derive the best-fit evolutionary mass for the stripped star.}
    \label{fig: mcmc no evol sed + hr}
\end{figure}

Table \ref{table: full mcmc no evol} shows the best-fit parameters for $A_V = 0.25-0.35$ mag using the SED and spectral fitting method described in Section \ref{sec: mcmc setup no evol}.
\begin{table*}[h!]
    \renewcommand{\arraystretch}{1.1}
    \begin{tabular}{c|c c c c c c c}
    \hline
     $A_V$ & $T_{\rm eff}$ & $\log(g / \rm{cm\,s^{-2}})$ & $X_{\rm H,s}$ & $R$ &  $\log(L / \rm L_\odot)$ & $M_{\rm evol}$ & $M_{\rm spec}$\\
     $[$mag$]$ & [K] & & & [$\rm R_\odot$]& & [$\rm M_\odot$] & [$\rm M_\odot$]\\\hline
     0.25 & $68206^{+752}_{-807}$ & $4.755^{+0.002}_{-0.002}$ & $0.626^{+0.011}_{-0.011}$ & $0.949^{+0.010}_{-0.010}$ & $4.244^{+0.013}_{-0.014}$ & $3.711^{+0.046}_{-0.049}$ & $1.868^{+0.047}_{-0.044}$\\
     0.26 & $68237^{+749}_{-807}$ & $4.755^{+0.002}_{-0.002}$ & $0.625^{+0.010}_{-0.010}$ & $0.957^{+0.010}_{-0.010}$ & $4.252^{+0.013}_{-0.014}$ & $3.737^{+0.046}_{-0.049}$ & $1.899^{+0.046}_{-0.044}$\\
     0.27 & $68222^{+743}_{-791}$ & $4.755^{+0.002}_{-0.002}$ & $0.626^{+0.012}_{-0.012}$ & $0.964^{+0.010}_{-0.010}$ & $4.259^{+0.013}_{-0.014}$ & $3.761^{+0.046}_{-0.048}$ & $1.930^{+0.047}_{-0.044}$\\
     0.28 & $68279^{+735}_{-804}$ & $4.755^{+0.002}_{-0.002}$ & $0.625^{+0.011}_{-0.011}$ & $0.972^{+0.010}_{-0.010}$ & $4.267^{+0.013}_{-0.014}$ & $3.789^{+0.047}_{-0.048}$ & $1.960^{+0.048}_{-0.044}$\\
     0.29 & $68300^{+718}_{-790}$ & $4.755^{+0.002}_{-0.002}$ & $0.625^{+0.011}_{-0.011}$ & $0.980^{+0.010}_{-0.010}$ & $4.275^{+0.013}_{-0.014}$ & $3.816^{+0.045}_{-0.048}$ & $1.993^{+0.048}_{-0.045}$\\
     0.30 & $68288^{+731}_{-794}$ & $4.755^{+0.002}_{-0.002}$ & $0.625^{+0.012}_{- 0.011}$ & $0.988^{+0.011}_{- 0.010}$ & $4.281^{+0.013}_{- 0.014}$ & $3.840^{+0.046}_{- 0.048}$ & $2.025^{+0.050}_{- 0.046}$\\
     0.31 & $68299^{+714}_{-777}$ & $4.755^{+0.002}_{-0.002}$ & $0.625^{+0.011}_{-0.011}$ & $0.996^{+0.010}_{-0.010}$ & $4.289^{+0.013}_{-0.014}$ & $3.866^{+0.045}_{-0.048}$ & $2.059^{+0.049}_{-0.046}$\\
     0.32 & $68307^{+738}_{-776}$ & $4.755^{+0.002}_{-0.002}$ & $0.625^{+0.011}_{-0.012}$ & $1.005^{+0.010}_{-0.010}$ & $4.296^{+0.013}_{-0.014}$ & $3.891^{+0.047}_{-0.048}$ & $2.093^{+0.050}_{-0.047}$\\
     0.33 & $73654^{+651}_{-410}$ & $4.815^{+0.003}_{-0.003}$ & $0.604^{+0.009}_{-0.009}$ & $0.982^{+0.008}_{-0.008}$ & $4.408^{+0.009}_{-0.008}$ & $4.279^{+0.032}_{-0.028}$ & $2.298^{+0.042}_{-0.044}$\\
     0.34 & $73663^{+559}_{-415}$ & $4.815^{+0.003}_{-0.003}$ & $0.604^{+0.009}_{-0.009}$ & $0.990^{+0.008}_{-0.008}$ & $4.416^{+0.009}_{-0.008}$ & $4.305^{+0.031}_{-0.029}$ & $2.336^{+0.044}_{-0.045}$\\
     0.35 & $73663^{+568}_{-425}$ & $4.815^{+0.003}_{-0.003}$ & $0.604^{+0.009}_{-0.009}$ & $0.998^{+0.008}_{-0.008}$ & $4.423^{+0.009}_{-0.008}$ & $4.330^{+0.032}_{-0.029}$ & $2.375^{+0.044}_{-0.045}$\\
     \hline
    \end{tabular}
    \caption{Best-fit parameters constrained for J0040-7333 from SED and spectral fits without constraints from evolutionary models (Section \ref{sec: mcmc setup no evol}). $A_V$ is fixed for each run. We report the median, with the 16th and 84th percentiles from the MCMC posteriors as the lower and upper statistical uncertainties, respectively.}
    \label{table: full mcmc no evol}
\end{table*}
Figure \ref{fig: mcmc no evol sed + hr} shows a summary plot of all fits.

The inferred stellar parameters are broadly consistent across the full range of explored $A_V$ values. Generally, higher extinction yields
higher inferred luminosities. This is because a higher intrinsic luminosity is required to compensate for the increased extinction to reproduce the observed photometry. This, in turn, affects the inferred stripped-star mass, which we derive directly from the mass--luminosity relation (see the right panel of Figure \ref{fig: mcmc no evol sed + hr}). Additionally, we find a significant discrepancy between the inferred evolutionary and spectroscopic masses. This suggests that the uncertainties on the atmospheric parameters -- most importantly, $\log g$ -- are likely underestimated. This indicates that the systematic uncertainties likely dominate over the statistical uncertainties inferred from the MCMC posteriors.

\subsection{SED and spectral fit with evolutionary tracks}\label{apdx: with evol model}
\begin{figure}[h!]
    \centering
    \includegraphics[width= \linewidth]{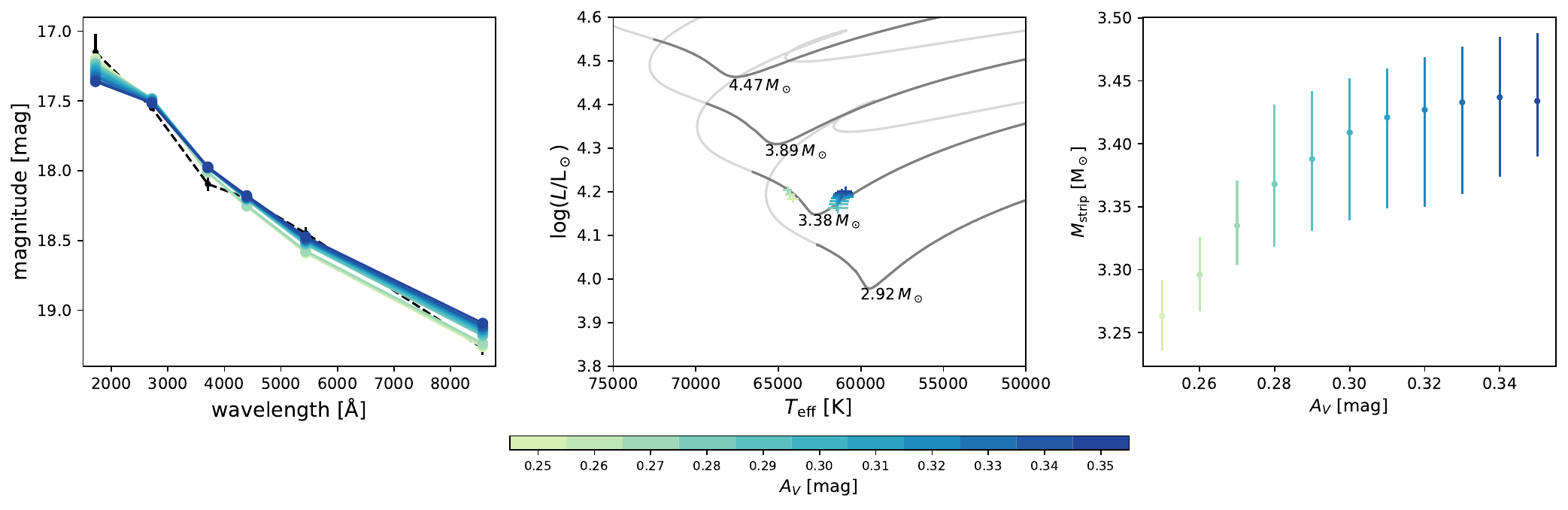}
    \caption{Results for the SED and spectral fits with evolutionary model constraints for a range of $A_V$ values (Section \ref{sec: mcmc sampling setup}; Table \ref{table: full mcmc}). Left: best-fit SED models for each $A_V$.
    Middle: HR diagram, with the colored points marking the best-fit $T_{\rm eff}$ and $\log(L)$. The error bars here represent the statistical uncertainties from the MCMC posterior distributions. Right: best-fit stripped star mass, $M_{\rm stripped}$, as a function of reddening, $A_V$.}
    \label{fig: spec sed all}
\end{figure}

We also repeat the fitting routine from Section \ref{sec: mcmc sampling setup} for a range of reddening values: $A_V = 0.25-0.35$ mag. We report the full list of best-fit parameters in Table \ref{table: full mcmc}.
\begin{table*}[h!]
    \centering
    \renewcommand{\arraystretch}{1.1}
    \begin{tabular}{c|c c c c c c c c}
    \hline
     $A_V$& $M_i$ & $X_{\rm He,c}$ & $X_{\rm H,s}$ & $T_{\rm eff}$ & $\log(g / \rm{cm\,s^{-2}})$ & $\log(L / \rm L_\odot)$ & $R$& $M_{\rm evol}$\\
     $[$mag$]$ & [$\rm M_\odot$] &  &  & [K] &  & & [$\rm R_\odot$]& [$\rm M_\odot$]\\\hline
     0.25 & $ 9.784^{+0.057}_{-0.056}$ &  $0.130 ^{+ 0.032}_{-0.025}$ &  $0.655 ^{+0.010}_{-0.010}$&  $ 64125^{+365}_{-345}$&  $4.949 ^{+ 0.002}_{-0.001}$&  $4.184 ^{+0.010}_{-0.010}$&  $ 1.002^{+ 0.006}_{-0.006}$&  $ 3.263^{+0.029}_{-0.027}$\\
     0.26 & $9.853 ^{+ 0.060}_{-0.057}$ &  $ 0.137^{+ 0.032}_{-0.025}$ &  $ 0.653^{+ 0.009}_{-0.009}$&  $64308 ^{+ 300}_{-323}$&  $ 4.984^{+0.001}_{- 0.001}$&  $4.194 ^{+ 0.009}_{- 0.010}$&  $ 1.008^{+0.006 }_{- 0.006}$&  $3.296 ^{+ 0.030}_{-0.029}$\\
     0.27 & $ 9.932^{+ 0.071}_{-0.062}$ &  $ 0.150^{+0.031}_{-0.027}$ &  $ 0.652^{+ 0.009}_{-0.009}$&  $ 64434^{+ 267}_{-259}$&  $ 4.948^{+ 0.001}_{-0.001}$&  $ 4.204^{+ 0.008}_{- 0.009}$&  $ 1.015^{+ 0.007}_{- 0.006}$&  $3.335 ^{+ 0.036}_{-0.031}$\\
     0.28 & $ 9.912^{+ 0.133}_{-0.105}$ &  $0.667 ^{+ 0.016}_{-0.023}$ &  $ 0.674^{+ 0.009}_{-0.010}$&  $ 61350^{+ 627}_{-582}$&  $ 4.907^{+ 0.012}_{- 0.012}$&  $ 4.163^{+ 0.014}_{- 0.013}$&  $ 1.069^{+ 0.009}_{- 0.009}$&  $ 3.368^{+ 0.063}_{-0.050}$\\
     0.29 & $9.953 ^{+ 0.114}_{-0.121}$ &  $ 0.670^{+ 0.017}_{-0.020}$ &  $ 0.672^{+ 0.010}_{-0.008}$&  $ 61433^{+ 514}_{-665}$&  $ 4.904^{+ 0.011}_{-0.013}$&  $4.171 ^{+ 0.011}_{-0.015}$&  $ 1.077^{+ 0.009}_{-0.009}$&  $ 3.388^{+ 0.054}_{-0.057}$\\
     0.30 & $9.994^{+ 0.090}_{-0.147}$ &  $0.674 ^{+ 0.018}_{-0.017}$ &  $ 0.660^{+ 0.011}_{-0.007}$&  $ 61451^{+ 426}_{-733}$&  $ 4.899^{+ 0.010}_{-0.014}$&  $ 4.178^{+0.010}_{-0.016}$&  $ 1.085^{+ 0.009}_{-0.008}$&  $ 3.409^{+ 0.043}_{-0.070}$\\
     0.31 & $ 10.020^{+ 0.081}_{-0.152}$ &  $ 0.679^{+0.019}_{-0.015}$ &  $0.669 ^{+ 0.011}_{-0.006}$&  $ 61426^{+375}_{-729}$&  $ 4.894^{+ 0.010}_{-0.004}$&  $ 4.184^{+ 0.009}_{-0.016}$&  $1.094 ^{+0.010}_{-0.008}$ &$ 3.421^{+ 0.039}_{-0.072}$\\
     0.32 & $ 10.030^{+ 0.087}_{-0.163}$ &  $ 0.684^{+ 0.022}_{-0.015}$ &  $ 0.669^{+ 0.011}_{-0.006}$&  $ 61332^{+ 402}_{-877}$&  $ 4.888^{+ 0.010}_{-0.018}$&  $ 4.188^{+ 0.009}_{-0.016}$&  $ 1.103^{+0.010}_{-0.009}$&  $ 3.427^{+0.042}_{-0.077}$\\
     0.33 & $ 10.041^{+ 0.092}_{-0.154}$ &  $ 0.690^{+ 0.019}_{-0.014}$ &  $ 0.669^{+0.010}_{-0.006}$&  $ 61244^{+ 420}_{-797}$&  $ 4.881^{+ 0.011}_{-0.017}$&  $ 4.193^{+ 0.009}_{- 0.015}$&  $ 1.112^{+ 0.010}_{- 0.009}$&  $ 3.433^{+ 0.044}_{-0.073}$\\
     0.34 & $ 10.048^{+ 0.100}_{-0.132}$ &  $ 0.695^{+ 0.016}_{-0.014}$ &  $ 0.669^{+ 0.008}_{-0.007}$&  $ 61140^{+ 443}_{-656}$&  $ 4.875^{+ 0.011}_{- 0.015}$&  $ 4.197^{+ 0.010}_{- 0.013}$&  $ 1.121^{+0.010}_{-0.009}$&  $3.437^{+0.048}_{-0.063}$\\
     0.35 & $10.041 ^{+ 0.114}_{-0.091}$ &  $ 0.705^{+0.009}_{-0.017}$ &  $ 0.669^{+ 0.007}_{-0.006}$&  $60900^{+ 559}_{-364}$&  $ 4.865^{+ 0.014}_{-0.009}$&  $ 4.208^{+ 0.010}_{-0.010}$&  $ 1.132^{+ 0.009}_{-0.010}$&  $ 3.434^{+ 0.054}_{-0.044}$\\
     \hline
    \end{tabular}
    \caption{Best-fit parameters constrained for J0040-7333 from SED and spectra fits with evolutionary track constraints (Section \ref{sec: mcmc sampling setup}). $A_V$ is fixed for each run.}
    \label{table: full mcmc}
\end{table*}
Figure \ref{fig: spec sed all} shows the best-fit SED and evolutionary models for each $A_V$; we also show the inferred stripped star mass as a function of the chosen reddening value.

Similar to the fits without  evolutionary tracks, we find that higher reddening values generally produce higher inferred luminosities, and therefore higher inferred evolutionary masses. 
We also find that because different evolutionary stages can produce similar observable properties, the evolutionary models introduce degeneracies between the core helium fraction and effective temperature. We find that models with lower $A_V$ favor later evolutionary stages with higher temperatures, while models with higher $A_V$ favor earlier evolutionary stages with lower temperatures. Despite this difference in evolutionary stage, the models return broadly consistent luminosities, radii, and stripped-star masses.

\section{Surface hydrogen fraction}\label{apdx: hydrogen abundance}

\begin{figure}[h!]
    \centering
    \includegraphics[width=\linewidth]{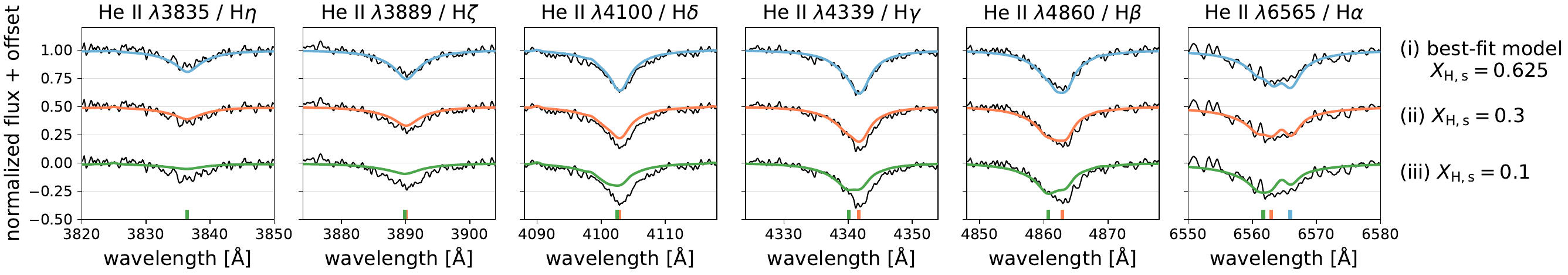}
    \caption{Black curve shows the observed spectrum, and the blue curve shows the best-fit spectral model. Orange and green curves show models with the best-fit effective temperature and surface gravity, but with surface hydrogen fraction fixed to $X_{\rm H, s} = 0.3$ and $X_{\rm H, s} = 0.1$, respectively. Tick marks along the x-axis mark the effective line center.
    Only He\,{\sc ii} $\lambda4100$ / H\,$\delta$, He\,{\sc ii} $\lambda4339$ / H\,$\gamma$, and He\,{\sc ii} $\lambda4860$ / H\,$\beta$ were included in the spectral fitting, but we find that the best-fit model provides good agreement with the observed spectrum across all wavelengths.}
    \label{fig: He-Balmer all}
\end{figure}

If the He\,{\sc ii}/Balmer blends were significantly affected by flux from a luminous companion, we expect that the quality of the spectral fits differ across wavelength, because the relative contribution of each stellar component is wavelength dependent. To check this, we compare the observed spectrum to the best-fit synthetic spectra across all He\,{\sc ii}/Balmer blends covered by our MagE observation. Figure \ref{fig: He-Balmer all} shows all He\,{\sc ii}/Balmer blends between $\lambda = 3800-6000\,{\AA}$.
Although we only used He\,{\sc ii} $\lambda4100$ / H\,$\delta$, He\,{\sc ii} $\lambda4339$ / H\,$\gamma$, and He\,{\sc ii} $\lambda4860$ / H\,$\beta$ for our fits, we find that the remaining He\,{\sc ii} / Balmer blends also show good agreement between the observed and the best-fit spectra. Models with lower $X_{\rm H, s}$ show poorer agreement with the observed spectrum regardless of wavelength. The agreement across all optical wavelengths disfavors the argument that the inferred high hydrogen abundance is caused by contamination from a luminous companion.

\section{Fitting $A_V$ as a free parameter}\label{apdx: free AV}
\begin{figure}[h!]
    \centering
    \includegraphics[width=0.8\linewidth]{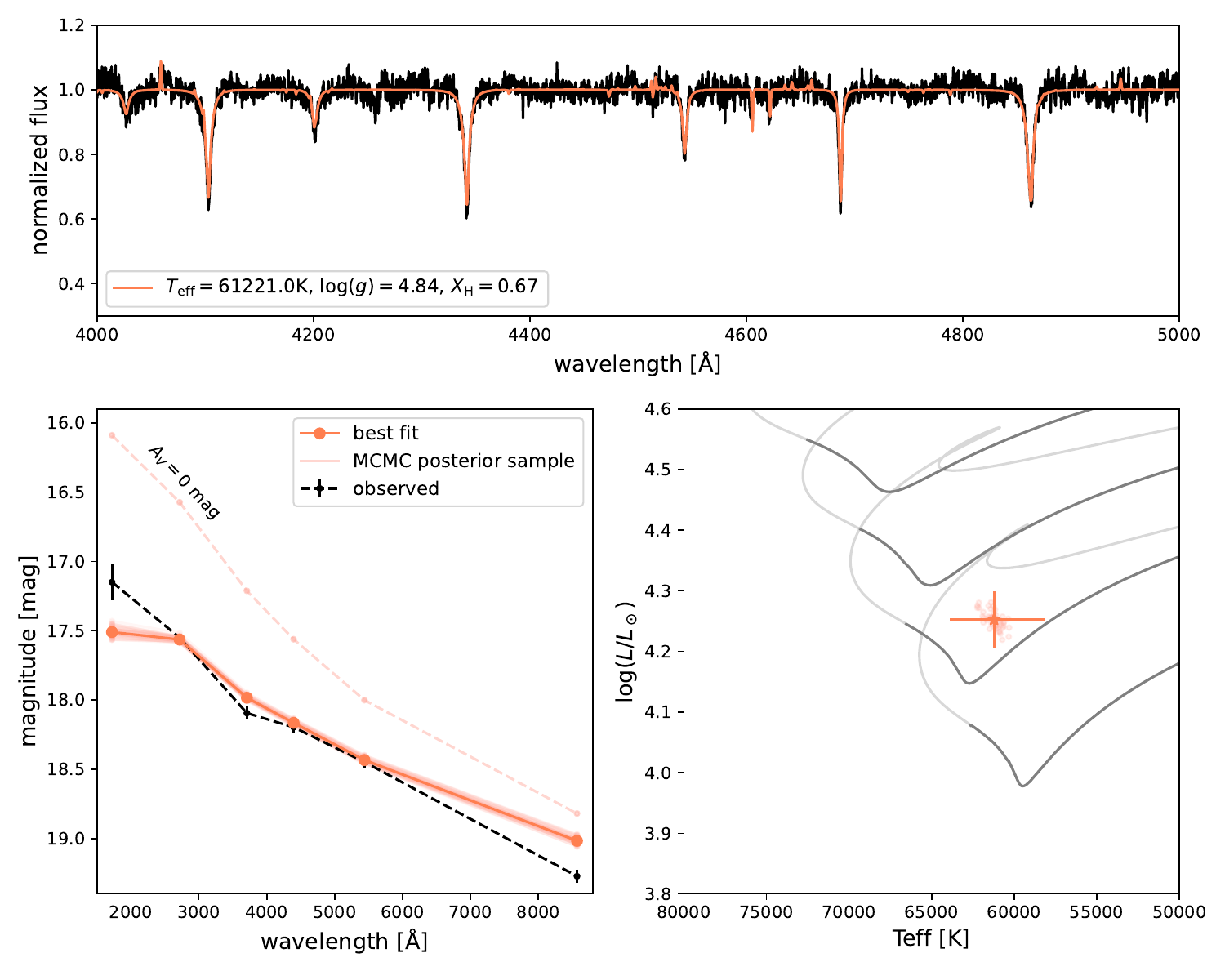}
    \caption{Top: observed spectrum (black) and the best-fit spectral model (orange). Bottom left: observed photometry (black) and the best-fit SED model (solid purple curve). Light purple curves show a set of SED models randomly sampled from the MCMC posterior distribution. Dashed orange curve shows the intrinsic best-fit SED model before reddening is applied. 
    Bottom right: HR diagram with stripped star evolutionary tracks shown in gray. Thick segments show the core helium-burning stage. The orange star marks the best-fit $T_{\rm eff}$ and $\log(L)$, and the light purple markers show a random sample drawn from the MCMC posterior distribution.}
    \label{fig: free AV results}
\end{figure}

We repeat the spectral and SED fitting, while treating reddening, $A_V$, as a free parameter. We report the inferred best-fit parameters in Table \ref{table: free AV} and show the corresponding best-fit spectra, SED, and evolutionary models in Figure \ref{fig: free AV results}.
\begin{table}[h!]
    \centering
    \renewcommand{\arraystretch}{1.3}
    \begin{tabular}{c c c c c c c c c}
    \hline
     $A_V$& $M_i$ & $X_{\rm He,c}$ & $X_{\rm H,s}$ & $T_{\rm eff}$ & $\log(g / \rm{cm\,s^{-2}})$ & $\log(L / \rm L_\odot)$ & $R$& $M_{\rm stripped}$\\
     $[$mag$]$ & [$\rm M_\odot$] &  &  & [K] &  & & [$\rm R_\odot$]& [$\rm M_\odot$]\\\hline
     $0.43^{+0.02}_{-0.02}$ & $ 10.35^{+0.14}_{-0.13}$ &  $0.711 ^{+ 0.01}_{-0.01}$ &  $0.67 ^{+0.01}_{-0.01}$&  $ 61220^{+522}_{-410}$&  $4.84 ^{+ 0.01}_{-0.01}$&  $4.253 ^{+0.02}_{-0.02}$&  $ 1.19^{+ 0.02}_{-0.02}$&  $ 3.586^{+0.07}_{-0.07}$\\
     \hline
    \end{tabular}
    \caption{Best-fit parameters inferred from spectral and SED fitting with $A_V$ as a free parameter.}
    \label{table: free AV}
\end{table}

We recover stripped star properties that are broadly consistent with the fitting routines where $A_V$ is fixed (Section \ref{sec: mcmc setup no evol} and Section \ref{sec: mcmc sampling setup}). However, we find that the best-fit SED model does not match the observed photometry very well. We recover a significantly larger extinction $A_V$ than predicted by simple color checks. This is likely due to the fact there are several orders of magnitude more data points from the spectrum compared to the SED, which leads to the spectral data dominating the likelihood function. Since the spectrum more strongly constrains the atmospheric parameters with little impact on the reddening, $A_V$ becomes less constrained in this setup.





\begin{thebibliography}{}
\expandafter\ifx\csname natexlab\endcsname\relax\def\natexlab#1{#1}\fi
\providecommand{\url}[1]{\href{#1}{#1}}
\providecommand{\dodoi}[1]{doi:~\href{http://doi.org/#1}{\nolinkurl{#1}}}
\providecommand{\doeprint}[1]{\href{http://ascl.net/#1}{\nolinkurl{http://ascl.net/#1}}}
\providecommand{\doarXiv}[1]{\href{https://arxiv.org/abs/#1}{\nolinkurl{https://arxiv.org/abs/#1}}}

\bibitem[{{Abbott} {et~al.}(2017){Abbott}, {Abbott}, {Abbott}, {Acernese}, {Ackley}, {Adams}, {Adams}, {Addesso}, {Adhikari}, {Adya}, {Affeldt}, {Afrough}, {Agarwal}, {Agathos}, {Agatsuma}, {Aggarwal}, {Aguiar}, {Aiello}, {Ain}, {Ajith}, {Allen}, {Allen}, {Allocca}, {Altin}, {Amato}, {Ananyeva}, {Anderson}, {Anderson}, {Angelova}, {Antier}, {Appert}, {Arai}, {Araya}, {Areeda}, {Arnaud}, {Arun}, {Ascenzi}, {Ashton}, {Ast}, {Aston}, {Astone}, {Atallah}, {Aufmuth}, {Aulbert}, {AultONeal}, {Austin}, {Avila-Alvarez}, {Babak}, {Bacon}, {Bader}, {Bae}, {Bailes}, {Baker}, {Baldaccini}, {Ballardin}, {Ballmer}, {Banagiri}, {Barayoga}, {Barclay}, {Barish}, {Barker}, {Barkett}, {Barone}, {Barr}, {Barsotti}, {Barsuglia}, {Barta}, {Barthelmy}, {Bartlett}, {Bartos}, {Bassiri}, {Basti}, {Batch}, {Bawaj}, {Bayley}, {Bazzan}, {B{\'e}csy}, {Beer}, {Bejger}, {Belahcene}, {Bell}, {Berger}, {Bergmann}, {Bernuzzi}, {Bero}, {Berry}, {Bersanetti}, {Bertolini}, {Betzwieser}, {Bhagwat}, {Bhandare}, {Bilenko}, {Billingsley}, {Billman},
  {Birch}, {Birney}, {Birnholtz}, {Biscans}, {Biscoveanu}, {Bisht}, {Bitossi}, {Biwer}, {Bizouard}, {Blackburn}, {Blackman}, {Blair}, {Blair}, {Blair}, {Bloemen}, {Bock}, {Bode}, {Boer}, {Bogaert}, {Bohe}, {Bondu}, {Bonilla}, {Bonnand}, {Boom}, {Bork}, {Boschi}, {Bose}, {Bossie}, {Bouffanais}, {Bozzi}, {Bradaschia}, {Brady}, {Branchesi}, {Brau}, {Briant}, {Brillet}, {Brinkmann}, {Brisson}, {Brockill}, {Broida}, {Brooks}, {Brown}, {Brown}, {Brunett}, {Buchanan}, {Buikema}, {Bulik}, {Bulten}, {Buonanno}, {Buskulic}, {Buy}, {Byer}, {Cabero}, {Cadonati}, {Cagnoli}, {Cahillane}, {Calder{\'o}n Bustillo}, {Callister}, {Calloni}, {Camp}, {Canepa}, {Canizares}, {Cannon}, {Cao}, {Cao}, {Capano}, {Capocasa}, {Carbognani}, {Caride}, {Carney}, {Carullo}, {Casanueva Diaz}, {Casentini}, {Caudill}, {Cavagli{\`a}}, {Cavalier}, {Cavalieri}, {Cella}, {Cepeda}, {Cerd{\'a}-Dur{\'a}n}, {Cerretani}, {Cesarini}, {Chamberlin}, {Chan}, {Chao}, {Charlton}, {Chase}, {Chassande-Mottin}, {Chatterjee}, {Chatziioannou}, {Cheeseboro},
  {Chen}, {Chen}, {Chen}, {Cheng}, {Chia}, {Chincarini}, {Chiummo}, {Chmiel}, {Cho}, {Cho}, {Chow}, {Christensen}, {Chu}, {Chua}, \& {Chua}}]{Abbott_2017}
{Abbott}, B.~P., {Abbott}, R., {Abbott}, T.~D., {et~al.} 2017, \prl, 119, 161101, \dodoi{10.1103/PhysRevLett.119.161101}

\bibitem[{{Ahmad} {et~al.}(2004){Ahmad}, {Jeffery}, \& {Fullerton}}]{Ahmad_2004}
{Ahmad}, A., {Jeffery}, C.~S., \& {Fullerton}, A.~W. 2004, \aap, 418, 275, \dodoi{10.1051/0004-6361:20035917}

\bibitem[{{Bhattacharya} \& {van den Heuvel}(1991)}]{Bhattacharya1991}
{Bhattacharya}, D., \& {van den Heuvel}, E.~P.~J. 1991, \physrep, 203, 1, \dodoi{10.1016/0370-1573(91)90064-S}

\bibitem[{{Blomberg} {et~al.}(2026){Blomberg}, {El-Badry}, {Ludwig}, {Drout}, \& {G{\"o}tberg}}]{Blomberg_2026}
{Blomberg}, L., {El-Badry}, K., {Ludwig}, B., {Drout}, M.~R., \& {G{\"o}tberg}, Y. 2026, \pasp, 138, 024202, \dodoi{10.1088/1538-3873/ae3a19}

\bibitem[{{Carvalho} \& {Johns-Krull}(2023)}]{Carvalho_2023}
{Carvalho}, A., \& {Johns-Krull}, C.~M. 2023, Research Notes of the American Astronomical Society, 7, 91, \dodoi{10.3847/2515-5172/acd37e}

\bibitem[{{Choudhury} {et~al.}(2018){Choudhury}, {Subramaniam}, {Cole}, \& {Sohn}}]{Choudhury_2018}
{Choudhury}, S., {Subramaniam}, A., {Cole}, A.~A., \& {Sohn}, Y.-J. 2018, \mnras, 475, 4279, \dodoi{10.1093/mnras/sty087}

\bibitem[{{Claret}(1998)}]{Claret_1998}
{Claret}, A. 1998, \aaps, 131, 395, \dodoi{10.1051/aas:1998278}

\bibitem[{Claret(2001)}]{Claret_2001}
Claret, A. 2001, \mnras, 327, 989, \dodoi{10.1046/j.1365-8711.2001.04783.x}

\bibitem[{{Conroy} {et~al.}(2020){Conroy}, {Kochoska}, {Hey}, {Pablo}, {Hambleton}, {Jones}, {Giammarco}, {Abdul-Masih}, \& {Pr{\v{s}}a}}]{phoebe_2020}
{Conroy}, K.~E., {Kochoska}, A., {Hey}, D., {et~al.} 2020, \apjs, 250, 34, \dodoi{10.3847/1538-4365/abb4e2}

\bibitem[{{Cotton} {et~al.}(2024){Cotton}, {Filipovi{\'c}}, {Camilo}, {Indebetouw}, {Alsaberi}, {Anih}, {Baker}, {Bastian}, {Boji{\v{c}}i{\'c}}, {Carli}, {Cavallaro}, {Crawford}, {Dai}, {Haberl}, {Levin}, {Luken}, {Pennock}, {Rajabpour}, {Stappers}, {van Loon}, {Zijlstra}, {Buchner}, {Geyer}, {Goedhart}, \& {Serylak}}]{Cotton2024}
{Cotton}, W.~D., {Filipovi{\'c}}, M.~D., {Camilo}, F., {et~al.} 2024, \mnras, 529, 2443, \dodoi{10.1093/mnras/stae277}

\bibitem[{{Crawford} {et~al.}(2001){Crawford}, {Kaspi}, {Manchester}, {Lyne}, {Camilo}, \& {D'Amico}}]{Crawford_2001}
{Crawford}, F., {Kaspi}, V.~M., {Manchester}, R.~N., {et~al.} 2001, \apj, 553, 367, \dodoi{10.1086/320635}

\bibitem[{{Crowther}(2007)}]{Crowther2007}
{Crowther}, P.~A. 2007, \araa, 45, 177, \dodoi{10.1146/annurev.astro.45.051806.110615}

\bibitem[{{{\c{S}}ener} \& {Jeffery}(2014)}]{scener_2014}
{{\c{S}}ener}, H.~T., \& {Jeffery}, C.~S. 2014, \mnras, 440, 2676, \dodoi{10.1093/mnras/stu397}

\bibitem[{{Dewi} {et~al.}(2006){Dewi}, {Podsiadlowski}, \& {Sena}}]{Dewi_2006}
{Dewi}, J.~D.~M., {Podsiadlowski}, P., \& {Sena}, A. 2006, \mnras, 368, 1742, \dodoi{10.1111/j.1365-2966.2006.10233.x}

\bibitem[{{Dewi} {et~al.}(2002){Dewi}, {Pols}, {Savonije}, \& {van den Heuvel}}]{2002_Dewi}
{Dewi}, J.~D.~M., {Pols}, O.~R., {Savonije}, G.~J., \& {van den Heuvel}, E.~P.~J. 2002, \mnras, 331, 1027, \dodoi{10.1046/j.1365-8711.2002.05257.x}

\bibitem[{{Dickey} \& {Lockman}(1990)}]{Dickey_1990}
{Dickey}, J.~M., \& {Lockman}, F.~J. 1990, \araa, 28, 215, \dodoi{10.1146/annurev.aa.28.090190.001243}

\bibitem[{{Drout} {et~al.}(2023){Drout}, {G{\"o}tberg}, {Ludwig}, {Groh}, {de Mink}, {O'Grady}, \& {Smith}}]{Drout_2023}
{Drout}, M.~R., {G{\"o}tberg}, Y., {Ludwig}, B.~A., {et~al.} 2023, Science, 382, 1287, \dodoi{10.1126/science.ade4970}

\bibitem[{{Drout} {et~al.}(2013){Drout}, {Soderberg}, {Mazzali}, {Parrent}, {Margutti}, {Milisavljevic}, {Sanders}, {Chornock}, {Foley}, {Kirshner}, {Filippenko}, {Li}, {Brown}, {Cenko}, {Chakraborti}, {Challis}, {Friedman}, {Ganeshalingam}, {Hicken}, {Jensen}, {Modjaz}, {Perets}, {Silverman}, \& {Wong}}]{Drout_2013}
{Drout}, M.~R., {Soderberg}, A.~M., {Mazzali}, P.~A., {et~al.} 2013, \apj, 774, 58, \dodoi{10.1088/0004-637X/774/1/58}

\bibitem[{{Dutta} \& {Klencki}(2024)}]{Dutta_2024}
{Dutta}, D., \& {Klencki}, J. 2024, \aap, 687, A215, \dodoi{10.1051/0004-6361/202349065}

\bibitem[{Edenhofer {et~al.}(2024)Edenhofer, Zucker, Frank, Saydjari, Speagle, Finkbeiner, \& En{\ss}lin}]{edenhofer_2024}
Edenhofer, G., Zucker, C., Frank, P., {et~al.} 2024, Astronomy \& Astrophysics, 685, A82

\bibitem[{Eldridge {et~al.}(2013)Eldridge, Fraser, Smartt, Maund, \& Crockett}]{eldridge_2013}
Eldridge, J.~J., Fraser, M., Smartt, S.~J., Maund, J.~R., \& Crockett, R.~M. 2013, Monthly Notices of the Royal Astronomical Society, 436, 774

\bibitem[{{Eldridge} {et~al.}(2017){Eldridge}, {Stanway}, {Xiao}, {McClelland}, {Taylor}, {Ng}, {Greis}, \& {Bray}}]{Eldridge2017}
{Eldridge}, J.~J., {Stanway}, E.~R., {Xiao}, L., {et~al.} 2017, \pasa, 34, e058, \dodoi{10.1017/pasa.2017.51}

\bibitem[{{Ercolino} {et~al.}(2025){Ercolino}, {Jin}, {Langer}, \& {Dessart}}]{Ercolino_2025}
{Ercolino}, A., {Jin}, H., {Langer}, N., \& {Dessart}, L. 2025, \aap, 696, A103, \dodoi{10.1051/0004-6361/202453426}

\bibitem[{{Evans} \& {Howarth}(2008)}]{Evans_2008}
{Evans}, C.~J., \& {Howarth}, I.~D. 2008, \mnras, 386, 826, \dodoi{10.1111/j.1365-2966.2008.13012.x}

\bibitem[{{Fadeyev} {et~al.}(2025){Fadeyev}, {Kuranov}, \& {Yungelson}}]{Fadeyev_2025}
{Fadeyev}, Y.~A., {Kuranov}, A.~G., \& {Yungelson}, L.~R. 2025, Astronomy Letters, 51, 25, \dodoi{10.1134/S106377372570015X}

\bibitem[{{Flaugher} {et~al.}(2015){Flaugher}, {Diehl}, {Honscheid}, {Abbott}, {Alvarez}, {Angstadt}, {Annis}, {Antonik}, {Ballester}, {Beaufore}, {Bernstein}, {Bernstein}, {Bigelow}, {Bonati}, {Boprie}, {Brooks}, {Buckley-Geer}, {Campa}, {Cardiel-Sas}, {Castander}, {Castilla}, {Cease}, {Cela-Ruiz}, {Chappa}, {Chi}, {Cooper}, {da Costa}, {Dede}, {Derylo}, {DePoy}, {de Vicente}, {Doel}, {Drlica-Wagner}, {Eiting}, {Elliott}, {Emes}, {Estrada}, {Fausti Neto}, {Finley}, {Flores}, {Frieman}, {Gerdes}, {Gladders}, {Gregory}, {Gutierrez}, {Hao}, {Holland}, {Holm}, {Huffman}, {Jackson}, {James}, {Jonas}, {Karcher}, {Karliner}, {Kent}, {Kessler}, {Kozlovsky}, {Kron}, {Kubik}, {Kuehn}, {Kuhlmann}, {Kuk}, {Lahav}, {Lathrop}, {Lee}, {Levi}, {Lewis}, {Li}, {Mandrichenko}, {Marshall}, {Martinez}, {Merritt}, {Miquel}, {Mu{\~n}oz}, {Neilsen}, {Nichol}, {Nord}, {Ogando}, {Olsen}, {Palaio}, {Patton}, {Peoples}, {Plazas}, {Rauch}, {Reil}, {Rheault}, {Roe}, {Rogers}, {Roodman}, {Sanchez}, {Scarpine}, {Schindler}, {Schmidt},
  {Schmitt}, {Schubnell}, {Schultz}, {Schurter}, {Scott}, {Serrano}, {Shaw}, {Smith}, {Soares-Santos}, {Stefanik}, {Stuermer}, {Suchyta}, {Sypniewski}, {Tarle}, {Thaler}, {Tighe}, {Tran}, {Tucker}, {Walker}, {Wang}, {Watson}, {Weaverdyck}, {Wester}, {Woods}, {Yanny}, \& {DES Collaboration}}]{DeCam}
{Flaugher}, B., {Diehl}, H.~T., {Honscheid}, K., {et~al.} 2015, \aj, 150, 150, \dodoi{10.1088/0004-6256/150/5/150}

\bibitem[{{Foreman-Mackey} {et~al.}(2013){Foreman-Mackey}, {Hogg}, {Lang}, \& {Goodman}}]{emcee_2013}
{Foreman-Mackey}, D., {Hogg}, D.~W., {Lang}, D., \& {Goodman}, J. 2013, \pasp, 125, 306, \dodoi{10.1086/670067}

\bibitem[{{Gaia Collaboration} {et~al.}(2016){Gaia Collaboration}, {Prusti}, {de Bruijne}, {Brown}, {Vallenari}, {Babusiaux}, {Bailer-Jones}, {Bastian}, {Biermann}, {Evans}, {Eyer}, {Jansen}, {Jordi}, {Klioner}, {Lammers}, {Lindegren}, {Luri}, {Mignard}, {Milligan}, {Panem}, {Poinsignon}, {Pourbaix}, {Randich}, {Sarri}, {Sartoretti}, {Siddiqui}, {Soubiran}, {Valette}, {van Leeuwen}, {Walton}, {Aerts}, {Arenou}, {Cropper}, {Drimmel}, {H{\o}g}, {Katz}, {Lattanzi}, {O'Mullane}, {Grebel}, {Holland}, {Huc}, {Passot}, {Bramante}, {Cacciari}, {Casta{\~n}eda}, {Chaoul}, {Cheek}, {De Angeli}, {Fabricius}, {Guerra}, {Hern{\'a}ndez}, {Jean-Antoine-Piccolo}, {Masana}, {Messineo}, {Mowlavi}, {Nienartowicz}, {Ord{\'o}{\~n}ez-Blanco}, {Panuzzo}, {Portell}, {Richards}, {Riello}, {Seabroke}, {Tanga}, {Th{\'e}venin}, {Torra}, {Els}, {Gracia-Abril}, {Comoretto}, {Garcia-Reinaldos}, {Lock}, {Mercier}, {Altmann}, {Andrae}, {Astraatmadja}, {Bellas-Velidis}, {Benson}, {Berthier}, {Blomme}, {Busso}, {Carry}, {Cellino}, {Clementini},
  {Cowell}, {Creevey}, {Cuypers}, {Davidson}, {De Ridder}, {de Torres}, {Delchambre}, {Dell'Oro}, {Ducourant}, {Fr{\'e}mat}, {Garc{\'\i}a-Torres}, {Gosset}, {Halbwachs}, {Hambly}, {Harrison}, {Hauser}, {Hestroffer}, {Hodgkin}, {Huckle}, {Hutton}, {Jasniewicz}, {Jordan}, {Kontizas}, {Korn}, {Lanzafame}, {Manteiga}, {Moitinho}, {Muinonen}, {Osinde}, {Pancino}, {Pauwels}, {Petit}, {Recio-Blanco}, {Robin}, {Sarro}, {Siopis}, {Smith}, {Smith}, {Sozzetti}, {Thuillot}, {van Reeven}, {Viala}, {Abbas}, {Abreu Aramburu}, {Accart}, {Aguado}, {Allan}, {Allasia}, {Altavilla}, {{\'A}lvarez}, {Alves}, {Anderson}, {Andrei}, {Anglada Varela}, {Antiche}, {Antoja}, {Ant{\'o}n}, {Arcay}, {Atzei}, {Ayache}, {Bach}, {Baker}, {Balaguer-N{\'u}{\~n}ez}, {Barache}, {Barata}, {Barbier}, {Barblan}, {Baroni}, {Barrado y Navascu{\'e}s}, {Barros}, {Barstow}, {Becciani}, {Bellazzini}, {Bellei}, {Bello Garc{\'\i}a}, {Belokurov}, {Bendjoya}, {Berihuete}, {Bianchi}, {Bienaym{\'e}}, {Billebaud}, {Blagorodnova}, {Blanco-Cuaresma}, {Boch},
  {Bombrun}, {Borrachero}, {Bouquillon}, {Bourda}, {Bouy}, {Bragaglia}, {Breddels}, {Brouillet}, {Br{\"u}semeister}, {Bucciarelli}, {Budnik}, {Burgess}, {Burgon}, {Burlacu}, {Busonero}, {Buzzi}, {Caffau}, {Cambras}, {Campbell}, {Cancelliere}, {Cantat-Gaudin}, {Carlucci}, {Carrasco}, {Castellani}, {Charlot}, {Charnas}, {Charvet}, {Chassat}, {Chiavassa}, {Clotet}, {Cocozza}, {Collins}, {Collins}, \& {Costigan}}]{Gaia}
{Gaia Collaboration}, {Prusti}, T., {de Bruijne}, J.~H.~J., {et~al.} 2016, \aap, 595, A1, \dodoi{10.1051/0004-6361/201629272}

\bibitem[{{Gaia Collaboration} {et~al.}(2023){Gaia Collaboration}, {Vallenari}, {Brown}, {Prusti}, {de Bruijne}, {Arenou}, {Babusiaux}, {Biermann}, {Creevey}, {Ducourant}, {Evans}, {Eyer}, {Guerra}, {Hutton}, {Jordi}, {Klioner}, {Lammers}, {Lindegren}, {Luri}, {Mignard}, {Panem}, {Pourbaix}, {Randich}, {Sartoretti}, {Soubiran}, {Tanga}, {Walton}, {Bailer-Jones}, {Bastian}, {Drimmel}, {Jansen}, {Katz}, {Lattanzi}, {van Leeuwen}, {Bakker}, {Cacciari}, {Casta{\~n}eda}, {De Angeli}, {Fabricius}, {Fouesneau}, {Fr{\'e}mat}, {Galluccio}, {Guerrier}, {Heiter}, {Masana}, {Messineo}, {Mowlavi}, {Nicolas}, {Nienartowicz}, {Pailler}, {Panuzzo}, {Riclet}, {Roux}, {Seabroke}, {Sordo}, {Th{\'e}venin}, {Gracia-Abril}, {Portell}, {Teyssier}, {Altmann}, {Andrae}, {Audard}, {Bellas-Velidis}, {Benson}, {Berthier}, {Blomme}, {Burgess}, {Busonero}, {Busso}, {C{\'a}novas}, {Carry}, {Cellino}, {Cheek}, {Clementini}, {Damerdji}, {Davidson}, {de Teodoro}, {Nu{\~n}ez Campos}, {Delchambre}, {Dell'Oro}, {Esquej},
  {Fern{\'a}ndez-Hern{\'a}ndez}, {Fraile}, {Garabato}, {Garc{\'\i}a-Lario}, {Gosset}, {Haigron}, {Halbwachs}, {Hambly}, {Harrison}, {Hern{\'a}ndez}, {Hestroffer}, {Hodgkin}, {Holl}, {Jan{\ss}en}, {Jevardat de Fombelle}, {Jordan}, {Krone-Martins}, {Lanzafame}, {L{\"o}ffler}, {Marchal}, {Marrese}, {Moitinho}, {Muinonen}, {Osborne}, {Pancino}, {Pauwels}, {Recio-Blanco}, {Reyl{\'e}}, {Riello}, {Rimoldini}, {Roegiers}, {Rybizki}, {Sarro}, {Siopis}, {Smith}, {Sozzetti}, {Utrilla}, {van Leeuwen}, {Abbas}, {{\'A}brah{\'a}m}, {Abreu Aramburu}, {Aerts}, {Aguado}, {Ajaj}, {Aldea-Montero}, {Altavilla}, {{\'A}lvarez}, {Alves}, {Anders}, {Anderson}, {Anglada Varela}, {Antoja}, {Baines}, {Baker}, {Balaguer-N{\'u}{\~n}ez}, {Balbinot}, {Balog}, {Barache}, {Barbato}, {Barros}, {Barstow}, {Bartolom{\'e}}, {Bassilana}, {Bauchet}, {Becciani}, {Bellazzini}, {Berihuete}, {Bernet}, {Bertone}, {Bianchi}, {Binnenfeld}, {Blanco-Cuaresma}, {Blazere}, {Boch}, {Bombrun}, {Bossini}, {Bouquillon}, {Bragaglia}, {Bramante}, {Breedt},
  {Bressan}, {Brouillet}, {Brugaletta}, {Bucciarelli}, {Burlacu}, {Butkevich}, {Buzzi}, {Caffau}, {Cancelliere}, {Cantat-Gaudin}, {Carballo}, {Carlucci}, {Carnerero}, {Carrasco}, {Casamiquela}, {Castellani}, {Castro-Ginard}, {Chaoul}, {Charlot}, {Chemin}, {Chiaramida}, {Chiavassa}, {Chornay}, {Comoretto}, {Contursi}, {Cooper}, {Cornez}, {Cowell}, {Crifo}, {Cropper}, {Crosta}, {Crowley}, {Dafonte}, {Dapergolas}, {David}, {David}, {de Laverny}, {De Luise}, \& {De March}}]{Gaia_dr3}
{Gaia Collaboration}, {Vallenari}, A., {Brown}, A.~G.~A., {et~al.} 2023, \aap, 674, A1, \dodoi{10.1051/0004-6361/202243940}

\bibitem[{{Gehrels} {et~al.}(2004){Gehrels}, {Chincarini}, {Giommi}, {Mason}, {Nousek}, {Wells}, {White}, {Barthelmy}, {Burrows}, {Cominsky}, {Hurley}, {Marshall}, {M{\'e}sz{\'a}ros}, {Roming}, {Angelini}, {Barbier}, {Belloni}, {Campana}, {Caraveo}, {Chester}, {Citterio}, {Cline}, {Cropper}, {Cummings}, {Dean}, {Feigelson}, {Fenimore}, {Frail}, {Fruchter}, {Garmire}, {Gendreau}, {Ghisellini}, {Greiner}, {Hill}, {Hunsberger}, {Krimm}, {Kulkarni}, {Kumar}, {Lebrun}, {Lloyd-Ronning}, {Markwardt}, {Mattson}, {Mushotzky}, {Norris}, {Osborne}, {Paczynski}, {Palmer}, {Park}, {Parsons}, {Paul}, {Rees}, {Reynolds}, {Rhoads}, {Sasseen}, {Schaefer}, {Short}, {Smale}, {Smith}, {Stella}, {Tagliaferri}, {Takahashi}, {Tashiro}, {Townsley}, {Tueller}, {Turner}, {Vietri}, {Voges}, {Ward}, {Willingale}, {Zerbi}, \& {Zhang}}]{Gehrels_2004}
{Gehrels}, N., {Chincarini}, G., {Giommi}, P., {et~al.} 2004, \apj, 611, 1005, \dodoi{10.1086/422091}

\bibitem[{{Gitika} {et~al.}(2023){Gitika}, {Bailes}, {Shannon}, {Reardon}, {Cameron}, {Shamohammadi}, {Miles}, {Flynn}, {Corongiu}, \& {Kramer}}]{Gitika_2023}
{Gitika}, P., {Bailes}, M., {Shannon}, R.~M., {et~al.} 2023, \mnras, 526, 3370, \dodoi{10.1093/mnras/stad2841}

\bibitem[{{Gordon} {et~al.}(2024){Gordon}, {Fitzpatrick}, {Massa}, {Bohlin}, {Chastenet}, {Murray}, {Clayton}, {Lennon}, {Misselt}, \& {Sandstrom}}]{Gordon_2024}
{Gordon}, K.~D., {Fitzpatrick}, E.~L., {Massa}, D., {et~al.} 2024, \apj, 970, 51, \dodoi{10.3847/1538-4357/ad4be1}

\bibitem[{{G{\"o}tberg} {et~al.}(2017){G{\"o}tberg}, {de Mink}, \& {Groh}}]{Gotberg2017}
{G{\"o}tberg}, Y., {de Mink}, S.~E., \& {Groh}, J.~H. 2017, \aap, 608, A11, \dodoi{10.1051/0004-6361/201730472}

\bibitem[{{G{\"o}tberg} {et~al.}(2018){G{\"o}tberg}, {de Mink}, {Groh}, {Kupfer}, {Crowther}, {Zapartas}, \& {Renzo}}]{gotberg_2018}
{G{\"o}tberg}, Y., {de Mink}, S.~E., {Groh}, J.~H., {et~al.} 2018, \aap, 615, A78, \dodoi{10.1051/0004-6361/201732274}

\bibitem[{{G{\"o}tberg} {et~al.}(2019){G{\"o}tberg}, {de Mink}, {Groh}, {Leitherer}, \& {Norman}}]{gotberg_2019}
{G{\"o}tberg}, Y., {de Mink}, S.~E., {Groh}, J.~H., {Leitherer}, C., \& {Norman}, C. 2019, \aap, 629, A134, \dodoi{10.1051/0004-6361/201834525}

\bibitem[{{G{\"o}tberg} {et~al.}(2020){G{\"o}tberg}, {Korol}, {Lamberts}, {Kupfer}, {Breivik}, {Ludwig}, \& {Drout}}]{gotberg_2020}
{G{\"o}tberg}, Y., {Korol}, V., {Lamberts}, A., {et~al.} 2020, \apj, 904, 56, \dodoi{10.3847/1538-4357/abbda5}

\bibitem[{{G{\"o}tberg} {et~al.}(2023){G{\"o}tberg}, {Drout}, {Ji}, {Groh}, {Ludwig}, {Crowther}, {Smith}, {de Koter}, \& {de Mink}}]{gotberg_2023}
{G{\"o}tberg}, Y., {Drout}, M.~R., {Ji}, A.~P., {et~al.} 2023, \apj, 959, 125, \dodoi{10.3847/1538-4357/ace5a3}

\bibitem[{{Graczyk} {et~al.}(2020){Graczyk}, {Pietrzy{\'n}ski}, {Thompson}, {Gieren}, {Zgirski}, {Villanova}, {G{\'o}rski}, {Wielg{\'o}rski}, {Karczmarek}, {Narloch}, {Pilecki}, {Taormina}, {Smolec}, {Suchomska}, {Gallenne}, {Nardetto}, {Storm}, {Kudritzki}, {Ka{\l}uszy{\'n}ski}, \& {Pych}}]{Graczyk_2020}
{Graczyk}, D., {Pietrzy{\'n}ski}, G., {Thompson}, I.~B., {et~al.} 2020, \apj, 904, 13, \dodoi{10.3847/1538-4357/abbb2b}

\bibitem[{Götberg {et~al.}(2023)Götberg, Drout, Ji, Groh, Ludwig, Crowther, Smith, de~Koter, \& de~Mink}]{gotberg_2023_models}
Götberg, Y., Drout, M., Ji, A., {et~al.} 2023, Stellar properties of observed stars stripped in binaries in the Magellanic Clouds - Spectral Models, 1,  Zenodo, \dodoi{10.5281/zenodo.7976200}

\bibitem[{{Hachinger} {et~al.}(2012){Hachinger}, {Mazzali}, {Taubenberger}, {Hillebrandt}, {Nomoto}, \& {Sauer}}]{Hachinger2012}
{Hachinger}, S., {Mazzali}, P.~A., {Taubenberger}, S., {et~al.} 2012, \mnras, 422, 70, \dodoi{10.1111/j.1365-2966.2012.20464.x}

\bibitem[{{Hagen} {et~al.}(2017){Hagen}, {Siegel}, {Hoversten}, {Gronwall}, {Immler}, \& {Hagen}}]{Hagen_2017}
{Hagen}, L. M.~Z., {Siegel}, M.~H., {Hoversten}, E.~A., {et~al.} 2017, \mnras, 466, 4540, \dodoi{10.1093/mnras/stw2954}

\bibitem[{{Han} {et~al.}(2003){Han}, {Podsiadlowski}, {Maxted}, \& {Marsh}}]{Han2003}
{Han}, Z., {Podsiadlowski}, P., {Maxted}, P.~F.~L., \& {Marsh}, T.~R. 2003, \mnras, 341, 669, \dodoi{10.1046/j.1365-8711.2003.06451.x}

\bibitem[{{Han} {et~al.}(2002){Han}, {Podsiadlowski}, {Maxted}, {Marsh}, \& {Ivanova}}]{Han2002}
{Han}, Z., {Podsiadlowski}, P., {Maxted}, P.~F.~L., {Marsh}, T.~R., \& {Ivanova}, N. 2002, \mnras, 336, 449, \dodoi{10.1046/j.1365-8711.2002.05752.x}

\bibitem[{He {et~al.}(2025)He, Meng, \& Lei}]{he_2025}
He, R., Meng, X., \& Lei, Z. 2025, Monthly Notices of the Royal Astronomical Society, 543, 2243

\bibitem[{{Heber}(2009)}]{Heber2009}
{Heber}, U. 2009, \araa, 47, 211, \dodoi{10.1146/annurev-astro-082708-101836}

\bibitem[{Heber(2016)}]{Heber2016}
Heber, U. 2016, \pasp, 128, 082001, \dodoi{10.1088/1538-3873/128/966/082001}

\bibitem[{{Hook} {et~al.}(2004){Hook}, {J{\o}rgensen}, {Allington-Smith}, {Davies}, {Metcalfe}, {Murowinski}, \& {Crampton}}]{GMOS}
{Hook}, I.~M., {J{\o}rgensen}, I., {Allington-Smith}, J.~R., {et~al.} 2004, \pasp, 116, 425, \dodoi{10.1086/383624}

\bibitem[{{Horvat} {et~al.}(2018){Horvat}, {Conroy}, {Pablo}, {Hambleton}, {Kochoska}, {Giammarco}, \& {Pr{\v{s}}a}}]{phoebe_2018}
{Horvat}, M., {Conroy}, K.~E., {Pablo}, H., {et~al.} 2018, \apjs, 237, 26, \dodoi{10.3847/1538-4365/aacd0f}

\bibitem[{{Hota} {et~al.}(2024{\natexlab{a}}){Hota}, {Subramaniam}, {Dhanush}, {Cioni}, \& {Subramanian}}]{Hota_smc_shell}
{Hota}, S., {Subramaniam}, A., {Dhanush}, S.~R., {Cioni}, M.-R.~L., \& {Subramanian}, S. 2024{\natexlab{a}}, \mnras, 532, 322, \dodoi{10.1093/mnras/stae1438}

\bibitem[{{Hota} {et~al.}(2024{\natexlab{b}}){Hota}, {Subramaniam}, {Nayak}, \& {Subramanian}}]{Hota_smc}
{Hota}, S., {Subramaniam}, A., {Nayak}, P.~K., \& {Subramanian}, S. 2024{\natexlab{b}}, \aj, 168, 255, \dodoi{10.3847/1538-3881/ad7de2}

\bibitem[{{Hovis-Afflerbach} {et~al.}(2025){Hovis-Afflerbach}, {G{\"o}tberg}, {Schootemeijer}, {Klencki}, {Strom}, {Ludwig}, \& {Drout}}]{hovis_2025}
{Hovis-Afflerbach}, B., {G{\"o}tberg}, Y., {Schootemeijer}, A., {et~al.} 2025, \aap, 697, A239, \dodoi{10.1051/0004-6361/202453185}

\bibitem[{{Ivanova} {et~al.}(2013){Ivanova}, {Justham}, {Chen}, {De Marco}, {Fryer}, {Gaburov}, {Ge}, {Glebbeek}, {Han}, {Li}, {Lu}, {Marsh}, {Podsiadlowski}, {Potter}, {Soker}, {Taam}, {Tauris}, {van den Heuvel}, \& {Webbink}}]{Ivanova_2013}
{Ivanova}, N., {Justham}, S., {Chen}, X., {et~al.} 2013, \aapr, 21, 59, \dodoi{10.1007/s00159-013-0059-2}

\bibitem[{{Janka}(2017)}]{Janka2017}
{Janka}, H.-T. 2017, \apj, 837, 84, \dodoi{10.3847/1538-4357/aa618e}

\bibitem[{{Jansen} {et~al.}(2001){Jansen}, {Lumb}, {Altieri}, {Clavel}, {Ehle}, {Erd}, {Gabriel}, {Guainazzi}, {Gondoin}, {Much}, {Munoz}, {Santos}, {Schartel}, {Texier}, \& {Vacanti}}]{Jansen_2001}
{Jansen}, F., {Lumb}, D., {Altieri}, B., {et~al.} 2001, \aap, 365, L1, \dodoi{10.1051/0004-6361:20000036}

\bibitem[{{Jeffreys}(1946)}]{Jeffreys_1946}
{Jeffreys}, H. 1946, Proceedings of the Royal Society of London Series A, 186, 453, \dodoi{10.1098/rspa.1946.0056}

\bibitem[{{Jermyn} {et~al.}(2023){Jermyn}, {Bauer}, {Schwab}, {Farmer}, {Ball}, {Bellinger}, {Dotter}, {Joyce}, {Marchant}, {Mombarg}, {Wolf}, {Sunny Wong}, {Cinquegrana}, {Farrell}, {Smolec}, {Thoul}, {Cantiello}, {Herwig}, {Toloza}, {Bildsten}, {Townsend}, \& {Timmes}}]{jermyn_2023}
{Jermyn}, A.~S., {Bauer}, E.~B., {Schwab}, J., {et~al.} 2023, \apjs, 265, 15, \dodoi{10.3847/1538-4365/acae8d}

\bibitem[{{Jiang} {et~al.}(2021){Jiang}, {Tauris}, {Chen}, \& {Fuller}}]{Jiang_2021}
{Jiang}, L., {Tauris}, T.~M., {Chen}, W.-C., \& {Fuller}, J. 2021, \apjl, 920, L36, \dodoi{10.3847/2041-8213/ac2cc9}

\bibitem[{{Jonas} \& {MeerKAT Team}(2016)}]{Meerkat}
{Jonas}, J., \& {MeerKAT Team}. 2016, in MeerKAT Science: On the Pathway to the SKA, 1, \dodoi{10.22323/1.277.0001}

\bibitem[{{Jones} {et~al.}(2020){Jones}, {Conroy}, {Horvat}, {Giammarco}, {Kochoska}, {Pablo}, {Brown}, {Sowicka}, \& {Pr{\v{s}}a}}]{phoebe_2020b}
{Jones}, D., {Conroy}, K.~E., {Horvat}, M., {et~al.} 2020, \apjs, 247, 63, \dodoi{10.3847/1538-4365/ab7927}

\bibitem[{{Kippenhahn} \& {Weigert}(1967)}]{Kippenhahn_1967}
{Kippenhahn}, R., \& {Weigert}, A. 1967, \zap, 65, 251

\bibitem[{{Klencki} {et~al.}(2022){Klencki}, {Istrate}, {Nelemans}, \& {Pols}}]{Klencki_2022}
{Klencki}, J., {Istrate}, A., {Nelemans}, G., \& {Pols}, O. 2022, \aap, 662, A56, \dodoi{10.1051/0004-6361/202142701}

\bibitem[{{Kramer} {et~al.}(1998){Kramer}, {Xilouris}, {Lorimer}, {Doroshenko}, {Jessner}, {Wielebinski}, {Wolszczan}, \& {Camilo}}]{Kramer_1998}
{Kramer}, M., {Xilouris}, K.~M., {Lorimer}, D.~R., {et~al.} 1998, \apj, 501, 270, \dodoi{10.1086/305790}

\bibitem[{{Kumar} {et~al.}(2012){Kumar}, {Ghosh}, {Hutchings}, {Kamath}, {Kathiravan}, {Mahesh}, {Murthy}, {Nagbhushana}, {Pati}, {Rao}, {Rao}, {Sriram}, \& {Tandon}}]{UVIT}
{Kumar}, A., {Ghosh}, S.~K., {Hutchings}, J., {et~al.} 2012, in Society of Photo-Optical Instrumentation Engineers (SPIE) Conference Series, Vol. 8443, Space Telescopes and Instrumentation 2012: Ultraviolet to Gamma Ray, ed. T.~{Takahashi}, S.~S. {Murray}, \& J.-W.~A. {den Herder}, 84431N, \dodoi{10.1117/12.924507}

\bibitem[{{Labrie} {et~al.}(2023){Labrie}, {Simpson}, {Cardenes}, {Turner}, {Soraisam}, {Quint}, {Oberdorf}, {Placco}, {Berke}, {Smirnova}, {Conseil}, {Vacca}, \& {Thomas-Osip}}]{dragons}
{Labrie}, K., {Simpson}, C., {Cardenes}, R., {et~al.} 2023, Research Notes of the American Astronomical Society, 7, 214, \dodoi{10.3847/2515-5172/ad0044}

\bibitem[{{Lah} {et~al.}(2024){Lah}, {Colless}, {D'Eugenio}, {Groves}, \& {Gelfand}}]{Lah_2024}
{Lah}, P., {Colless}, M., {D'Eugenio}, F., {Groves}, B., \& {Gelfand}, J.~D. 2024, \mnras, 535, 2598, \dodoi{10.1093/mnras/stae2534}

\bibitem[{{Lamers} {et~al.}(1995){Lamers}, {Snow}, \& {Lindholm}}]{Lamers_1995}
{Lamers}, H. J.~G.~L.~M., {Snow}, T.~P., \& {Lindholm}, D.~M. 1995, \apj, 455, 269, \dodoi{10.1086/176575}

\bibitem[{{Lang} {et~al.}(2016){Lang}, {Hogg}, \& {Mykytyn}}]{lang_2016}
{Lang}, D., {Hogg}, D.~W., \& {Mykytyn}, D. 2016, {The Tractor: Probabilistic astronomical source detection and measurement}, Astrophysics Source Code Library, record ascl:1604.008.
\newblock \doeprint{1604.008}

\bibitem[{{Laplace} {et~al.}(2020){Laplace}, {G{\"o}tberg}, {de Mink}, {Justham}, \& {Farmer}}]{Laplace_2020}
{Laplace}, E., {G{\"o}tberg}, Y., {de Mink}, S.~E., {Justham}, S., \& {Farmer}, R. 2020, \aap, 637, A6, \dodoi{10.1051/0004-6361/201937300}

\bibitem[{{Laroche} {et~al.}(2026){Laroche}, {Drout}, {G{\"o}tberg}, {Ludwig}, {Wei}, {Senchyna}, {Renzo}, {Dutta}, {O'Grady}, \& {Pritzkuleit}}]{Laroche_2026}
{Laroche}, A., {Drout}, M.~R., {G{\"o}tberg}, Y., {et~al.} 2026, arXiv e-prints, arXiv:2609.30384, \dodoi{10.48550/arXiv.2609.30384}

\bibitem[{{Lomb}(1976)}]{Lomb_1976}
{Lomb}, N.~R. 1976, \apss, 39, 447, \dodoi{10.1007/BF00648343}

\bibitem[{{Lucy}(1967)}]{Lucy_1967}
{Lucy}, L.~B. 1967, \zap, 65, 89

\bibitem[{{Ludwig} {et~al.}(2026){Ludwig}, {Drout}, {G{\"o}tberg}, {Lang}, \& {Laroche}}]{ludwig_2026}
{Ludwig}, B., {Drout}, M.~R., {G{\"o}tberg}, Y., {Lang}, D., \& {Laroche}, A. 2026, \apj, 999, 73, \dodoi{10.3847/1538-4357/ae2d13}

\bibitem[{Ludwig(2025)}]{Ludwig_thesis}
Ludwig, B.~A. 2025, PhD thesis.
\newblock \url{https://go.openathens.net/redirector/caltech.edu?url=https://www.proquest.com/dissertations-theses/first-observations-intermediate-mass-stripped/docview/3323151593/se-2}

\bibitem[{Lyman {et~al.}(2016)Lyman, Bersier, James, Mazzali, Eldridge, Fraser, \& Pian}]{lyman_2016}
Lyman, J., Bersier, D., James, P., {et~al.} 2016, Monthly Notices of the Royal Astronomical Society, 457, 328

\bibitem[{{Ma} \& {Fuller}(2024)}]{Ma_2024}
{Ma}, L., \& {Fuller}, J. 2024, \apj, 975, 1, \dodoi{10.3847/1538-4357/ad7788}

\bibitem[{{Manchester} {et~al.}(2006){Manchester}, {Fan}, {Lyne}, {Kaspi}, \& {Crawford}}]{Manchester_2006}
{Manchester}, R.~N., {Fan}, G., {Lyne}, A.~G., {Kaspi}, V.~M., \& {Crawford}, F. 2006, \apj, 649, 235, \dodoi{10.1086/505461}

\bibitem[{{Manchester} {et~al.}(2005){Manchester}, {Hobbs}, {Teoh}, \& {Hobbs}}]{Manchester2005}
{Manchester}, R.~N., {Hobbs}, G.~B., {Teoh}, A., \& {Hobbs}, M. 2005, \aj, 129, 1993, \dodoi{10.1086/428488}

\bibitem[{{Marshall} {et~al.}(2008){Marshall}, {Burles}, {Thompson}, {Shectman}, {Bigelow}, {Burley}, {Birk}, {Estrada}, {Jones}, {Smith}, {Kowal}, {Castillo}, {Storts}, \& {Ortiz}}]{MagE}
{Marshall}, J.~L., {Burles}, S., {Thompson}, I.~B., {et~al.} 2008, in Society of Photo-Optical Instrumentation Engineers (SPIE) Conference Series, Vol. 7014, Ground-based and Airborne Instrumentation for Astronomy II, ed. I.~S. {McLean} \& M.~M. {Casali}, 701454, \dodoi{10.1117/12.789972}

\bibitem[{{McConnell} {et~al.}(1991){McConnell}, {McCulloch}, {Hamilton}, {Ables}, {Hall}, {Jacka}, \& {Hunt}}]{McConnell_1991}
{McConnell}, D., {McCulloch}, P.~M., {Hamilton}, P.~A., {et~al.} 1991, \mnras, 249, 654, \dodoi{10.1093/mnras/249.4.654}

\bibitem[{{M{\"u}ller-Horn} {et~al.}(2026){M{\"u}ller-Horn}, {El-Badry}, {Sander}, {Rix}, {Blomberg}, {Hermes}, {Nagarajan}, {Shahaf}, {Jin}, {Rowan}, {Dutta}, {Fern{\'a}ndez-Trincado}, {G{\"o}tberg}, {Ilyin}, {Maccarone}, {M{\'e}ndez Delgado}, {Stringfellow}, {Tkachenko}, {Villase{\~n}or}, \& {Zari}}]{mullerhorn_2026}
{M{\"u}ller-Horn}, J., {El-Badry}, K., {Sander}, A. A.~C., {et~al.} 2026, arXiv e-prints, arXiv:2608.05276.
\newblock \doarXiv{2608.05276}

\bibitem[{{Nagarajan} {et~al.}(2026){Nagarajan}, {El-Badry}, {Fuller}, {Guo}, \& {Tauris}}]{Nagarajan_2026}
{Nagarajan}, P., {El-Badry}, K., {Fuller}, J., {Guo}, Y., \& {Tauris}, T.~M. 2026, arXiv e-prints, arXiv:2604.03385, \dodoi{10.48550/arXiv.2604.03385}

\bibitem[{{Naz{\'e}} {et~al.}(2023){Naz{\'e}}, {Britavskiy}, {Rauw}, {Labadie-Bartz}, \& {Sim{\'o}n-D{\'\i}az}}]{naze_2023}
{Naz{\'e}}, Y., {Britavskiy}, N., {Rauw}, G., {Labadie-Bartz}, J., \& {Sim{\'o}n-D{\'\i}az}, S. 2023, \mnras, 525, 1641, \dodoi{10.1093/mnras/stad2280}

\bibitem[{{Naz{\'e}} {et~al.}(2025){Naz{\'e}}, {Rauw}, {Ko{\l}aczek-Szyma{\'n}ski}, {Britavskiy}, \& {Labadie-Bartz}}]{naze_2025}
{Naz{\'e}}, Y., {Rauw}, G., {Ko{\l}aczek-Szyma{\'n}ski}, P.~A., {Britavskiy}, N., \& {Labadie-Bartz}, J. 2025, \aap, 703, A239, \dodoi{10.1051/0004-6361/202556441}

\bibitem[{{Nidever} {et~al.}(2017){Nidever}, {Olsen}, {Walker}, {Vivas}, {Blum}, {Kaleida}, {Choi}, {Conn}, {Gruendl}, {Bell}, {Besla}, {Mu{\~n}oz}, {Gallart}, {Martin}, {Olszewski}, {Saha}, {Monachesi}, {Monelli}, {de Boer}, {Johnson}, {Zaritsky}, {Stringfellow}, {van der Marel}, {Cioni}, {Jin}, {Majewski}, {Martinez-Delgado}, {Monteagudo}, {No{\"e}l}, {Bernard}, {Kunder}, {Chu}, {Bell}, {Santana}, {Frechem}, {Medina}, {Parkash}, {Navarrete}, \& {Hayes}}]{smash}
{Nidever}, D.~L., {Olsen}, K., {Walker}, A.~R., {et~al.} 2017, \aj, 154, 199, \dodoi{10.3847/1538-3881/aa8d1c}

\bibitem[{{Paxton} {et~al.}(2011){Paxton}, {Bildsten}, {Dotter}, {Herwig}, {Lesaffre}, \& {Timmes}}]{paxton_2010}
{Paxton}, B., {Bildsten}, L., {Dotter}, A., {et~al.} 2011, \apjs, 192, 3, \dodoi{10.1088/0067-0049/192/1/3}

\bibitem[{{Paxton} {et~al.}(2013){Paxton}, {Cantiello}, {Arras}, {Bildsten}, {Brown}, {Dotter}, {Mankovich}, {Montgomery}, {Stello}, {Timmes}, \& {Townsend}}]{paxton_2013}
{Paxton}, B., {Cantiello}, M., {Arras}, P., {et~al.} 2013, \apjs, 208, 4, \dodoi{10.1088/0067-0049/208/1/4}

\bibitem[{{Paxton} {et~al.}(2015){Paxton}, {Marchant}, {Schwab}, {Bauer}, {Bildsten}, {Cantiello}, {Dessart}, {Farmer}, {Hu}, {Langer}, {Townsend}, {Townsley}, \& {Timmes}}]{paxton_2015}
{Paxton}, B., {Marchant}, P., {Schwab}, J., {et~al.} 2015, \apjs, 220, 15, \dodoi{10.1088/0067-0049/220/1/15}

\bibitem[{{Paxton} {et~al.}(2018){Paxton}, {Schwab}, {Bauer}, {Bildsten}, {Blinnikov}, {Duffell}, {Farmer}, {Goldberg}, {Marchant}, {Sorokina}, {Thoul}, {Townsend}, \& {Timmes}}]{Paxton_2018}
{Paxton}, B., {Schwab}, J., {Bauer}, E.~B., {et~al.} 2018, \apjs, 234, 34, \dodoi{10.3847/1538-4365/aaa5a8}

\bibitem[{{Paxton} {et~al.}(2019){Paxton}, {Smolec}, {Schwab}, {Gautschy}, {Bildsten}, {Cantiello}, {Dotter}, {Farmer}, {Goldberg}, {Jermyn}, {Kanbur}, {Marchant}, {Thoul}, {Townsend}, {Wolf}, {Zhang}, \& {Timmes}}]{paxton_2019}
{Paxton}, B., {Smolec}, R., {Schwab}, J., {et~al.} 2019, \apjs, 243, 10, \dodoi{10.3847/1538-4365/ab2241}

\bibitem[{{Pecaut} \& {Mamajek}(2013)}]{Pecaut_2013}
{Pecaut}, M.~J., \& {Mamajek}, E.~E. 2013, \apjs, 208, 9, \dodoi{10.1088/0067-0049/208/1/9}

\bibitem[{{Poole} {et~al.}(2008){Poole}, {Breeveld}, {Page}, {Landsman}, {Holland}, {Roming}, {Kuin}, {Brown}, {Gronwall}, {Hunsberger}, {Koch}, {Mason}, {Schady}, {vanden Berk}, {Blustin}, {Boyd}, {Broos}, {Carter}, {Chester}, {Cucchiara}, {Hancock}, {Huckle}, {Immler}, {Ivanushkina}, {Kennedy}, {Marshall}, {Morgan}, {Pandey}, {de Pasquale}, {Smith}, \& {Still}}]{Poole_2008}
{Poole}, T.~S., {Breeveld}, A.~A., {Page}, M.~J., {et~al.} 2008, \mnras, 383, 627, \dodoi{10.1111/j.1365-2966.2007.12563.x}

\bibitem[{{Price-Whelan} {et~al.}(2017){Price-Whelan}, {Hogg}, {Foreman-Mackey}, \& {Rix}}]{joker_2017}
{Price-Whelan}, A.~M., {Hogg}, D.~W., {Foreman-Mackey}, D., \& {Rix}, H.-W. 2017, \apj, 837, 20, \dodoi{10.3847/1538-4357/aa5e50}

\bibitem[{{Prochaska} {et~al.}(2020{\natexlab{a}}){Prochaska}, {Hennawi}, {Westfall}, {Cooke}, {Wang}, {Hsyu}, {Davies}, {Farina}, \& {Pelliccia}}]{Prochaska_2020a}
{Prochaska}, J., {Hennawi}, J., {Westfall}, K., {et~al.} 2020{\natexlab{a}}, The Journal of Open Source Software, 5, 2308, \dodoi{10.21105/joss.02308}

\bibitem[{{Prochaska} {et~al.}(2020{\natexlab{b}}){Prochaska}, {Hennawi}, {Cooke}, {Westfall}, {Wang}, {EmAstro}, {Tiffanyhsyu}, {Wasserman}, {Villaume}, {Marijana777}, {Schindler}, {Young}, {Simha}, {Wilde}, {Tejos}, {Isbell}, {Fl{\"o}rs}, {Sandford}, {Vasovi{\'c}}, {Betts}, \& {Holden}}]{Prochaska_2020b}
{Prochaska}, J.~X., {Hennawi}, J., {Cooke}, R., {et~al.} 2020{\natexlab{b}}, {pypeit/PypeIt: Release 1.0.0}, v1.0.0,  Zenodo, \dodoi{10.5281/zenodo.3743493}

\bibitem[{{Pr{\v{s}}a} \& {Zwitter}(2005)}]{phoebe_2005}
{Pr{\v{s}}a}, A., \& {Zwitter}, T. 2005, \apj, 628, 426, \dodoi{10.1086/430591}

\bibitem[{{Pr{\v{s}}a} {et~al.}(2016){Pr{\v{s}}a}, {Conroy}, {Horvat}, {Pablo}, {Kochoska}, {Bloemen}, {Giammarco}, {Hambleton}, \& {Degroote}}]{phoebe_2016}
{Pr{\v{s}}a}, A., {Conroy}, K.~E., {Horvat}, M., {et~al.} 2016, \apjs, 227, 29, \dodoi{10.3847/1538-4365/227/2/29}

\bibitem[{{Ramachandran} {et~al.}(2023){Ramachandran}, {Klencki}, {Sander}, {Pauli}, {Shenar}, {Oskinova}, \& {Hamann}}]{Ramachandran_2023}
{Ramachandran}, V., {Klencki}, J., {Sander}, A.~A.~C., {et~al.} 2023, \aap, 674, L12, \dodoi{10.1051/0004-6361/202346818}

\bibitem[{{Rodrigo} \& {Solano}(2020)}]{Rodrigo_2020}
{Rodrigo}, C., \& {Solano}, E. 2020, in XIV.0 Scientific Meeting (virtual) of the Spanish Astronomical Society, 182

\bibitem[{{Rodrigo} {et~al.}(2012){Rodrigo}, {Solano}, \& {Bayo}}]{Rodrigo_2012}
{Rodrigo}, C., {Solano}, E., \& {Bayo}, A. 2012, {SVO Filter Profile Service Version 1.0}, IVOA Working Draft 15 October 2012, \dodoi{10.5479/ADS/bib/2012ivoa.rept.1015R}

\bibitem[{{Rodrigo} {et~al.}(2024){Rodrigo}, {Cruz}, {Aguilar}, {Aller}, {Solano}, {G{\'a}lvez-Ortiz}, {Jim{\'e}nez-Esteban}, {Mas-Buitrago}, {Bayo}, {Cort{\'e}s-Contreras}, {Murillo-Ojeda}, {Bonoli}, {Cenarro}, {Dupke}, {L{\'o}pez-Sanjuan}, {Mar{\'\i}n-Franch}, {de Oliveira}, {Moles}, {Taylor}, {Varela}, \& {Rami{\'o}}}]{rodrigo_2024}
{Rodrigo}, C., {Cruz}, P., {Aguilar}, J.~F., {et~al.} 2024, \aap, 689, A93, \dodoi{10.1051/0004-6361/202449998}

\bibitem[{{Roming} {et~al.}(2005){Roming}, {Kennedy}, {Mason}, {Nousek}, {Ahr}, {Bingham}, {Broos}, {Carter}, {Hancock}, {Huckle}, {Hunsberger}, {Kawakami}, {Killough}, {Koch}, {McLelland}, {Smith}, {Smith}, {Soto}, {Boyd}, {Breeveld}, {Holland}, {Ivanushkina}, {Pryzby}, {Still}, \& {Stock}}]{swift_UVOT}
{Roming}, P. W.~A., {Kennedy}, T.~E., {Mason}, K.~O., {et~al.} 2005, \ssr, 120, 95, \dodoi{10.1007/s11214-005-5095-4}

\bibitem[{{Sana} {et~al.}(2012){Sana}, {de Mink}, {de Koter}, {Langer}, {Evans}, {Gieles}, {Gosset}, {Izzard}, {Le Bouquin}, \& {Schneider}}]{Sana_2012}
{Sana}, H., {de Mink}, S.~E., {de Koter}, A., {et~al.} 2012, Science, 337, 444, \dodoi{10.1126/science.1223344}

\bibitem[{{Scargle}(1982)}]{Scargle_1982}
{Scargle}, J.~D. 1982, \apj, 263, 835, \dodoi{10.1086/160554}

\bibitem[{{Schaffenroth} {et~al.}(2023){Schaffenroth}, {Barlow}, {Pelisoli}, {Geier}, \& {Kupfer}}]{schaffenroth_2023}
{Schaffenroth}, V., {Barlow}, B.~N., {Pelisoli}, I., {Geier}, S., \& {Kupfer}, T. 2023, \aap, 673, A90, \dodoi{10.1051/0004-6361/202244697}

\bibitem[{{Shenar}(2026)}]{Shenar2026}
{Shenar}, T. 2026, in Encyclopedia of Astrophysics, Volume 2, Vol.~2, 569--583, \dodoi{10.1016/B978-0-443-21439-4.00031-6}

\bibitem[{Simpson {et~al.}(2025)Simpson, Smirnova, Labrie, Hirst, Berke, Rawlings, Turner, \& Vacca}]{DRAGONS_software}
Simpson, C., Smirnova, O., Labrie, K., {et~al.} 2025, {DRAGONS (4.1.0)},  Zenodo, \dodoi{10.5281/zenodo.15226659}

\bibitem[{{Singh} {et~al.}(2014){Singh}, {Tandon}, {Agrawal}, {Antia}, {Manchanda}, {Yadav}, {Seetha}, {Ramadevi}, {Rao}, {Bhattacharya}, {Paul}, {Sreekumar}, {Bhattacharyya}, {Stewart}, {Hutchings}, {Annapurni}, {Ghosh}, {Murthy}, {Pati}, {Rao}, {Stalin}, {Girish}, {Sankarasubramanian}, {Vadawale}, {Bhalerao}, {Dewangan}, {Dedhia}, {Hingar}, {Katoch}, {Kothare}, {Mirza}, {Mukerjee}, {Shah}, {Shah}, {Mohan}, {Sangal}, {Nagabhusana}, {Sriram}, {Malkar}, {Sreekumar}, {Abbey}, {Hansford}, {Beardmore}, {Sharma}, {Murthy}, {Kulkarni}, {Meena}, {Babu}, \& {Postma}}]{astrosat}
{Singh}, K.~P., {Tandon}, S.~N., {Agrawal}, P.~C., {et~al.} 2014, in Society of Photo-Optical Instrumentation Engineers (SPIE) Conference Series, Vol. 9144, Space Telescopes and Instrumentation 2014: Ultraviolet to Gamma Ray, ed. T.~{Takahashi}, J.-W.~A. {den Herder}, \& M.~{Bautz}, 91441S, \dodoi{10.1117/12.2062667}

\bibitem[{{Smith} {et~al.}(2011){Smith}, {Li}, {Filippenko}, \& {Chornock}}]{smith_2011}
{Smith}, N., {Li}, W., {Filippenko}, A.~V., \& {Chornock}, R. 2011, \mnras, 412, 1522, \dodoi{10.1111/j.1365-2966.2011.17229.x}

\bibitem[{{Tandon} {et~al.}(2020){Tandon}, {Postma}, {Joseph}, {Devaraj}, {Subramaniam}, {Barve}, {George}, {Ghosh}, {Girish}, {Hutchings}, {Kamath}, {Kathiravan}, {Kumar}, {Lancelot}, {Leahy}, {Mahesh}, {Mohan}, {Nagabhushana}, {Pati}, {Rao}, {Sankarasubramanian}, {Sriram}, \& {Stalin}}]{Tandon_2020}
{Tandon}, S.~N., {Postma}, J., {Joseph}, P., {et~al.} 2020, \aj, 159, 158, \dodoi{10.3847/1538-3881/ab72a3}

\bibitem[{{Tauris} {et~al.}(2015){Tauris}, {Langer}, \& {Podsiadlowski}}]{Tauris2015}
{Tauris}, T., {Langer}, N., \& {Podsiadlowski}, P. 2015, arXiv e-prints, arXiv:1510.07874, \dodoi{10.48550/arXiv.1510.07874}

\bibitem[{{Tauris} {et~al.}(2012){Tauris}, {Langer}, \& {Kramer}}]{Tauris2012}
{Tauris}, T.~M., {Langer}, N., \& {Kramer}, M. 2012, \mnras, 425, 1601, \dodoi{10.1111/j.1365-2966.2012.21446.x}

\bibitem[{{Tauris} {et~al.}(2013){Tauris}, {Langer}, {Moriya}, {Podsiadlowski}, {Yoon}, \& {Blinnikov}}]{Tauris2013}
{Tauris}, T.~M., {Langer}, N., {Moriya}, T.~J., {et~al.} 2013, \apjl, 778, L23, \dodoi{10.1088/2041-8205/778/2/L23}

\bibitem[{{Tauris} \& {Sennels}(2000)}]{Tauris_2000}
{Tauris}, T.~M., \& {Sennels}, T. 2000, \aap, 355, 236, \dodoi{10.48550/arXiv.astro-ph/9909149}

\bibitem[{{Tauris} {et~al.}(2017){Tauris}, {Kramer}, {Freire}, {Wex}, {Janka}, {Langer}, {Podsiadlowski}, {Bozzo}, {Chaty}, {Kruckow}, {van den Heuvel}, {Antoniadis}, {Breton}, \& {Champion}}]{tauris_2017}
{Tauris}, T.~M., {Kramer}, M., {Freire}, P.~C.~C., {et~al.} 2017, \apj, 846, 170, \dodoi{10.3847/1538-4357/aa7e89}

\bibitem[{{Titus} {et~al.}(2019){Titus}, {Stappers}, {Morello}, {Caleb}, {Filipovi{\'c}}, {}, {McBride}, {Ho}, \& {Buckley}}]{Titus_2019}
{Titus}, N., {Stappers}, B.~W., {Morello}, V., {et~al.} 2019, \mnras, 487, 4332, \dodoi{10.1093/mnras/stz1578}

\bibitem[{{Toonen} {et~al.}(2018){Toonen}, {Perets}, {Igoshev}, {Michaely}, \& {Zenati}}]{Toonen_2018}
{Toonen}, S., {Perets}, H.~B., {Igoshev}, A.~P., {Michaely}, E., \& {Zenati}, Y. 2018, \aap, 619, A53, \dodoi{10.1051/0004-6361/201833164}

\bibitem[{{Udalski} {et~al.}(1997){Udalski}, {Kubiak}, \& {Szymanski}}]{ogle_ii}
{Udalski}, A., {Kubiak}, M., \& {Szymanski}, M. 1997, \actaa, 47, 319, \dodoi{10.48550/arXiv.astro-ph/9710091}

\bibitem[{{Udalski} {et~al.}(1992){Udalski}, {Szymanski}, {Kaluzny}, {Kubiak}, \& {Mateo}}]{ogle_1}
{Udalski}, A., {Szymanski}, M., {Kaluzny}, J., {Kubiak}, M., \& {Mateo}, M. 1992, \actaa, 42, 253

\bibitem[{{Udalski} {et~al.}(2015){Udalski}, {Szyma{\'n}ski}, \& {Szyma{\'n}ski}}]{ogle_iv}
{Udalski}, A., {Szyma{\'n}ski}, M.~K., \& {Szyma{\'n}ski}, G. 2015, \actaa, 65, 1, \dodoi{10.48550/arXiv.1504.05966}

\bibitem[{{Valli} {et~al.}(2025){Valli}, {de Mink}, {Justham}, {Callister}, {Johnston}, {Kresse}, {Langer}, {Rubio}, {Vigna-G{\'o}mez}, \& {Wang}}]{Valli_2025}
{Valli}, R., {de Mink}, S.~E., {Justham}, S., {et~al.} 2025, arXiv e-prints, arXiv:2505.08857, \dodoi{10.48550/arXiv.2505.08857}

\bibitem[{{van Kerkwijk} \& {Kulkarni}(1999)}]{vanKerkwijk_1999}
{van Kerkwijk}, M.~H., \& {Kulkarni}, S.~R. 1999, \apjl, 516, L25, \dodoi{10.1086/311991}

\bibitem[{{Vigna-G{\'o}mez} {et~al.}(2018){Vigna-G{\'o}mez}, {Neijssel}, {Stevenson}, {Barrett}, {Belczynski}, {Justham}, {de Mink}, {M{\"u}ller}, {Podsiadlowski}, {Renzo}, {Sz{\'e}csi}, \& {Mandel}}]{Vigna_2018}
{Vigna-G{\'o}mez}, A., {Neijssel}, C.~J., {Stevenson}, S., {et~al.} 2018, \mnras, 481, 4009, \dodoi{10.1093/mnras/sty2463}

\bibitem[{{Vink}(2017)}]{Vink2017}
{Vink}, J.~S. 2017, \aap, 607, L8, \dodoi{10.1051/0004-6361/201731902}

\bibitem[{{von Zeipel}(1924)}]{von_Zeipel_1924}
{von Zeipel}, H. 1924, \mnras, 84, 665, \dodoi{10.1093/mnras/84.9.665}

\bibitem[{{Wilson}(1990)}]{Wilson_1990}
{Wilson}, R.~E. 1990, \apj, 356, 613, \dodoi{10.1086/168867}

\bibitem[{{Wu} \& {Fuller}(2022)}]{Wu_2022}
{Wu}, S.~C., \& {Fuller}, J. 2022, \apjl, 940, L27, \dodoi{10.3847/2041-8213/ac9b3d}

\bibitem[{{Wu} {et~al.}(2020){Wu}, {Chen}, {Chen}, {Li}, \& {Han}}]{Wu_2020}
{Wu}, Y., {Chen}, X., {Chen}, H., {Li}, Z., \& {Han}, Z. 2020, \aap, 634, A126, \dodoi{10.1051/0004-6361/201935792}

\bibitem[{{Wu} {et~al.}(2018){Wu}, {Chen}, {Li}, \& {Han}}]{Wu_2018}
{Wu}, Y., {Chen}, X., {Li}, Z., \& {Han}, Z. 2018, \aap, 618, A14, \dodoi{10.1051/0004-6361/201832686}

\bibitem[{{Yang} {et~al.}(2025){Yang}, {Han}, {Zhou}, {Jing}, {Chen}, {Wang}, {Li}, {Wang}, {Wang}, {Ge}, {Guo}, {Li}, {Shao}, {Liu}, {Su}, {Hou}, {Huang}, {Jiang}, {Jiang}, {Sun}, {Wang}, {Wang}, {Wang}, {Wang}, {Wang}, {Wang}, {Wang}, {Xu}, {Xu}, {Xu}, {Yan}, {Yan}, {You}, {Yu}, {Yuan}, \& {Zhang}}]{Yang_2025}
{Yang}, Z.~L., {Han}, J.~L., {Zhou}, D.~J., {et~al.} 2025, Science, 388, 859, \dodoi{10.1126/science.ado0769}

\bibitem[{{Yungelson} {et~al.}(2024){Yungelson}, {Kuranov}, {Postnov}, {Kuranova}, {Oskinova}, \& {Hamann}}]{Yungelson_2024}
{Yungelson}, L., {Kuranov}, A., {Postnov}, K., {et~al.} 2024, \aap, 683, A37, \dodoi{10.1051/0004-6361/202347806}

\bibitem[{{Zapartas} {et~al.}(2026){Zapartas}, {Fox}, {Su}, {Souropanis}, {Drout}, {Rocha}, {van Dyk}, {Williams}, {Briel}, {Renzo}, {Andrews}, {Fragos}, {Gossage}, {Kruckow}, {Liotine}, {Ryder}, {Srivastava}, \& {Teng}}]{Zapartas_2026}
{Zapartas}, E., {Fox}, O.~D., {Su}, J., {et~al.} 2026, \mnras, 546, staf2208, \dodoi{10.1093/mnras/staf2208}

\bibitem[{Zaritsky {et~al.}(1997)Zaritsky, Harris, \& Thompson}]{MCPS}
Zaritsky, D., Harris, J., \& Thompson, I. 1997, arXiv preprint astro-ph/9706206

\bibitem[{{Zaritsky} {et~al.}(1996){Zaritsky}, {Schectman}, \& {Bredthauer}}]{zaritsky_1996}
{Zaritsky}, D., {Schectman}, S.~A., \& {Bredthauer}, G. 1996, \pasp, 108, 104, \dodoi{10.1086/133698}

\bibitem[{Zivick {et~al.}(2017)Zivick, Kallivayalil, Linden, Fritz, Besla, van~der Marel, Koz{\l}owski, Geha, Kochanek, Anderson, {et~al.}}]{zivick_2017}
Zivick, P., Kallivayalil, N., Linden, S., {et~al.} 2017, Proceedings of the International Astronomical Union, 13, 394

\end{thebibliography}
\end{document}